\documentclass[namedreferences]{solarphysics}
\usepackage[hyperref,optionalrh,solaromanenum]{spr-sola-addons} 
\usepackage{graphicx}                    
\usepackage{color}                       
\usepackage{breakurl}                         
\usepackage{hyperref}

\begin{document}

\begin{article}

\begin{opening}

\title{Critical Science Plan for the Daniel K. Inouye Solar Telescope (DKIST)}

%
\author[addressref={aff1,aff2},corref,email={mark.rast@lasp.colorado.edu}]{\inits{M.P. }\fnm{Mark P. }\lnm{Rast}\orcid{0000-0002-9232-9078}}
\author[addressref={aff3},corref,email={}]{\inits{N. }\fnm{Nazaret }\lnm{ Bello Gonz\'alez}}
\author[addressref={aff4},corref,email={}]{\inits{L. }\fnm{Luis }\lnm{Bellot Rubio}\orcid{0000-0001-8669-8857}}
\author[addressref={aff5},corref,email={}]{\inits{W. }\fnm{Wenda }\lnm{Cao}\orcid{0000-0003-2427-6047}}
\author[addressref={aff6},corref,email={}]{\inits{G. }\fnm{Gianna }\lnm{Cauzzi}\orcid{0000-0002-6116-7301}}
\author[addressref={aff7},corref,email={}]{\inits{E. }\fnm{Edward }\lnm{DeLuca}\orcid{0000-0001-7416-2895}}
\author[addressref={aff8,aff9},corref,email={}]{\inits{B. }\fnm{Bart }\lnm{De Pontieu}\orcid{0000-0002-8370-952X}}
\author[addressref={aff10,aff9},corref,email={}]{\inits{L. }\fnm{Lyndsay }\lnm{Fletcher}\orcid{0000-0001-9315-7899}}
\author[addressref={aff11},corref,email={}]{\inits{S.E. }\fnm{Sarah E. }\lnm{Gibson}\orcid{0000-0001-9831-2640}}
\author[addressref={aff11},corref,email={}]{\inits{P.G. }\fnm{Philip G. }\lnm{Judge}\orcid{0000-0001-5174-0568}}
\author[addressref={aff12,aff13},corref,email={}]{\inits{Y. }\fnm{Yukio }\lnm{Katsukawa}\orcid{0000-0002-5054-8782}}
\author[addressref={aff1,aff6},corref,email={}]{\inits{M.D. }\fnm{Maria D. }\lnm{Kazachenko}\orcid{0000-0001-8975-7605}}
\author[addressref={aff14},corref,email={}]{\inits{E. }\fnm{Elena }\lnm{Khomenko}\orcid{0000-0003-3812-620X}}
\author[addressref={aff15},corref,email={}]{\inits{E. }\fnm{Enrico }\lnm{Landi}\orcid{0000-0002-9325-9884}}
\author[addressref={aff6},corref,email={}]{\inits{V. }\fnm{Valentin }\lnm{Mart\'inez Pillet}\orcid{0000-0001-7764-6895}}
\author[addressref={aff6},corref,email={}]{\inits{G.J.D. }\fnm{Gordon J.D. }\lnm{Petrie}\orcid{0000-0001-8462-9161}}
\author[addressref={aff16},corref,email={}]{\inits{J. }\fnm{Jiong }\lnm{Qiu}\orcid{}}
\author[addressref={aff17},corref,email={}]{\inits{L.A. }\fnm{Laurel A. }\lnm{Rachmeler}\orcid{0000-0002-3770-009X}}
\author[addressref={aff11},corref,email={}]{\inits{M. }\fnm{Matthias }\lnm{Rempel}\orcid{0000-0001-5850-3119}}
\author[addressref={aff3},corref,email={}]{\inits{W. }\fnm{Wolfgang }\lnm{Schmidt}\orcid{}}
\author[addressref={aff18},corref,email={}]{\inits{E. }\fnm{Eamon }\lnm{Scullion}\orcid{0000-0001-9590-6427}}
\author[addressref={aff19},corref,email={}]{\inits{X. }\fnm{Xudong }\lnm{Sun}\orcid{0000-0003-4043-616X}}
\author[addressref={aff20},corref,email={}]{\inits{B.T. }\fnm{Brian T. }\lnm{Welsch}\orcid{0000-0003-2244-641X}}
\author[addressref={aff21},corref,email={}]{\inits{V. }\fnm{Vincenzo }\lnm{Andretta}\orcid{0000-0003-1962-9741}}
\author[addressref={aff18},corref,email={}]{\inits{P. }\fnm{Patrick }\lnm{Antolin}\orcid{0000-0003-1529-4681}}
\author[addressref={aff22},corref,email={}]{\inits{T.R. }\fnm{Thomas R. }\lnm{Ayres}\orcid{0000-0002-1242-5124}}
\author[addressref={aff23},corref,email={}]{\inits{K.S. }\fnm{K.S. }\lnm{Balasubramaniam}\orcid{}}
\author[addressref={aff24},corref,email={}]{\inits{I. }\fnm{Istvan }\lnm{Ballai}\orcid{0000-0002-3066-7653}}
\author[addressref={aff2},corref,email={}]{\inits{T.E. }\fnm{Thomas E. }\lnm{Berger}\orcid{}}
\author[addressref={aff25},corref,email={}]{\inits{S.J. }\fnm{Stephen J. }\lnm{Bradshaw}\orcid{0000-0002-3300-6041}}
\author[addressref={aff9},corref,email={}]{\inits{M. }\fnm{Mats }\lnm{Carlsson}\orcid{0000-0003-3079-8095}}
\author[addressref={aff11},corref,email={}]{\inits{R. }\fnm{Roberto }\lnm{Casini}\orcid{0000-0001-6990-513X}}
\author[addressref={aff11},corref,email={}]{\inits{R. }\fnm{Rebecca }\lnm{Centeno}\orcid{0000-0002-1327-1278}}
\author[addressref={aff1,aff2},corref,email={}]{\inits{S.R. }\fnm{Steven R. }\lnm{Cranmer}\orcid{0000-0002-3699-3134}}
\author[addressref={aff26},corref,email={}]{\inits{C. }\fnm{Craig }\lnm{DeForest}\orcid{0000-0002-7164-2786}}
\author[addressref={aff27},corref,email={}]{\inits{Y. }\fnm{Yuanyong }\lnm{Deng}\orcid{}}
\author[addressref={aff24},corref,email={}]{\inits{R. }\fnm{Robertus }\lnm{Erd\'elyi}\orcid{0000-0003-3439-4127}}
\author[addressref={aff28},corref,email={}]{\inits{V. }\fnm{Viktor }\lnm{Fedun}\orcid{}}
\author[addressref={aff3},corref,email={}]{\inits{C.E. }\fnm{Catherine E. }\lnm{Fischer}\orcid{0000-0001-9352-3027}}
\author[addressref={aff29},corref,email={}]{\inits{S.J. }\fnm{Sergio J. }\lnm{Gonz\'alez Manrique}\orcid{0000-0002-6546-5955}}
\author[addressref={aff30},corref,email={}]{\inits{M. }\fnm{Michael }\lnm{Hahn}\orcid{0000-0001-7748-4179}}
\author[addressref={aff31},corref,email={}]{\inits{L. }\fnm{Louise }\lnm{Harra}\orcid{0000-0001-9457-6200}}
\author[addressref={aff9},corref,email={}]{\inits{V.M.J. }\fnm{Vasco M.J. }\lnm{Henriques}\orcid{0000-0002-4024-7732}}
\author[addressref={aff8},corref,email={}]{\inits{N.E. }\fnm{Neal E. }\lnm{Hurlburt}\orcid{0000-0003-3323-7488}}
\author[addressref={aff6},corref,email={}]{\inits{S. }\fnm{Sarah }\lnm{Jaeggli}\orcid{0000-0001-5459-2628}}
\author[addressref={aff9},corref,email={}]{\inits{S. }\fnm{Shahin }\lnm{Jafarzadeh}\orcid{0000-0002-7711-5397}}
\author[addressref={aff24},corref,email={}]{\inits{R. }\fnm{Rekha }\lnm{Jain}\orcid{}}
\author[addressref={aff32},corref,email={}]{\inits{S.M. }\fnm{Stuart M. }\lnm{Jefferies}\orcid{}}
\author[addressref={aff33},corref,email={}]{\inits{P.H. }\fnm{Peter H. }\lnm{Keys}\orcid{0000-0003-1400-8356}}
\author[addressref={aff1,aff6},corref,email={}]{\inits{A.F. }\fnm{Adam F. }\lnm{Kowalski}\orcid{0000-0001-7458-1176}}
\author[addressref={aff34},corref,email={}]{\inits{C. }\fnm{Christoph }\lnm{Kuckein}\orcid{0000-0002-3242-1497}}
\author[addressref={aff19},corref,email={}]{\inits{J.R. }\fnm{Jeffrey R. }\lnm{Kuhn}\orcid{0000-0003-1361-9104}}
\author[addressref={aff33},corref,email={}]{\inits{J. }\fnm{Jiajia }\lnm{Liu}\orcid{0000-0003-2569-1840}}
\author[addressref={aff8,aff35},corref,email={}]{\inits{W. }\fnm{Wei }\lnm{Liu}\orcid{0000-0001-8794-3420}}
\author[addressref={aff16},corref,email={}]{\inits{D. }\fnm{Dana }\lnm{Longcope}\orcid{}}
\author[addressref={aff36},corref,email={}]{\inits{R.T.J. }\fnm{R.T. James }\lnm{McAteer}\orcid{0000-0003-1493-101X}}
\author[addressref={aff11},corref,email={}]{\inits{S.W. }\fnm{Scott W. }\lnm{McIntosh}\orcid{0000-0002-7369-1776}}
\author[addressref={aff37},corref,email={}]{\inits{D.E. }\fnm{David E. }\lnm{McKenzie}\orcid{}}
\author[addressref={aff7},corref,email={}]{\inits{M.P. }\fnm{Mari Paz }\lnm{Miralles}\orcid{}}
\author[addressref={aff18},corref,email={}]{\inits{R.J. }\fnm{Richard J. }\lnm{Morton}\orcid{0000-0001-5678-9002}}
\author[addressref={aff38,aff39},corref,email={}]{\inits{K. }\fnm{Karin }\lnm{Muglach}\orcid{0000-0002-5547-9683}}
\author[addressref={aff33},corref,email={}]{\inits{C.J. }\fnm{Chris J. }\lnm{Nelson}\orcid{0000-0003-1400-8356}}
\author[addressref={aff8,aff35},corref,email={}]{\inits{N.K. }\fnm{Navdeep K. }\lnm{Panesar}\orcid{0000-0001-7620-362X}}
\author[addressref={aff40},corref,email={}]{\inits{S. }\fnm{Susanna }\lnm{Parenti}\orcid{0000-0003-1438-1310}}
\author[addressref={aff41},corref,email={}]{\inits{C.E. }\fnm{Clare E. }\lnm{Parnell}\orcid{0000-0002-5694-9069}}
\author[addressref={aff42},corref,email={}]{\inits{B. }\fnm{Bala }\lnm{Poduval}\orcid{0000-0003-1258-0308}}
\author[addressref={aff6},corref,email={}]{\inits{K.P.. }\fnm{Kevin P. }\lnm{Reardon}\orcid{0000-0001-8016-0001}}
\author[addressref={aff43},corref,email={}]{\inits{J.W. }\fnm{Jeffrey W. }\lnm{Reep}\orcid{0000-0003-4739-1152}}
\author[addressref={aff6},corref,email={}]{\inits{T.A. }\fnm{Thomas A. }\lnm{Schad}\orcid{0000-0002-7451-9804}}
\author[addressref={aff2},corref,email={}]{\inits{D. }\fnm{Donald }\lnm{Schmit}\orcid{0000-0002-9654-0815}}
\author[addressref={aff44,aff18},corref,email={}]{\inits{R. }\fnm{Rahul }\lnm{Sharma}\orcid{0000-0002-0197-9041}}
\author[addressref={aff14},corref,email={}]{\inits{H. }\fnm{Hector }\lnm{Socas-Navarro}\orcid{0000-0001-9896-4622}}
\author[addressref={aff45},corref,email={}]{\inits{A.K. }\fnm{Abhishek K. }\lnm{Srivastava}\orcid{}}
\author[addressref={aff37},corref,email={}]{\inits{A.C. }\fnm{Alphonse C. }\lnm{Sterling}\orcid{0000-0003-1281-897X}}
\author[addressref={aff13},corref,email={}]{\inits{Y. }\fnm{Yoshinori }\lnm{Suematsu}\orcid{0000-0003-4452-858X}}
\author[addressref={aff6},corref,email={}]{\inits{L.A. }\fnm{Lucas A. }\lnm{Tarr}\orcid{0000-0002-8259-8303}}
\author[addressref={aff8,aff35},corref,email={}]{\inits{S. }\fnm{Sanjiv }\lnm{Tiwari}\orcid{0000-0001-7817-2978}}
\author[addressref={aff6},corref,email={}]{\inits{A. }\fnm{Alexandra }\lnm{Tritschler}\orcid{0000-0003-3147-8026}}
\author[addressref={aff24},corref,email={}]{\inits{G. }\fnm{Gary }\lnm{Verth}\orcid{}}
\author[addressref={aff46},corref,email={}]{\inits{A. }\fnm{Angelos }\lnm{Vourlidas}\orcid{0000-0002-8164-5948}}
\author[addressref={aff5},corref,email={}]{\inits{H. }\fnm{Haimin }\lnm{Wang}\orcid{0000-0002-5233-565X}}
\author[addressref={aff43},corref,email={}]{\inits{Y.-M. }\fnm{Yi-Ming  }\lnm{Wang}\orcid{0000-0002-3527-5958}}
\author[addressref={aff47},corref,email={}]{\inits{}\fnm{}\lnm{NSO, DKIST project, and DKIST instrument scientists}}
\author[addressref={aff47},corref,email={}]{\inits{}\fnm{}\lnm{the DKIST Science Working Group}}
\author[addressref={aff47},corref,email={}]{\inits{}\fnm{}\lnm{the DKIST Critical Science Plan Community}}
%
\runningauthor{Rast, M.P. et al.}
\runningtitle{DKIST Critical Science Plan}

\address[id=aff1]{Department of Astrophysical and Planetary Sciences, University of Colorado, Boulder CO 80309, USA}
\address[id=aff2]{Laboratory for Atmospheric and Space Physics, University of Colorado, Boulder CO 80303, USA}
\address[id=aff3]{Kiepenheuer-Institut f\"ur Sonnenphysik, 79104 Freiburg, Germany}
\address[id=aff4]{Instituto de Astrof\'isica de Andaluc\'ia (CSIC), E-18080 Granada, Spain}
\address[id=aff5]{Big Bear Solar Observatory, New Jersey Institute of Technology, Big Bear City, CA 92314, USA}
\address[id=aff6]{National Solar Observatory, Boulder CO 80303, USA}
\address[id=aff7]{Harvard-Smithsonian Center for Astrophysics, Cambridge, MA 02138, USA}
\address[id=aff8]{Lockheed Martin Solar \& Astrophysics Laboratory, Palo Alto, CA 94304, USA}
\address[id=aff9]{Institute of Theoretical Astrophysics, Rosseland Centre for Solar Physics, University of Oslo, N-0315 Oslo, Norway}
\address[id=aff10]{School of Physics and Astronomy, University of Glasgow, Glasgow, G12 8QQ, UK}
\address[id=aff11]{High Altitude Observatory, National Center for Atmospheric Research, Boulder CO 80307, USA}
\address[id=aff12]{Department of Astronomical Science, School of Physical Sciences, The Graduate University for Advanced Studies, SOKENDAI, Tokyo 181-8588, Japan}
\address[id=aff13]{National Astronomical Observatory of Japan, National Institutes of Natural Science, Tokyo 181-8588, Japan}
\address[id=aff14]{Instituto de Astrof\'isica de Canarias, 38205 La Laguna, Tenerife, Spain}
\address[id=aff15]{Department of Climate and Space Sciences and Engineering, University of Michigan, Ann Arbor, MI 48109, USA}
\address[id=aff16]{Department of Physics, Montana State University, Bozeman, MT 59717, USA}
\address[id=aff17]{NOAA National Centers for Environmental Information, Boulder, CO 80305, USA}
\address[id=aff18]{Department of Mathematics, Physics and Electrical Engineering, Northumbria University, Newcastle upon Tyne NE1 8ST, UK}
\address[id=aff19]{Institute for Astronomy, University of Hawaii, Pukalani, HI 96768, USA}
\address[id=aff20]{Department of Natural and Applied Sciences, University of Wisconsin, Green Bay, WI 54311, USA}
\address[id=aff21]{INAF - Osservatorio Astronomico di Capodimonte, 80131 Naples, Italy}
\address[id=aff22]{Center for Astrophysics and Space Astronomy, University of Colorado, Boulder CO 80309, USA}
\address[id=aff23]{AFRL Battlespace Environment Division, Albuquerque, NM 87117, USA}
\address[id=aff24]{School of Mathematics and Statistics, The University of Sheffield, Sheffield, S3 7RH, United Kingdom}
\address[id=aff25]{Department of Physics and Astronomy, Rice University, Houston, TX 77005, USA}
\address[id=aff26]{Southwest Research Institute, Boulder, CO 80302, USA}
\address[id=aff27]{Key Laboratory of Solar Activity, National Astronomical Observatories, Chinese Academy of Sciences, Beijing 100012, People's Republic of China}
\address[id=aff28]{Department of Automatic Control and Systems Engineering, The University of Sheffield, Sheffield, S1 3JD, United Kingdom}
\address[id=aff29]{Astronomical Institute of the Slovak Academy of Sciences, 05960 Tatransk\'a Lomnica, Slovak Republic}
\address[id=aff30]{Columbia Astrophysics Laboratory, Columbia University, New York, NY 10027, USA}
\address[id=aff31]{PMOD/WRC, 7260 Davos Dorf, Switzerland}
\address[id=aff32]{Department of Physics and Astronomy, Georgia State University, GA 30303, USA}
\address[id=aff33]{Astrophysics Research Centre, School of Mathematics and Physics, Queen's University, Belfast, BT7 1NN, UK}
\address[id=aff34]{Leibniz-Institut f\"ur Astrophysik, 14482 Potsdam, Germany}
\address[id=aff35]{Bay Area Environmental Research Institute, NASA Research Park, Moffett Field, CA 94035, USA}
\address[id=aff36]{Department of Astronomy, New Mexico State University, NM 88003, USA}
\address[id=aff37]{NASA Marshall Space Flight Center, Huntsville, AL 35812, USA}
\address[id=aff38]{Catholic University of America, Washington, DC 20064, USA}
\address[id=aff39]{NASA Goddard Space Flight Center, Greenbelt, MD 20771, USA}
\address[id=aff40]{Institut d'Astrophysique Spatiale, CNRS, Univ. Paris-Sud, Universit\'e Paris-Saclay, 91405, Orsay, France}
\address[id=aff41]{School of Mathematics and Statistics, University of St Andrews, St Andrews, Fife, KY16 9SS, UK}
\address[id=aff42]{Space Science Center, University of New Hampshire, Durham, NH 03824, USA}
\address[id=aff43]{Space Science Division, Naval Research Laboratory, Washington, DC 20375, USA}
\address[id=aff44]{Departamento de F\'isica y Matem\'aticas, Universidad de Alcal\'a, 28871 Alcal\'a de Henares, Madrid, Spain}
\address[id=aff45]{Department of Physics, Indian Institute of Technology (BHU), Varanasi-221005, India}
\address[id=aff46]{Applied Physics Laboratory, Johns Hopkins University, Laurel, MD 20723, USA}
\address[id=aff47]{see https://www.nso.edu/telescopes/dki-solar-telescope-2-2/csp/}

\begin{abstract}
The Daniel K. Inouye Solar Telescope (DKIST) will revolutionize our ability to measure, understand and model the basic physical processes that control the structure and dynamics of the Sun and its atmosphere.  The first-light DKIST images, released publicly on 29 January 2020, only hint at the extraordinary capabilities which will accompany full commissioning of the five facility instruments.  With this Critical Science Plan (CSP) we attempt to anticipate some of what those capabilities will enable, providing a  snapshot of some of the scientific pursuits that the Daniel K. Inouye Solar Telescope hopes to engage as start-of-operations nears.  The work builds on the combined contributions of the DKIST Science Working Group (SWG) and CSP Community members$^{46}$, who generously shared their experiences, plans, knowledge and dreams.  Discussion is primarily focused on those issues to which DKIST will uniquely contribute.
\end{abstract}

\vfill\eject
%
\keywords{solar photosphere, chromosphere, corona}

\end{opening}

%
 \section{Introduction}
 \label{s:intro}
The primary goal of the Daniel K. Inouye Solar Telescope (DKIST) is to address long-standing problems in solar physics, such as the operation of the solar dynamo and the heating and acceleration of the solar chromospheric and coronal plasma, but its scientific impact will extend well beyond the Sun. 
DKIST data will contribute to our understanding of fundamental physical processes, such as the generation and annihilation of magnetic field in plasmas of very high electrical conductivity, the role of turbulence under extreme conditions not achievable in terrestrial laboratories and the quantum mechanical underpinnings of polarization spectroscopy essential to the interpretation of a broad range of astrophysical observations. 
The anticipated high spatial and temporal resolution high-precision spectropolarimetric observations of the continuously reorganizing and reconfiguring solar magnetic field will allow detailed study of the underlying impulsive energy release and particle acceleration mechanisms responsible for the formation of particle beams and plasma ejecta. These processes are ubiquitous in astrophysics, critical to the stability of laboratory plasmas and directly impact our ability to robustly extend human technology into the Earth's space environment. 
 
With a post-focus suite of five instruments (see W\"oger et al. 2020, DeWijn et al. 2020, Jaeggli et al. 2020, Fehlmann et al. 2020 and von der L\"uhe et al. 2020 in this Topical Volume), the DKIST's novel capabilities come with extreme flexibility and consequent complexity. 
Significant effort is required to understand how to best leverage that flexibility to achieve the rich scientific goals uniquely accessible by DKIST soon after the start of operations.  The strategy of the National Solar Observatory (NSO) has been to actively engage a large cross-section of the US and international solar and space physics community in defining these goals and how to achieve them.  This was done not only to expand the range of science to be pursued, but also to ensure that the early critical science can indeed be addressed using the DKIST telescope and the anticipated post-focus instrument suite. 

At the heart of the DKIST Critical Science Plan, described here, are scientific goals formulated by the DKIST Science Working Group after considering the Science Use Cases contributed by the community via an Atlassian\textsuperscript{\textregistered} Jira\textsuperscript{\textregistered} development interface.  Science Use Case development was partially facilitated by a series of Critical Science Plan Workshops hosted jointly by the NSO and community partners (\url{https://www.nso.edu/telescopes/dki-solar-telescope/csp/dkist-csp-workshops/}). Though participation in those workshops was not mandatory for Science Use Case development, they served as an efficient way to acquaint the community with the telescope and instrument capabilities, the range of operational possibilities and some of the challenges unique to ground-based observing.  In turn, the DKIST project acquired a sense of the range and popularity of different observing configurations and an understanding of what will be required to meet the critical early scientific objectives.

This DKIST Critical Science Plan (CSP) is a snapshot of some of the scientific pursuits that the Daniel K. Inouye Solar Telescope hopes to enable as start-of-operations nears.  The first-light DKIST images,  
released publicly on 29 January 2020 (\url{https://www.nso.edu/inouye-solar-telescope-first-light/}), only hint at the extraordinary capabilities which will accompany full commissioning of the five facility instruments (see Rimmele et al. 2020 in this Topical Volume).  The CSP is an attempt to anticipate some of what those capabilities will enable.  The discussion is divided into four broad research areas:  Magnetoconvection and Dynamo processes, Flares and Eruptive Activity, Magnetic Connectivity through the Non-Eruptive Solar Atmosphere, and Long-Term Studies of the Sun, Special Topics and Broader Implications.  Each of these includes an introductory discussion followed more detailed exposition of specific research topics.  Discussion is primarily focused on issues to which DKIST will uniquely contribute.  References are necessarily incomplete; they are exempli gratia only, even where not explicitly so noted.

\section{Unique DKIST Capabilities}\label{s:dkist}

This section provides a very brief overview of the unique capabilities DKIST will contribute.  The full capabilities are discussed in detail in the accompanying papers in this Topical Volume (Rimmele et al. 2020, W\"oger et al. 2020, DeWijn et al. 2020, Jaeggli et al. 2020, Fehlmann et al. 2020, von der L\"uhe et al. 2020, Tritschler et al. 2020, Harrington et al. 2020, Davey et al. 2020).  

The DKIST primary mirror has a diameter of 4m, a size chosen so that small-scale plasma dynamics in the solar photosphere can be resolved while simultaneously making polarization measurements of weak magnetic fields.  The all-reflective clear-aperture off-axis optical configuration of the telescope allows broad wavelength access and minimizes scattered light, yielding the dynamic range sensitivity necessary for studies of the full solar atmosphere from the deep photosphere to 1.5 solar radii (0.5 solar radii above the photosphere).  

DKIST's first-light instrument suite consists of the Cryogenic Near-Infrared Spectro-Polarimeter (CryoNIRSP, Fehlmann et al. this volume), the Diffraction-Limited Near-Infrared Spectropolarimeter (DL-NIRSP, Jaeggli et al. this volume), the Visible Broadband Imager (VBI, W\"oger et al. this volume), the Visible Spectro-Polarimeter (ViSP, DeWijn et al. this volume), and the Visible Tunable Filter (VTF, von der L\"uhe et al. this volume).
These five instruments will observe over the wavelength range 380 to 5000 nm.  With the exception of CryoNIRSP, the instruments can be used individually or in combination, and will employ a common integrated high-order adaptive optics system (Rimmele et al. 2020 this volume) which enables diffraction-limited observations with resolution ranging from 0.02 arcseconds at the shortest wavelengths to 0.09 arcseconds at 1800 nm. The stand-alone cryogenically cooled CryoNIRSP, on the other hand, will make seeing-limited observations with 0.15 arcsecond critical spatial sampling along the spectrograph slit over the spectral range 1000 to 5000 nm.  Scattering by the Earth's atmosphere, and thus the background sky brightness, is greatly reduced at longer wavelengths making regular observations of the dim corona possible, and all instruments employ specially developed high-frame-rate large-format cameras capable of non-destructive detector reads to improve signal to noise with longer exposures.   Importantly, all instruments, with the exception of VBI (a narrow-band diffraction-limited filtergraph) are highly sensitive spectropolarimeters. Careful calibration of the instrument and telescope polarimetric contributions \citep[][Harrington et al. this volume, and references therein]{2019JATIS...5c8001H} will allow measurement of magnetic flux densities of less than 1G both on and off the solar disk.  

 \section{Magnetoconvection and Dynamo Processes}\label{s:mhd} 
 
Magnetic fields on the Sun are highly structured and multi-scale.  Field is found, not only on the scale of active regions and sunspots, but down to the smallest spatial scales observed.  The small-scale fields may have their origin on larger-scales as the endpoint of a turbulent cascade acting on field produced by a global-scale dynamo, or they may originate with small-scale motions, perhaps in the photosphere itself, as the result of a turbulent fluctuation dynamo (also called a small-scale or local dynamo).  Many questions remain about the field distributions observed, about how and why the field is organized as it is in the solar photosphere. 
 
On the largest scale, the global solar magnetic activity cycle reflects dynamo processes that operate with remarkable regularity.  The most conspicuous aspect of the global-scale dynamo is the cyclic appearance and disappearance of large regions of strong magnetic field organized into sunspots and active regions.  These come and go with the solar cycle on spatial and temporal scales that suggest an origin in the deep convection zone or in the overshoot region just below it.  But the solar convection zone is in a highly turbulent state and spans many pressure scale heights, with the density changing by a factor of over a million between the bottom and the top.  If formed near the bottom of the convection zone these magnetic structures must survive transit across this highly stratified and vigorously convecting region.  Alternatively active regions could re-form in the upper convection zone or photosphere itself.  Though the vertical scale height in the photosphere is much smaller than the characteristic size of an active region, sunspot or even a pore, some aspects of active region evolution support this possibility, with the local dynamics displaying a complex interplay between the magnetic field and magneto-convective motions at all stages including formation and dissolution.  
 
Sunspot umbrae are the most strongly magnetized regions of the Sun accessible to spectropolarimetric observations.  They are cool and dark, characterized by magnetic fields that are close to vertical with respect to the photosphere, and exhibit constrained columnar convective motions that are only poorly understood.  The surrounding penumbral field is highly inclined, with multiple interleaved magnetic components permeated by strong plasma flows.  It is unknown how this multicomponent sunspot structure arises.  For example, reproduction of the observed penumbral properties by state-of-the-art radiative magnetohydrodynamic simulations requires upper boundary conditions on the magnetic field whose counterparts on the Sun have yet to be identified.  DKIST's large photon flux and high spatial and temporal resolution are critical to advancing our understanding of sunspot dynamics.  
 
The quiet (non-active region) Sun is magnetized almost everywhere.  In the quietest regions (internetwork) the fields appear to be weak and highly inclined.  Determining the origin of these small-scale fields is difficult and DKIST will make critical contributions to this problem as well.  It will allow simultaneous observations in Zeeman-sensitive and Hanle-sensitive lines, to determine if the Hanle polarization signals measured in the internetwork regions of the Sun are consistent with the field deduced from the Zeeman diagnostics.  These comparisons will allow assessment of the amount of unresolved flux present as function of resolution, and careful exploitation of these techniques will lead to precise measurement of the internetwork field coverage and its spatial distribution as a function of global solar magnetic activity.  The resulting statistics and evolution of the smallest scale fields on the Sun can then be compared to those seen in dynamo simulations to untangle the signatures of its origin. 
 
Beyond its origin, knowledge of the small-scale magnetic field distribution on the Sun is key to assessing the role of photospherically generated waves in atmospheric heating.  Waves propagate from the photosphere upward into the atmosphere, guided by local magnetic structures.  Acoustic modes steepen, shock and dissipate in the chromosphere or undergo mode conversion and make it to greater heights. The underlying magnetic field distribution is critical to these processes and thus to the heat and momentum budget of the atmosphere.  Small-scale fields also play an important role in modulating the solar radiative output.  Small concentrations of magnetic field are sites of reduced gas pressure, density, and thus opacity.   An amalgam of small-scale field elements reduces the average temperature gradient of a region, locally changing the radiative output and thus contributing to variations in global solar spectral irradiance.  Many small-scale magnetic structures remain unresolved by current instruments, and changes in their size and contrast distributions with the solar cycle are unknown.  Since weakly magnetized internetwork regions cover the bulk of the solar surface, currently unresolved field elements may play a significant role in global spectral irradiance trends. DKIST will play a critical role in unraveling their contributions.

Several critical science topics in this research area are discussed in detail below, including 1) the formation, structure and dynamics of small-scale photospheric magnetic fields, 2) wave generation and propagation, 3) magneto-convective modulation of the solar luminosity and 4) active region evolution and sunspot fine structure.

 \subsection{Small-Scale Photospheric Magnetic Fields:  Formation, Structure, Dynamics}
 \label{ss:smallscale}
 
 \vskip -0.2in
 \begin{figure}[h]
\includegraphics[width=0.7\textwidth,clip=]{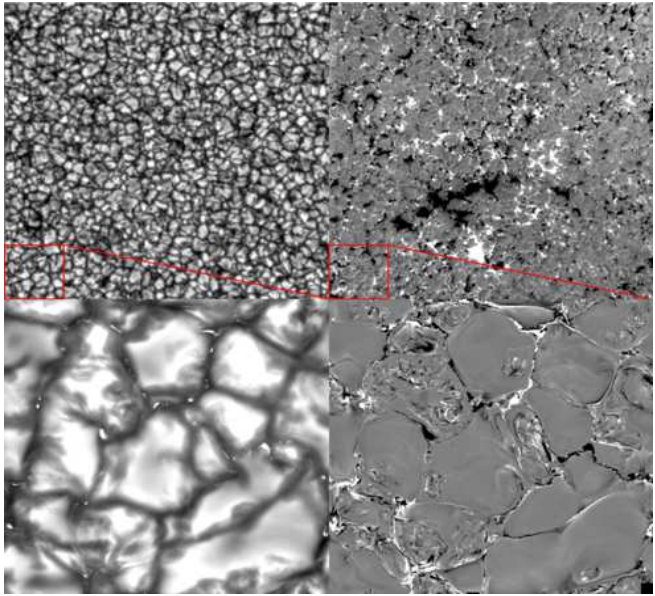}
\caption{Top: SST CRISP observation of the quiet Sun with 0.15" resolution ($\sim$100 km), displaying Stokes I and Stokes V.  Images courtesy L. Bellot Rubio (IAA-CSIC), S. Esteban Pozuelo (IAA-CSIC) and A. Ortiz (ITA, University of Oslo) \url{http://www.est-east.eu/est/index.php?option=com_content&view=article&id=774&lang=en?quiet-sun-magnetic-fields-2-2-2}.  Bottom: quiet Sun simulation with 4-km grid spacing (comparable to DKIST effective resolution of ~20 km) displaying bolometric  intensity  and  vertical  magnetic  field strength (in the range $+/-$ 200 G) at optical depth 0.1.  Simulation from~\cite{2014ApJ...789..132R}.}
\label{f:fig3}
\end{figure}

\noindent
{\it Is there a small scale dynamo operating in the solar photosphere?  What is the relative importance of local amplification, flux emergence, flux cancelation and non-local transport in small scale field evolution?  How turbulent are granular flows?  How dependent is quiet-sun magnetism on the solar cycle?}
\vskip 0.1in
\noindent
Convection in the solar photosphere is driven by rapid radiative losses from a thin layer that is less than a pressure scale height in vertical extent.  In this layer, the internal, kinetic, and magnetic energy of the plasma are all comparable (within a factor of a few).  This leads to vigorous dynamics with strong coupling between these three energy reservoirs.  While the solar convection zone is generally believed to be a highly turbulent medium, this is less than obvious in observations of the solar photosphere~\citep{1997A&A...328..229N}.  Plasma that is transported upward into the photosphere, and is observed as bright hot granules, has undergone significant horizontal expansion and is expected to show nearly laminar flow structure even at DKIST resolution.  A higher degree of turbulence may be found in the intergranular downflow lanes, but in these regions it likely develops as an advected instability at the interface between upflows and downflows as the fluid moves downward out of the photospheric boundary layer.  Turbulence may thus only be apparent in high spatial resolution measurements of the deep photosphere, and then only with the anticipated high contrast achievable by DKIST.

Small-scale magnetic field is ubiquitous on the solar surface (Figure~\ref{f:fig3}) and, in terms of unsigned flux density, exceeds the total active region magnetic field at all phases of the solar cycle~\citep[e.g.,][]{2008ApJ...672.1237L, 2009SSRv..144..275D, 2011ApJ...737...52L, 2019LRSP...16....1B}.  The quiet-Sun magnetic field in the solar photosphere has an average strength of about one third equipartition with the kinetic energy of the flows, about 600 G.  Since most of the field is concentrated in downflow lanes, the field strength reaches much higher local values and strong feedback on the convection is expected, but a detailed understanding of the interaction between the flow and small-scale turbulent magnetic field, and the consequent potential suppression of turbulence within intergranular downflows, is still elusive due to the resolution of current observations.  This has broader significance, as the development of turbulence in the shear layers of intergranular downflow lanes may contribute to spectral line broadening (\S\ref{ss:trp}).  
Furthermore, even in the quiet-Sun, a small fraction of the magnetic field is locally amplified to a strength of a few kilogauss.  Only the largest of the kilogauss elements are resolved with current instrumentation.  DKIST will allow dynamical studies of the formation and evolution of individual kilogauss elements and detailed statistical analysis of the quiet-Sun flux concentration size and strength distributions and their solar cycle dependencies.  Since modulation of the solar radiative output by magnetic flux elements is the primary cause of solar irradiance variations, these studies will be critically important in that context (see \S\ref{ss:irradiance} below).

Numerous small-scale magnetic loop structures are also found inside regions that are unipolar on larger scales, including 
both active region plage~\citep{2016ApJ...820L..13W, 2019ApJ...885...34W} and coronal holes~\citep{2016ApJ...818..203W}.  While difficult 
to discern in magnetograms, where the minority-polarity signatures are often absent, they are readily apparent in EUV images, 
have horizontal scales of $\sim$2--5 Mm, and evolve on time scales of minutes.  
In SDO/AIA images of active region plage, the structures sometimes show an inverted-Y configuration, suggestive of magnetic reconnection.  Small  loop structures are also found in the cores of bright coronal plumes, even when no minority-polarity flux is visible.  These appear to cluster there, suggesting a role in coronal plume energization and emission, with flow convergence causing interchange reconnection at the plume footpoints and consequent energy deposition.  Reflecting this dynamic, coronal plume emission changes on timescales of hours to a day as the supergranular flow 
field evolves~\citep{2016ApJ...818..203W, 2018ApJ...861..111A, 2019SoPh..294...92Q}.
Previous magnetogram based studies of ephemeral regions~\citep[e.g.,][]{2008ApJ...678..541H}  
measured ephemeral region emergence rates inside unipolar regions that were at least a factor of three lower 
than those in the quiet Sun.  The discovery of ubiquitous small-scale loops 
with very weak, if any, minority-polarity signatures, suggests that  
magnetograms may significantly underestimate the amount of mixed-polarity flux 
in plage, strong network and coronal hole regions. That mixed polarity flux 
may play an important role in coronal heating as it reconnects with 
the large-scale overlying fields, both inside active regions~\citep{2016ApJ...820L..13W, 2019ApJ...885...34W}
and inside coronal holes, where it may also help to drive the solar wind~\citep{2016ApJ...818..203W}. High resolution magnetic field measurements 
with DKIST will be able quantitatively assess this suggestion.

The small-scale magnetic field in the solar photosphere is maintained by a combination of several processes:  (1) dispersal of active region flux; (2) turbulent amplification by photospheric flows; and (3) small-scale flux emergence from deeper regions.  While process (1) reflects contributions from the large-scale global dynamo, processes (2) and (3) are linked to a turbulent fluctuation dynamo that relies on the chaotic nature of flows in the uppermost layers of the convection zone.  Some numerical simulations of turbulent fluctuation dynamos in the solar context have been successful in producing small-scale magnetic field with flux densities similar to that suggested by current observations, but these and all numerical fluctuation dynamo models operate at magnetic Prandtl numbers (ratio of viscosity and magnetic diffusivity) close to unity~\citep[e.g.,][]{1999ApJ...515L..39C, 2007A&A...465L..43V, 2014ApJ...789..132R, 2017A&A...604A..66K}.  By contrast, the magnetic Prandtl number in the solar photosphere, if based on molecular values of the diffusivities, may be as low as $10^{-5}$.   Producing a fluctuation dynamo at small magnetic Prandtl number and large Reynolds number, the parameter regime most relevant to the Sun, is an unmet challenge for both numerical simulations and laboratory experiments.  This makes the solar photosphere a unique plasma laboratory.
   
Moreover, Prandtl number unity models of the solar fluctuation dynamo can generate total unsigned magnetic flux densities close to those observed only when flux advection from deeper layers is allowed~\citep{2014ApJ...789..132R}, not if the dynamo is strictly local, operating only in the photosphere.  The observed level of quiet-Sun magnetism implies a significant amount of recirculation within the convection zone, with convective motions bringing substantial amounts of field up into the photosphere from below~\citep{2003ASPC..286..121S, 2014ApJ...789..132R, 2018ApJ...859..161R}.  Simulations thus suggest that a more complete understanding of the solar fluctuation dynamo requires a quantitative assessment of the relative importance of local field amplification and non-local field transport~\citep[see e.g.,][and references therein for recent observational studies]{2016ApJ...820...35G}.

Finally, the motions of small-scale magnetic field elements in the photosphere may be critical in the operation of the global solar dynamo.  Global scale transport of small-scale field is a central ingredient of many global dynamo models~\citep[see reviews by][]{2014ARA&A..52..251C, 2017SSRv..210..367C}, with the accumulation of field with a dominant polarity in the polar regions, as a result of supergranular diffusion and meridional flow advection of small-scale field elements, critical to the reversal of the global dipolar field.  For this reason, the measured solar polar field strength is often cited as the most reliable indicator for solar cycle predictions~\citep[e.g.,][]{2018NatCo...9.5209B}.  The behavior of small scale fields in the solar polar regions can be directly assessed via synoptic DKIST observations (\S\ref{ss:synoptic}), 
and the value of the transport coefficients used in dynamo models can be dramatically improved by separating advective and diffusive contributions to magnetic displacement statistics in high-resolution DKIST observations~\citep{2018ApJ...854..118A}.  In these ways, DKIST will enable quantitative assessment of the small-scale field behaviors which may underlie the global dynamo. 

\subsection{Acoustic Gravity Wave Excitation}
\label{ss:agwave}

\vskip -0.29in
\begin{figure}[h]
\hbox{\hspace{-2.5em}\includegraphics[width=0.95\textwidth,clip=]{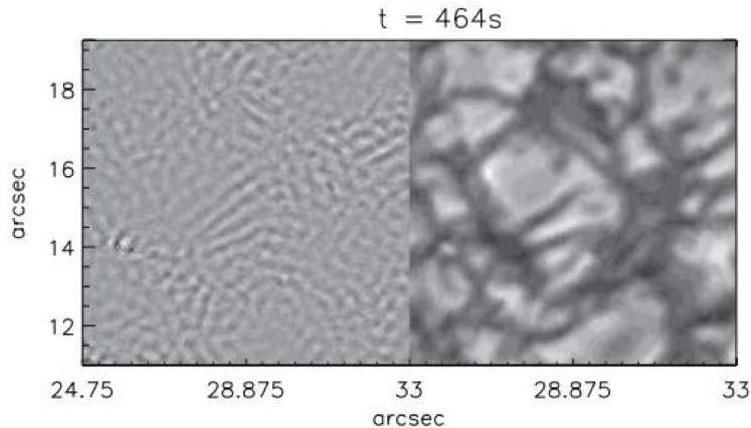}}
\caption{Filtered Doppler velocity (left) and simultaneous continuum image of the same area (right) from the first flight of the IMaX instrument on the SUNRISE balloon-borne observatory.  Time series of these suggest the launching of parallel wavefronts by new downflow lane formation during granule fragmentation.  From~\cite{2010ApJ...723L.175R}.}
\label{f:fig4}
\end{figure}

\noindent
{\it What excites the solar acoustic oscillations?  How do observed source properties effect helioseismic inferences?  What physics underlies flare induced acoustic wave emission? }
\vskip 0.1in
\noindent
Magnetohydrodynamic wave generation, propagation, mode conversion, and dissipation are central processes in the energy and momentum budget of the solar atmosphere.  Those issues are largely discussed in Section~\ref{ss:waves}.  Here we focus on the role of DKIST in determining the source of the solar $p$-modes and the potential for gravity wave observations and diagnostics in the lower solar atmosphere.   

Acoustic-mode excitation on the Sun and other stochastically excited stars likely results from small-scale dynamical processes associated with convection 
(Figure~\ref{f:fig4}), but the detailed source properties are not known.  Some studies point to excitation occurring within intergranular lanes by Reynolds stress induced pressure fluctuations, the Lighthill mechanism~\citep{1977ApJ...212..243G}, while others suggest that the dominant source is associated with pressure perturbations caused by local radiative cooling~\citep{1991LNP...388..195S, 1999ApJ...524..462R} and the sudden formation of new downflow plumes as often occurs during the fragmentation of large granules~\citep{1995ApJ...443..863R}.  Moreover, some flares can be strong acoustic wave sources~\citep[e.g.,][] {1998Natur.393..317K, 2003SoPh..218..151A, 2005ApJ...630.1168D}, though the mechanisms associated with acoustic emission during flaring is only partially understood~\citep{2014SoPh..289.1457L}.   

There is observational support for both of the hydrodynamic (non-flare) mechanisms suggested above~\citep{1995ApJ...444L.119R, 1998MNRAS.298L...7C, 1998ApJ...495L..27G, 1999ApJ...516..939S, 2000ApJ...535..464S, 2001ApJ...561..444S, 2010ApJ...723L.175R}, but it is unclear which process dominates, how much power is radiated by each, and how well that power is coupled to the solar $p$-modes.
Unambiguous characterization of the $p$-mode sources requires separating local wave motions from higher amplitude compressible convective flows and $p$-mode coherence patches.  This is theoretically challenging, and observationally likely requires high spatial and temporal resolution (on the order of tens of kilometers at 15-second cadence) observations over a range of heights from the deep to the middle photosphere.  A statistically significant number of individual events must be studied to assess their physical characteristics, frequency, and acoustic energy contributions, and thus their importance.  It is only with DKIST that these observational capabilities will become regularly available.  Moreover, direct measurement of pressure fluctuations along with the temperature and velocity fields are needed to fully characterize the excitation mechanisms, and pressure sensitive spectral diagnostics are only now being developed in preparation for DKIST start of operations. 

The precise nature of the $p$-mode excitation events, the efficiency, phasing and relative importance of the source contributions, is critical to both local and global helioseismic inferences.  For global helioseismology, and local methods that employ the mode spectrum such as ring-diagram analysis~\citep[][and references therein]{2002ApJ...570..855H, 2002ApJ...571..966G}, precise determination of the modal frequencies and frequency shifts depends on the assumed power-spectral line shape with which the modes are fit.  One source of systematic error in the measured solar $p$-mode frequencies is the unknown non-Lorentzian shape of the spectral lines.  Line profiles vary with wavenumber and frequency depending on the source depth and physical properties~\citep[e.g.,][]{1995A&A...299..245G, 1998ApJ...496..527R, 1998ApJ...495L.115N, 2000ApJ...535..464S}.  Similarly, local helioseismological deductions are sensitive to the phase relationship between the waves and their source.  For example, travel-time kernels used in time-distance helioseismology depend on the assumptions about the source phasing and characteristics~\citep[][]{2002ApJ...571..966G, 2004ApJ...608..580B}.  Source properties may be particularly critical for multi-height local helioseismology if the source is spatially and temporally extended, as it is likely to be.   

The photosphere is the lower boundary of the stably stratified solar atmosphere, and convective overshoot into the lower solar atmosphere drives internal acoustic-gravity waves.  These waves probe the physical properties of the layers above and may make a significant contribution to the solar chromosphere energy balance~\citep[e.g.,][]{1994ApJ...436..929H, 2008ApJ...681L.125S}.  The increased spatial resolution and temporal cadence of DKIST will allow higher spatial and temporal frequency waves to be observed, and the extended high quality observation periods anticipated will enhance the frequency resolution possible.  This may allow individual atmospheric gravity-wave mode identification and detailed study of their convective driving.  Gravity wave excitation is a process which is broadly important in regions more difficult to observe, such as the base of the solar convection zone and in the interior of other stars, and observations of the photosphere and lower solar atmosphere are critical to constraining the underlying excitation processes.  Moreover, since the solar atmosphere is highly magnetized, significant mode coupling occurs as the wave propagates upward, and as this coupling depends on the orientation and strength of the local field, such observations have immediate diagnostic value~\citep[][and Section~\ref{ss:waves} below]{2017ApJ...835..148V, 2019ApJ...872..166V}. 

\subsection{Magnetoconvective Modulation of the Solar Luminosity}
\label{ss:irradiance}

\vskip -0.2in
\begin{figure}[h]
\includegraphics[width=1.0\textwidth,clip=]{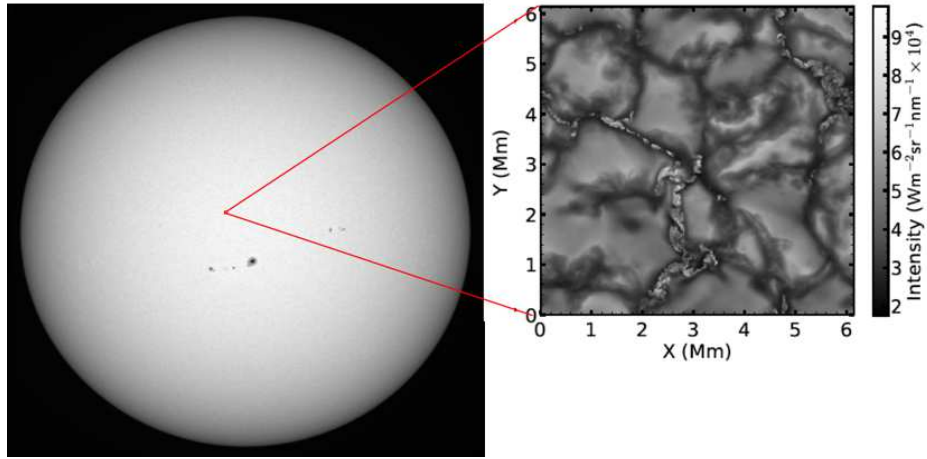}
\caption{Full-disk images are used in solar irradiance models because magnetic structure contributions depend on disk position.  There are significant differences in the magnetic morphology underlying pixels classified as the same structure at full-disk resolution. These influence solar spectral irradiance and its variability with solar cycle.  Image on left from the Precision Solar Photometric Telescope (image courtesy M.P. Rast, \url{http://lasp.colorado.edu/pspt_access/}).  Image on right from~\cite{2014ApJ...789..132R}.}
\label{f:fig5}
\end{figure}

\noindent
{\it How do small scale magnetic flux elements contribute to global solar irradiance variations?  How well do the observed temperature and pressure stratifications within flux elements agree with atmosphere models employed for irradiance reconstruction?}
\vskip 0.1in
\noindent
Observational evidence suggests that solar irradiance is modulated by changes in solar surface magnetism.  Based on this hypothesis, empirical techniques have been developed to reproduce total and spectral solar irradiance variations from observed changes in the coverage of magnetic features over the solar disk.  Depending on their size and field strength, different magnetic features show different center-to-limb variation in different spectral regions.  Thus both the disk location and disk coverage of the features are needed to model irradiance changes, and full-disk medium resolution (one to two arcsecond) observations are typically employed~\citep[e.g.,][]{1996JGR...10113541C, 2003A&A...399L...1K, 2017PhRvL.119i1102Y}.  

However, both high spatial resolution (sub-arcsec) observations and radiative magnetohydrodynamic simulations indicate the presence of magnetic structures too small to be detected at medium resolution (Figure~\ref{f:fig5}).  Spectra modeled on the basis of full-disk pixel average features do not necessarily capture the radiative output of the underlying highly structured atmosphere, nor do they distinguish between differently structured atmospheres with the same lower-resolution appearance~\citep{2011A&A...532A.140R, 2017ApJ...847...93C, 2019ApJ...870...89P}.  The mapping between full-disk imagery, the underlying small-scale structure, and spectral irradiance is only poorly understood, and this mapping is critical for accurate irradiance modeling, particularly spectral irradiance modeling~\citep[see][and references therein]{2013ACP....13.3945E}.  There thus remains significant uncertainty about whether full-disk magnetic structure based irradiance models can account for the observed spectral irradiance variations.  In particular, both the sign and the magnitude of the spectral irradiance trends with solar cycle are controversial, with some authors reporting out of phase irradiance variation in key wavelength bands~\citep{2009GeoRL..36.7801H}, while others report in phase variation across the spectrum~\citep{2013A&A...556L...3W}.  While out of phase variations have been explained in terms of a change in the mean photospheric temperature gradient with cycle, which magnetic component contributes most to that mean change remains unclear particularly when the contribution is disk integrated.  Addressing these uncertainties requires developing both an understanding of the statistical distributions of small scale elements and reliable techniques for measuring the photospheric temperature gradient~\citep{2016A&A...595A..71F, 2017ApJ...835...99C}. 

Unresolved magnetic elements are particularly important to the spectral output of the quiet Sun~\citep{2011A&A...532A.136S} and its center-to-limb profile~\citep{2015ApJ...808..192P}.  It is against this which the contrasts and contributions of magnetic structures are often measured.  The quiet Sun covers the majority of the solar photosphere and typically its integrated irradiance contribution is taken to be constant in time, but due to the presence of possibly time varying unresolved magnetic structures this may not be the case.  Measuring how the magnetic substructure of the quiet Sun actually changes with the solar cycle is a key DKIST capability.  Beyond this, the radiative contributions of magnetic structures such as plage and faculae depend on the fact that they are structured on unresolved scales~\citep{2005A&A...439..323O, 2009A&A...495..621C, 2011ApJ...736...69U}.  Some facular pixels show negative continuum contrast in full-disk images even close to the limb.  This may reflect uncertainty in the quiet-Sun profile against which the contrast is measured or underlying substructure in the faculae themselves~\citep[e.g.,][]{1997ApJ...484..479T, submittedH}.   Measuring the radiative properties of composite features and understanding the physical mechanisms that determine their spectral output is an important next step in the development of irradiance reconstruction and modeling techniques.  These must be capable of statistically accounting for contributions from a distribution of small-scale magnetic features unresolved in full-disk images.

Finally, the astrophysical implications of small-scale fields extend beyond the Sun.  The spectral energy distribution of stars is a key fundamental input in the modeling of planetary atmospheres~\citep[e.g.,][]{2012ApJ...761..166H, 2014ApJ...780..166M}.  In particular, UV radiation is responsible for the production and destruction of molecular species that are the anticipated biomarkers to be used in future NASA exoplanet atmospheric characterization missions.  More critically, UV variability is important in determining the climate and habitability of~\citep[e.g.,][]{2007AsBio...7...85S, 2013ApJ...763..149F, 2013ApJ...766...69L, 2019MNRAS.485.5598O} and biogenic processes on~\citep{2007Icar..192..582B} extrasolar planets.  While not sensitive to UV radiation directly, DKIST observations of the Sun can be used to develop, constrain and test models of stellar chromospheres and/or proxies that reliably reconstruct UV spectra from measurements obtained at longer wavelengths.  

\subsection{Active Region Evolution and Sunspot Fine Structure}
\label{ss:sunspots}
 
\vskip -0.15in
\begin{figure}[h] 
\includegraphics[width=0.65\textwidth,clip=]{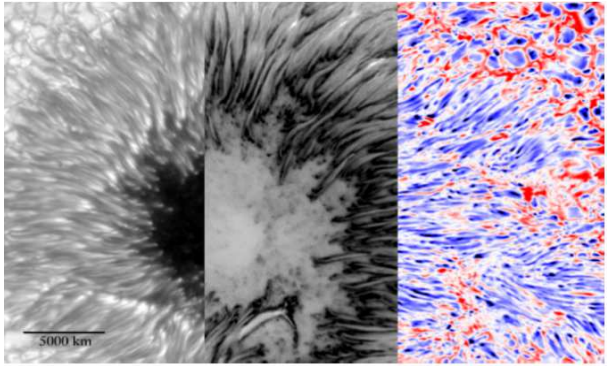}
\caption{Sunspot in active region 11302.  Observed in intensity (left), circular polarization (middle), and line-of-sight velocity (right) with the CRISP instrument at the Swedish 1-m Solar Telescope on 28 September 2011.  Penumbral fine structure is apparent in all three maps, showing details down to the diffraction limit of the telescope (0.15 arcsec).  Lateral edges of penumbral filaments appear to harbor localized short-lived vertical downflows ({\it red}).  Adapted from~\cite{2015ApJ...803...93E, 2016ApJ...832..170P}.}
\label{f:fig6}
\end{figure}

\noindent
{\it Why sunspots? Why is magnetic flux concentrated into sunspots?  How do they form?  Why do they have penumbrae?  What is the dynamical and magnetic substructure of sunspot umbrae and penumbrae, and what are the magnetic and dynamical links to the chromosphere, transition region and corona above?}
\vskip 0.1in

\noindent
Sunspots are the most prominent manifestation of strong magnetic fields in the solar atmosphere. They consist of a central dark umbra, within which the magnetic field is primarily vertically oriented, and a brighter penumbra, with weaker and more inclined fields.  Both regions exhibit small-scale structure, such as umbral dots and penumbral filaments, and rich dynamics with mass flows and wave motions spanning the photosphere and chromosphere.  The range of observed behaviors reflects the presence of magnetic fields of varying strengths and inclination angles, making sunspots an ideal laboratory for studying magneto-convective processes in complex magnetic field geometries.
 
However, despite significant advances over the past decades, observationally characterizing the field geometry and associated flow dynamics in sunspot umbrae and penumbrae presents formidable challenges.  These result from the small scales present, the low umbral photon counts and the difficulties in interpreting the distorted spectral line profiles observed~\citep{2011LRSP....8....4B, 2011LRSP....8....3R, 2019PASJ...71R...1H}.  As an example, we do not yet know the magnetic topology or flow structure of the penumbra at the smallest scales, nor how these vary with height in the atmosphere.  In state-of-the-art numerical models~\citep[e.g.,][]{2007ApJ...669.1390H, 2009ApJ...691..640R, 2011ApJ...740...15R, 2012RSPTA.370.3114R, 2012ApJ...750...62R}, penumbral filaments result from overturning
convection in highly inclined magnetic field regions, with the Evershed flow being the main flow component oriented in the radial direction.  These models predict the existence of vertical downflows at the edges of penumbral filaments where the overturning plasma descends below the solar surface, but systematic study of such downflows on the Sun is very difficult due to their intrinsically small sizes~\citep[e.g.,][and references therein]{2015ApJ...803...93E, 2016ApJ...832..170P}.  Supersonic vertical velocities at the outer end of penumbral filaments have been reported (Figure~\ref{f:fig6}) and can be associated with penumbral field lines dipping down below the solar surface~\citep{2013A&A...557A..25T, 2013A&A...557A..24V, 2016ApJ...832..170P}, but similar flows have also been detected in the inner penumbra without relation to the Evershed flow~\citep{2011ApJ...727...49L}.  In both regions, very large field strengths (on the order of 7 kG and 5 kG respectively, compared to typical peak umbral values of 2 - 3 kG) have been associated with the high-speed flows~\citep{2017A&A...607A..36S, 2018ApJ...852L..16O}, but because the observed Stokes profiles show anomalous shapes, these determinations are uncertain.  Multi-line, very high spatial resolution spectropolarimetric observations by DKIST will allow improved inference of both the field strengths and flow velocities and clarify the origin of the supersonic flows and the processes leading to local field amplification beyond umbral values. 

More fundamentally, the formation of the penumbra itself~\citep[e.g.,][]{2012A&A...537A..19R} remains a mystery. The process lasts only a few hours, making it difficult to capture, and observations have not yet identified the mechanism triggering the process around a naked pore.  There are indications that the chromospheric magnetic field may play a prominent role.  Current sunspot simulations~\citep{2012RSPTA.370.3114R, 2012ApJ...750...62R} only form an extended penumbra if the field inclination at the top boundary is artificially enhanced.  The nature and origin of a similar field on the Sun, if it exists, is unknown.  Penumbrae are a robust feature of sunspots and are present under a wide range of conditions, whereas the presence of penumbrae in sunspot simulations requires this special boundary condition.  This critical difference may be related to the fact that the dynamics of a penumbra during formation are observationally very different from those seen once it is mature. Instead of a regular Evershed outflow, 'counter' Evershed inflows, toward the umbra, are observed during formation.  These inward directed flows turn into the classical radial outward Evershed flow as the penumbra stabilizes~\citep{2011IAUS..273..134S}. This important transition and other chromospheric precursors~\citep{2012ApJ...747L..18S, 2017ApJ...834...76M} likely hold clues about the penumbral formation process, a process missing from the simulations.  High-sensitivity DKIST measurements of the vector magnetic field and flow in both the photosphere and the chromosphere during penumbra formation will allow an assessment of the relative importance of photospheric magneto-convective processes and the evolution of overlying chromospheric magnetic fields. 

In umbrae, strong magnetic fields severely constrain convective motions, and numerical simulations suggest that umbral dots are the signature of convective plumes~\citep{2006ApJ...641L..73S}.  Direct observational confirmation is difficult due to the small sizes and low contrast of the structures as well as the overall low photon counts in umbral regions.  These conspire to dramatically reduce the signal-to-noise ratio of the observations, particularly when spectropolarimetric measurements are needed to determine the magnetic field configuration along with the flow.  Interestingly, umbral dots may not be uniformly distributed throughout the umbra, with recent measurements suggesting the existence of very dark, fully non-convective umbral regions~\citep{2018A&A...617A..19L}.  DKIST's off axis design and consequent low scattered-light properties will allow high-quality high-resolution Doppler and spectropolarimetric measurements of even the darkest regions of the umbra.

The magnetic topology and dynamics in the chromosphere above sunspots are also poorly understood.  Few high-spatial resolution observations extending from the photosphere up into the chromosphere with the required sensitivity have been made, so the connectivity with height is uncertain.  None-the-less, intense chromospheric activity is observed on small spatial scales above sunspots.  Examples include short dynamic fibrils, umbral and penumbral microjets, penumbral bright dots, and jets at the edges of light bridges~\citep[e.g.,][]{2007Sci...318.1594K, 2008SoPh..252...43L, 2009ApJ...696L..66S, 2013ApJ...776...56R, 2014ApJ...787...58Y, 2014A&A...567A..96L, 2014ApJ...790L..29T, 2016ApJ...822...35A, 2016A&A...590A..57R, 2018ApJ...854...92T}.  These phenomena likely involve shocks and/or magnetic reconnection, are expected to be a source of chromospheric heating~\citep[e.g.,][]{2010ApJ...722..888B, 2011ApJ...735...65F, 2013A&A...556A.115D, 2017ApJ...845..102H, 2017A&A...605A..14N}, and may have far reaching effects, up into the overlying transition region and corona~\citep[e.g.,][and references therein]{2017ApJ...835L..19S, 2018ApJ...854...92T}, with some penumbral micro jets (perhaps those that are particularly large) displaying  signatures at transition region temperatures~\citep{2015ApJ...811L..33V, 2016ApJ...816...92T}.  

In general, these phenomena are very small scale, and distinguishing their origin requires high-resolution observations.  For example, much of the spatial and temporal structuring of umbral micro-jets that is thought to be reconnection driven may instead be the signature of a shock/compression front forming in the corrugated sunspot atmosphere, with local density/velocity structures enhancing or delaying the shock at different locations along a steepening compression front~\citep{2020arXiv200805482H}. 
Studying the dynamic chromosphere above sunspots in such detail is challenging.  The low-photon count issues, that as previously discussed restrict observations of the umbral photosphere, are even more of an issue for chromospheric features. Further, finely-structured faint absorption features, dark fibril-shaped inhomogeneities within umbral flashes, are observed in the chromospheric umbra when it is observed at the highest spatial resolution achievable by pre-DKIST telescopes~\citep[e.g.,][]{2009ApJ...696.1683S, 2015A&A...574A.131H}.  These, and even the emitting dynamic features such as umbral microjets~\citep{2013A&A...552L...1B, 2017A&A...605A..14N}, can be faint and extremely small-scale.  DKIST will allow high spatial and spectral resolution spectropolarimetric measurements of such dim small-scale features with sufficient polarimetric sensitivity to constrain their physical properties using tools such as inversion-based semi-empirical modeling~\citep[e.g.,][]{2013A&A...556A.115D}.  This will both enhance our understanding of these phenomena and allow their possible use as highly local diagnostics,   
helping us to understand the three-dimensional physical conditions at the small-scale event sights which can be compared directly with simulations.

The transition region above sunspot umbrae also poses significant observational challenges.  It too is highly dynamical at very small scales.  Strong downflows associated with strong brightenings have been observed with IRIS at transition region band passes and in upper chromospheric diagnostics~\citep[e.g.,][]{2016A&A...587A..20C, 2018ApJ...859..158S, 2020A&A...636A..35N}.  Fortunately, ground-based observations can be combined with these short-wavelength spaced-based measurements to the benefit of inversion efforts~\citep{2016ApJ...830L..30D, 2019A&A...627A..46B}.  Rapid DKIST multi-line time series along with simultaneous observations employing IRIS, Hinode, Solar Orbiter, and SDO, are essential in untangling the relationships between the rich chromospheric dynamics above sunspots and its role in the heating of the solar transition region and corona above sunspots.

Finally, the disappearance and dissolution of sunspots is not well understood.   High-cadence high-resolution magnetogram sequences are required to assess the role of small moving magnetic features which stream out from the penumbral border and merge with the surrounding network magnetic field~\citep[e.g.,][]{2005ApJ...635..659H}.  While these features seem to play a role in the slow disaggregation of the sunspot magnetic bundle, only limited measurements of their individual magnetic topologies and configurations and of their integrated contributions to sunspot decay have been made~\citep[e.g.,][]{2007ApJ...659..812K}.  
 
\section{Flares and Eruptive Activity}
\label{s:flare}

Solar flares and coronal mass ejections are spectacular phenomena that are not only critical to the space weather environment of Earth, but reflect processes ubiquitous in astrophysical plasmas.  The Sun, by its proximity, allows the unique possibility of high-resolution studies of the physics of magnetic reconnection, the related acceleration of particles to relativistic energies, the heating of the plasma to more than 10 million Kelvin, and the excitation of plasma waves.  Flares and mass ejection events are a template for magnetic activity on a variety of stars, some much more active than the Sun, with implications for the habitability of exoplanets.  Moreover, the underlying physical processes are common to even more energetic astrophysical events that produce relativistic jets, beams, and shocks.  

The solar atmosphere is a dynamic magnetized plasma of very high electrical conductivity that can form and sustain complex current systems.  When these currents abruptly reconfigure during magnetic field reconnection events, particle acceleration and bulk plasma ejection can result.   Intense electromagnetic radiation occurs when the energy of accelerated non-thermal particles or accompanying plasma waves is deposited and thermalized.  Observationally, solar flares are transient bursts of electromagnetic radiation over a wide range of wavelengths, from radio to X-ray (or even shorter in some cases).  Coronal mass ejections (CMEs) occur when a magnetic flux system with substantial magnetic free energy, initially confined by overlying field, evolves so that the outward magnetic pressure forces on the trapped plasma overcome the inward magnetic tension force of the overlying fields.  The energetic flux system can then escape into interplanetary space, with the consequent expulsion of magnetized coronal plasma into the heliosphere.  Subsequent flaring often occurs as the magnetic field in the wake of the coronal mass ejection relaxes back to a lower energy state.  Many flares occur without CMEs, and some CMEs occur without substantial enhancement of radiative emission, but it is common for both phenomena to occur as part of a solar eruptive event.

Much of the research on solar flares and CMEs over the last decades has focused on the evolution of temperature and density structures within the corona.  DKIST's ability to regularly and systematically measure coronal magnetic fields will enable more direct evaluation of the  connections between the local plasma properties, the radiative signatures of flares and CMEs, and the field evolution.  The ability of DKIST to readily probe the lower solar atmosphere at very high spatial resolution will allow determination of the physical conditions deep in the atmosphere during flaring and eruptive events, and DKIST measurements of subtle changes in the intensity and topology of the photospheric magnetic fields underlying flaring regions will help in assessing the amount of magnetic energy released during these events.   Spatially-resolved rapid imaging and spectroscopy of the chromospheric foot points of flares will allow determination of the actual extent of the sites of flare energy deposition, providing strong constraints on particle acceleration mechanisms.  Determination of the penetration depth of the flare disturbance into the lower atmosphere, facilitated by DKIST's multi-wavelength capabilities, will help define the roles of proposed energy transport mechanisms and the relative importance of waves and particle beams.  Together these capabilities will elucidate the mechanisms that underly the rapid conversion and deposition of the energy stored in stressed magnetic fields during flaring and plasma ejection events. 

Several critical science topics in this research area are discussed in detail below, including 1) flare and CME precursors, 2) changes in the magnetic field associated with flares and CMEs, 3) energy deposition during flares and 4) the fundamental structure and evolution of flare ribbons.

\subsection{Flare and Coronal Mass Ejection (CME) Precursors}
\label{ss:precursors}

\vskip -0.15in
\begin{figure}[h] 
\hbox{\hspace{-2.em}\includegraphics[width=0.6\textwidth,clip=]{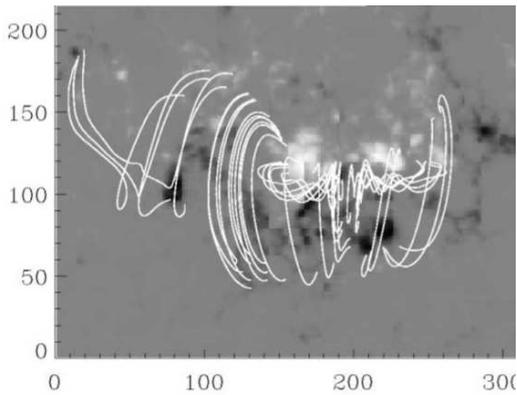}}
\caption{Nonlinear force-free magnetic field reconstruction of NOAA Active Region 9077 before the X5.7/3B (10:24 UT) flare on 14 July 2000, based on vector magneto-grams taken at the Huairou Station of the Beijing Astronomical Observatory. The photospheric longitudinal  magnetogram is shown in the background for reference.  Field lines trace the magnetic flux rope above the neutral line and the overlying arcade field.  From~\cite{2001ApJ...551L.115Y}.}
\label{f:fig8}
\end{figure}

\noindent
{\it What triggers coronal mass ejections and solar flares?  What are the relative importance of mechanisms proposed, such as flux emergence, current sheet instabilities and small scale flux cancellation?  When, where and how do large-scale flux ropes form?}
\vskip 0.1in

\noindent
Solar flares and coronal mass ejections result when magnetic energy stored in the solar corona is released.  The release is sudden (over a few to tens of minutes) compared to the timescale over which energy is built up and stored~\citep[many hours to days, e.g.,][]{2009AdSpR..43..739S}.  Despite decades of study, the physical mechanisms that trigger flares and/or CMEs remain elusive.  Understanding the conditions that lead up to a CME or flare and identifying the trigger mechanisms which initiate them are central to developing the predictive capabilities necessary for successful forecasting of their space weather impacts.

While no "necessary and sufficient" conditions for flaring or CME initiation have been identified, various precursor phenomena have been reported to occur in chromospheric and coronal observations in the minutes to hours preceding flares or CMEs~\citep[for reviews see e.g.,][]{2011LRSP....8....1C, 2011LRSP....8....6S, 2018SSRv..214...46G}.  These include multiple small-scale brightenings in H$\alpha$~\citep{1977A&A....59..255M, 2017NatCo...815905D} and/or soft x-ray~\citep[e.g.,][]{1991A&AS...87..277T}, plasmoid ejection~\citep[e.g.,][]{1994kofu.symp....1H, 2009ApJ...705.1721K}, S-shaped or inverse-S (sigmoidal) soft x-ray emission~\citep{1999GeoRL..26..627C}, and the slow rise and darkening of chromospheric filaments (filament activation)~\citep[e.g.,][]{1980SoPh...68..217M, 2004ApJ...610L.133S}.  There is also some evidence for spectral line broadening before flares~\citep[e.g.,][]{2009ApJ...691L..99H, 2017SoPh..292...38W} and pre-flare thermal X-ray enhancements without evidence of non-thermal particles~\citep{2017SoPh..292..151B}.   

The limited spatial resolution, cadence and wavelength coverage of many previous observations have hampered efforts to understand the physical nature of these potential precursor signals, but recent high-resolution spectropolarimetric observations with the 1.6-m Goode Solar Telescope suggest that very small-scale polarity inversions, currents and magnetic loops in the low atmosphere are associated with main flare initiation~\citep{2017NatAs...1E..85W}.  Magnetohydrodynamic simulations also show that the introduction of small opposite-polarity bipoles or reversed-magnetic-shear structures into a highly sheared larger-scale field can destabilize the field~\citep[e.g.,][]{2012ApJ...760...31K, 2017ApJ...842...86M}.   Reconnection between these small magnetic perturbations and the pre-existing sheared loops can result in flux rope eruption and large scale flaring.  

Resolving these small scale precursors is important for CME modeling.  It has been known for some time that, at least in some cases, the onset of a coronal mass ejection can precede flaring by several minutes, with the CME itself coinciding with precursor indicators.  This suggests that some precursors may serve as early signatures of the erupting fields.  Current CME models~\citep{2019LRSP...16....3T} can be grouped into two broad classes:  flux rope models~\citep{1996SoPh..167..217L, 2000JGR...105.2375L, 2007ApJ...668.1232F}, which assume that a flux rope exists before the eruption onset, in which case the precursors reflect the flux rope's sudden rise, and sheared magnetic arcade models~\citep{1989ApJ...343..971V, 1994ApJ...430..898M, 1999ApJ...510..485A, 2007ApJ...669..621L, 2012ApJ...760...81K, 2019ApJ...887..246D}, which postulate that the flux rope is formed during the initial stages of the flare, implying that precursor signatures are the manifestation of the flux rope formation process itself.  Not all models embrace this dichotomy.  For example, the flux cancellation model~\citep[e.g.,][]{2010ApJ...717L..26A} involves elements of both.  Careful DKIST study of small-scale precursors will thus inform CME eruption models.

\vfill\eject
\subsection{Changes in Magnetic Field Associated with Flares and CMEs}
\label{ss:bchange}

\vskip -0.15in
\begin{figure}[h] 
\includegraphics[width=0.85\textwidth,clip=]{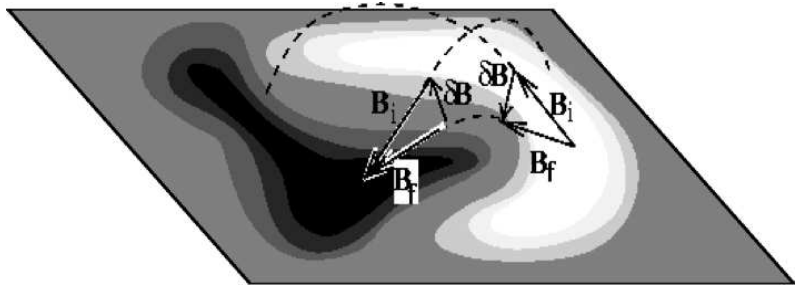}
\caption{Photospheric magnetic vector field changes associated with a flare. The initial photospheric field vectors $B_i$, the change in magnetic field $\delta B$ as a result of coronal restructuring during the flare/CME, and final state $B_f$.  The associated changes in the connectivity of the coronal field are represented by {\it dashed lines}. The changes agree with the expectation that the photospheric field should become more horizontal.  Adapted from~\cite{2008ASPC..383..221H}.}
\label{f:fig9}
\end{figure}

\noindent
{\it What are the differences between the pre- and post-flare/CME magnetic field configurations?  How do the magnetic field changes depend on height and time?  What are the theoretical and practical implications of these changes?  For example, how are flare emission and CME energy related to the magnetic restructuring?}
\vskip 0.1in

\noindent  
Flares and coronal mass ejections are one of the most spectacular manifestations of solar activity~\citep[e.g.,][]{2012LRSP....9....3W}.  Flares occur when stressed magnetic fields in the low corona abruptly reconfigure, accessing a lower energy state.  The rapid conversion of magnetic energy into plasma heating and bulk motion causes changes in coronal and chromospheric active region emission over a wide range of wavelengths.  The field reconfiguration propagates downward into the lower and denser atmospheric, with the resulting changes in the photospheric magnetic field being either transient or longer lived~\citep{2005ApJ...635..647S, 2015RAA....15..145W}.  Flares are often accompanied by CMEs and the onset of CMEs has been associated with flaring, although flares without CMEs and CMEs without flaring also occur.  

Direct measurement of the pre-flare/pre-CME magnetic field and the post-flare/post-CME reconfigured field has both conceptual and practical importance, contributing to our understanding of the underlying reconnection processes, with implications for plasma confinement.  CMEs are a key contributor to the interplanetary plasma dynamics and cause transient disturbances at Earth, where they interact with the Earth's magnetosphere.  They drive interplanetary shocks, an important source of solar energetic particles, and are a major contributor to space weather events.  While all CMEs are large plasma-containing magnetic structures expelled from the Sun, they exhibit a variety of forms.  Some have the classical "three-part" structure:  a bright front, interpreted as compressed plasma ahead of a flux rope, along with a dark cavity with a bright compact core~\citep[][and references therein]{2008ApJ...672.1221R}.  Others display a more complex geometry and can appear as narrow jets~\citep{2012LRSP....9....3W}.  The underlying magnetic field morphology plays a critical role in a CME's geoeffectiveness.

Direct measurement of the pre- and post- flare/CME magnetic field morphology is very challenging.  The magnetic field changes often occur on very short timescales, and propagate quickly through the layers of the solar atmosphere~\citep[e.g.,][]{2015RAA....15..145W, 2017NatAs...1E..85W, 2019ApJS..240...11P}.  Outside of the photosphere, low photon counts make it hard to achieve the needed sensitivity at the required cadence, and multi-wavelength full-Stokes flare studies are consequently difficult~\citep{2017ApJ...834...26K}.  It is such studies that are, however, key to quantitative knowledge of flare and CME magnetism.  While existing white-light CME observations provide information on the mass content of the CME and x-ray and EUV observations provide diagnostics of the flare and CME plasma thermal properties, regular measurements of the magnetic field strength and orientation are essential to the further development and validation of eruption models.  These require DKIST capabilities.

Importantly, DKIST on-disk, limb and off-limb capabilities will allow quantitative assessment of the relationship between plasma sheet observations and reconnection current sheet properties.  Current sheets are an important component of most solar flare models.  They are an essential element of post-CME flaring, and thin off-limb high-density high-temperature plasma structures are observed above limb flares following eruptions~\citep[e.g.,][]{2010ApJ...722..329S, 2013MNRAS.434.1309L, 2017ApJ...835..139S}.  Spectroscopic observations have led to detailed understanding of the plasma properties within these structures~\citep{2018ApJ...853L..15L, 2018ApJ...854..122W}, with implications for the underlying reconnection and field reconfiguration processes~\citep{2018ApJ...868..148L}.  Direct measurements, using DKIST instrumentation, of the magnetic field in and around these plasma sheets will allow more accurate assessments of the amount of flux opened by an eruption, better understanding of the evolution of the post-eruption field, and clarification of the mechanisms underlying the slow (sub-Alfv\'enic) outflows observed during flare reconnection events~\citep[e.g.,][and references therein]{2017ApJ...847L...1W}.

The DKIST instrument suite will enable regular multi-wavelength spectropolarimetric measurements at high temporal cadence and spatial resolution in the solar photosphere, chromosphere and low corona.  These will allow assessment of the local magnetic field properties, and determination of flare- and CME- induced changes in that field, simultaneously over several heights in the solar atmosphere.  When coupled to x-ray and EUV/UV observations from available and future space observatories, DKIST will, for the first time, allow us to directly connect coronal magnetic field evolution to flare and CME emission and plasma properties.  It will enable the careful measurement of the magnetic field reconfiguration that occurs at the origin of space weather events.

\vfill\eject
\subsection{Energy Deposition during Flares}
\label{ss:flareenergy}

\vskip -0.15in
\begin{figure}[h] 
\includegraphics[width=0.9\textwidth,clip=]{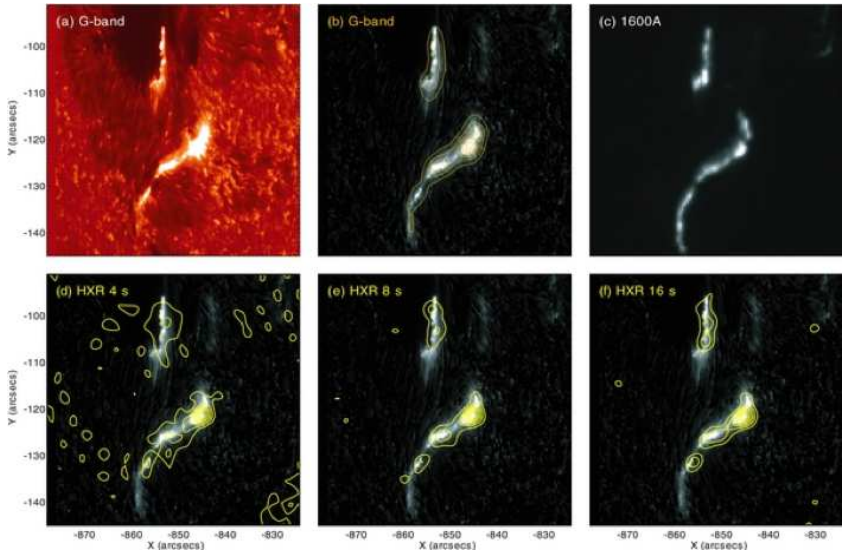}
\caption{Optical flare ribbons at the time of the hard x-ray peak. ($a$) Hinode/SOT G-band, ($b$) the same image as $a$ after pre-flare subtraction, ($c$) TRACE 160nm image taken nearly simultaneously, and ($d$) through ($f$) RHESSI contours (levels 15, 30, 45, 60, 75, and 90 percent of the peak flux) in the 25 -100 keV range for 4, 8 and 16 s time integrations respectively .  From~\cite{2011ApJ...739...96K}.}
\label{f:fig10}
\end{figure}

\noindent
{\it Waves or beams?  How is flare energy transported from the corona to the chromosphere?  Where are the nonthermal particles produced?  What is the origin of the flare optical continuum?  How compact are flare kernels?  What explains the large widths of chromospheric flare emission lines?}
\vskip 0.1in

\noindent
Flare energy, liberated in the corona, is transported to and dissipated in the chromosphere and upper photosphere, resulting in localized heating and intense bursts of radiation across the electromagnetic spectrum.  This leads to chromospheric expansion and drives mass motions, upward-moving chromospheric 'evaporation' and downward-flowing 'condensation'.  In some locations, hard x-ray (HXR) bremsstrahlung emission is observed, revealing the presence of non-thermal electrons containing a significant fraction of the total flare energy.  Ultimately, the majority of a flare's radiated energy is emitted in near-UV and optical lines and continua~\citep[e.g.,][]{2004GeoRL..3110802W}, and these provide rich diagnostics of the energy transport and deposition processes.

One of the longest-standing and most important questions in flare physics is the origin of the optical continuum, which was first observed during the famous Carrington-Hodgson flare of 1859 and generally accounts for a large fraction of the total flare emission~\citep{1989SoPh..121..261N}.  There are two primary candidates~\citep[e.g.,][]{2017ApJ...847...48H}:  the optical component of an enhanced black body (photospheric H$^-$ continuum enhancement) or hydrogen recombination continuum (Paschen and Balmer).  It is difficult to disentangle these two mechanisms~\citep{2014ApJ...783...98K}, but doing so would allow flare energy transport and deposition height discrimination, since the former implies penetration of a significant amount of energy down into the photosphere while the latter carries information about the ionization state evolution higher up in the 
chromosphere.  This in turn would significantly constrain the transport mechanisms.  Since the observed continuum emission carries away a large fraction of the flare energy making it unavailable to drive a dynamical response, it is also relevant to understanding why some flares are important acoustic sources while others are not~\citep{2014SoPh..289.1457L}.  

High time-cadence high spatial-resolution observations may help to distinguish these processes~\citep{2006SoPh..234...79H, 2006ApJ...641.1210X, 2007ApJ...656.1187F, 2011SoPh..269..269M, 2012ApJ...753L..26M, 2016ApJ...816...88K, 2017ApJ...838...32Y}.  The standard flare model suggests that flare energy is transported from the corona to the chromosphere by beams of non-thermal electrons.  If this is correct, the only way to significantly heat the photosphere is by reprocessing the electron beam energy via back-warming by Balmer and Paschen continuum radiation originating in the mid to high chromosphere~\citep{2005ApJ...630..573A}.  Radiation-hydrodynamical (RHD) models can predict the light curves of Balmer and Paschen continua consistent with time evolution of the electron-beam energy deposition~\citep{2014ApJ...794L..23H, 2015SoPh..290.3487K}.  The Balmer limit is not observable with the DKIST but the Paschen limit (820.5 nm) is.   As with the Balmer limit~\citep{2017ApJ...837..125K},  observations of Paschen line broadening near the Paschen limit and the presence or absence of a continuum jump there~\citep{1984SoPh...92..217N} can constrain the charge density and optical depth of the flaring solar chromosphere and be compared with RHD model predictions~\citep{2015SoPh..290.3487K}.  Achieving high spatial resolution and spectral resolution is critical.  Unequivocal evidence for dense chromospheric condensations during the impulsive phase of solar flares has been established~\citep{2017ApJ...836...12K} and intensive localized heating may be key to the interpretation of flare spectra.  

While the standard flare model invokes energy transported from the corona to the chromosphere by beams of non-thermal electrons, the dominant energy transport mechanism is still under debate.  Chromospheric heating to megakelvin temperatures can start before observable HXR emission begins~\citep{2013ApJ...771..104F}, and the apparent low height of HXR and optical sources during limb flares appears inconsistent with the expected penetration depths of accelerated electrons~\citep{2006SoPh..234...79H, 2010ApJ...714.1108K, 2012ApJ...753L..26M}.  Moreover, the electron beam densities implied by the HXR intensities observed require that a very large fraction of the pre-flare coronal thermal electron population be accelerated~\citep{2010ApJ...714.1108K, 2013ApJ...771..104F, 2014ApJ...780..107K}, so it is unclear whether the electron beam requirements can be met at coronal densities.  Finally, issues involving the origin, amplitude and stability of the return currents required to support these dense coronal beams are not fully resolved~\citep{2017ApJ...851...78A}.  Together these imply that additional modes of energy transport may be required, as do very recent modeling results which suggest that electron beams are incapable of producing coronal rain, which is often found shortly after the onset of loop flares~\citep{2020ApJ...890..100R}.  Possible additional transport mechanisms include ion beams~\citep{2006ApJ...644L..93H, 2007ApJ...664..573Z}, heat conduction~\citep{2014ApJ...795...10L, 2015A&A...584A...6G, 2016ApJ...833..211L} or Alfv\'en waves~\citep{2013ApJ...765...81R, 2016ApJ...818L..20R, 2018ApJ...853..101R}.  Each produces a different energy deposition rate as a function of height in the chromosphere, impacting the local atmospheric properties~\citep[e.g.,][]{2016ApJ...827..101K}, its temperature, density, bulk and turbulent velocities, ionization fractions and atomic level populations as a function of time, thus producing potentially unique spectral diagnostics.  DKIST's unique capabilities, working hand-in-hand with advanced radiation hydrodynamics simulations, will be critical to differentiating between these proposed energy transport and deposition mechanisms.

\subsection{The Fundamental Structure and Evolution of Flare Ribbons }
\label{ss:flareribbons}

\vskip -0.15in
\begin{figure}[h] 
\includegraphics[width=0.5\textwidth,clip=]{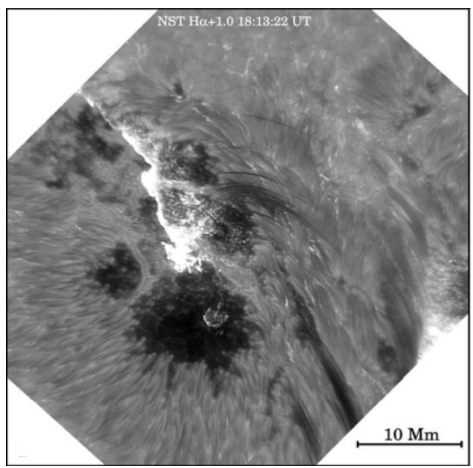}
\caption{Image of a flare ribbon taken with the 1.6 m Goode Solar Telescope's Visible Imaging Spectrometer (VIS) $+1 {\rm \AA}$  off-band the H$\alpha$ 6563 ${\rm \AA}$ line centre (0.07 ${\rm \AA}$ band pass), showing the ribbon crossing sunspots, coronal rain in the post-flare loops and fine-scale brightenings at the footpoints of the falling plasma in the chromosphere.  From~\cite{2016NatSR...624319J}.}
\label{f:fig11}
\end{figure}

\noindent
{\it What is the mapping between the evolution of a flare ribbon and the reconnection process?  What does flare ribbon evolution tell us about changes in the overlying three-dimensional field and the particle acceleration and heating processes?}
\vskip 0.1in

\noindent
Flare ribbons are localized elongated brightenings in the chromosphere and photosphere.  They are the sites of energy deposition at the  foot points of the magnetic loops involved in flare reconnection. The brightening, due to the increased temperature of the flaring chromospheric plasma, is believed to be primarily due to collisional heating by flare-accelerated non-thermal electrons, though questions about the viability of this mechanism remain (\S\ref{ss:flareenergy} above).  The brightest parts of a flare ribbon can be subject to localized heating of $10^{12}$ ergs cm$^{-2}$ s$^{-1}$~\citep{2011ApJ...739...96K}, on the order of 100,000 times the energy flux required to heat the quiet chromosphere.  The detailed structure and evolution of flare ribbons can provide insight into the fundamental length and timescales of flare energy deposition and the link between particle acceleration and the evolving coronal magnetic field.  

Flare ribbons occur in pairs that show a characteristic temporal evolution, reflecting changes in the overlying magnetic field.  Conjugate ribbons separate in time, mapping out the progress of coronal magnetic reconnection.  When observed simultaneously with the photospheric or chromospheric magnetic field, the motions of ribbons as a whole, and of the individual sources composing them, can be used to obtain the coronal magnetic reconnection rate~\citep{2002ApJ...566..528I, 2002ApJ...565.1335Q, 2009A&A...493..241F, 2018ApJ...869...21L}.  Distinct brightenings are often observed to run along the ribbons, and have been linked to properties of three-dimensional 'slip-running' reconnection~\citep{2014ApJ...784..144D}.  Moreover, since the evolution of the flare ribbon reflects changes in the large-scale coronal magnetic field, careful observations can inform our understanding of the dynamics of eruptive (CME) events~\citep[e.g.,][]{2017ApJ...839...67S, 2018SoPh..293...38H}.

At small scales, flare ribbons are composed of numerous small sources and show complex structure that appears differently at different wavelengths, reflecting structured energy deposition.  In line cores, flare ribbons exhibit a stranded structure, as if a small portion of numerous magnetic loop tops in the chromosphere are illuminated~\citep[e.g.,][]{2017ApJ...845...30M}.  In the visible and infrared continua or in the far wings of spectral lines, individual sources are very compact, with clear evidence for sub-arcsecond (100 km) structure suggesting very small scales for individual flaring loops or current dissipation sites~\citep{2014ApJ...788L..18S, 2016NatSR...624319J}.  The sizes of the sources vary with wavelength, and this has been interpreted as being due to the narrowing with depth of the chromospheric magnetic flux tube along which energy is being deposited~\citep{2012ApJ...750L...7X}.  Additionally, the individual source sites display a 'core-halo' continuum intensity sub-structure~\citep{1993ApJ...406..306N, 2006SoPh..234...79H, 2006ApJ...641.1210X, 2007PASJ...59S.807I}.  This may be a signature of spatially varying heating across a magnetic structure or of two heating components, direct heating by energetic particles in the core and indirect heating by radiative back-warming (one of the mechanisms proposed to account for near-UV and optical continuum emission in flare kernels) in the halo~\citep{1993ApJ...406..306N, 2006ApJ...641.1210X}.

Flare kernels are the very brightest sources in flare ribbons, and are the locations most clearly associated with strong hard X-ray emission generated by non-thermal electrons.  On arcsecond scales, flare ribbon sub-structure is well-aligned with regions of high vertical photospheric current density~\citep{2014ApJ...788...60J}.  In the very earliest stages of a flare, these kernels can expand significantly on timescales less than one second~\citep{2010AN....331..596X}, possibly as a result of reconnection spreading through a highly-stressed region of the coronal magnetic field.  As discussed above, flare ribbons themselves also spatially propagate as the site of reconnection changes.  At the leading edge of a propagating ribbon, sites of particularly large Doppler line broadening are observed~\citep{2016NatSR...624319J, 2018ApJ...861...62P}, suggesting that the strongest energy input into the lower chromosphere and photosphere occurs at the locations of the most recently reconnected field.  DKIST will allow regular high resolution (sub-arcsecond) measurements of these underlying spatial and temporal correlations, measurements critical to determining the flare energy flux into the solar chromosphere and the fundamental  scales at which discrete coronal reconnection occurs~\citep{2015ApJ...807L..22G}.  Moreover, the DKIST VISP spectral band pass will be wide enough to measure the exceptionally strong shock fronts that can accompany large flares.  The magnitude of the observed Doppler red shifts in current observations of C-class solar flare ribbons (such as those using the CRisp Imaging Spectro-Polarimeter~\citep{2008ApJ...689L..69S} on the Swedish 1-m Solar Telescope~\citep{2003SPIE.4853..341S}) can exceed 1~\AA~relative to the H$\alpha$ line center~\citep{2017NatCo...815905D}. 
These red-shifts are attributed to the development of a shock fronts in the upper chromosphere which propagate downwards towards the photosphere with velocities up to $50~{\rm km~s^{-1}}$ soon after flare onset and electron beam injection.  Supporting models of H$\alpha$ line formation indicate that stronger flares yield even stronger shock fronts with even greater Doppler red-shifts, up to 4~\AA, not yet observed with modern instruments.  DKIST's ability to capture these larger redshifts within the VISP spectral band pass will contribute significantly to a detailed understanding of solar flare electron beam properties and how they scale with flare energy.

\section{Magnetic Connectivity through the Non-Eruptive Solar Atmosphere}\label{s:magcon}

The solar magnetic field extends from the solar interior, across the photosphere, through the chromosphere and transition  region, and into the corona, permeating the entire volume and out into the heliosphere.  Plasma  motions  in  the  solar convective 
envelope produce a highly structured photospheric magnetic field distribution which can be considered the boundary condition for the magnetic field above.  As the photospheric boundary field evolves, the field above must reconfigure, and because the atmosphere extends over many scale heights, the energetics and dynamics of the response are diverse and multi-scaled.  A central goal of the DKIST science mission is a quantitative  understanding of the complex interplay between flows, radiation, conduction, wave propagation, dissipation and reconnection in the solar atmosphere.  This is critical to the resolution of long-standing problems in solar and stellar astrophysics such as chromospheric and coronal heating, the origin and acceleration of the solar wind, and the propagation of magnetic disturbances into the heliosphere. 

Understanding the complex connected solar atmosphere, with its widely varying ionized and partially ionized plasma regimes, requires diverse, flexible, and multi-spectral line polarimetric instrumentation capable of simultaneously probing many heights at high temporal cadence and over small spatial length scales.  DKIST is at the forefront of these observational challenges.  New capabilities in high precision spectropolarimetry will allow inference of magnetic field properties with spatial and temporal resolutions never before achieved.  For example, while the distribution of solar photospheric magnetic fields has been routinely measured since the invention of the Zeeman-effect based magnetograph,  measurement of the chromospheric field and its connections to the upper solar atmosphere requires interpretation of spectral lines formed under conditions of non-local thermodynamic equilibrium (NLTE) in which both Zeeman and atomic-level polarization processes are important.  DKIST aims to facilitate these with unprecedented precision.

DKIST is also at the forefront of coronal spectropolarimetry, providing new opportunities to understand the magnetic connectivity and energetics of the solar corona.  Extrapolation of the photospheric magnetic field into the solar corona is inherently limited by assumptions made regarding the distribution of currents in the magnetized volume.  To advance our understanding of, for example, the available free magnetic energy driving solar flares and eruptions, remote sensing of the magnetic field in the corona itself is necessary.  DKIST, as the world's largest coronagraphic polarimeter, will allow the measurement of the full Stokes spectra of the forbidden magnetic-dipole emission lines of highly ionized metals.  These will be used to determine the topology and evolution of the coronal field and, along with DKIST chromospheric diagnostics, its connectivity to the lower atmosphere.

Several critical science topics in this research area are discussed in detail below, including 1) the mass and energy cycle in the low solar atmosphere, 2) the origin and acceleration of the solar wind, 3) magnetic reconnection throughout the solar atmosphere, 4) waves in the solar atmosphere, 5) impact of flux emergence on the non-eruptive solar atmosphere, 6) multilayer magnetometry and 7) large-scale magnetic topology, helicity, and structures.

\subsection{Mass and Energy Cycle in Low Solar Atmosphere}
\label{ss:masscycle}

\vskip -0.15in
\begin{figure}[h] 
\includegraphics[width=0.955\textwidth,clip=]{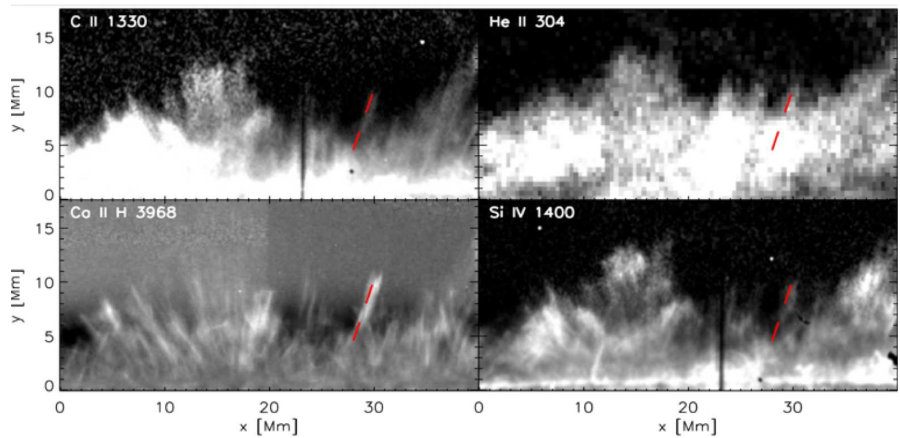}
\caption{Simultaneous images of spicules in Ca II H $3968{\rm\AA}$ (Hinode), C II $1330{\rm\AA}$ and Si IV $1400{\rm\AA}$ (IRIS) and He II $304{\rm\AA}$ (SDO/AIA).  Future coordinated observations with DKIST, ALMA and other observatories are key to addressing unresolved questions about spicule heating and their role in the mass and energy balance of the transition region and corona.  Image from~\cite{2015ApJ...806..170S}.}
\label{f:fig13}
\end{figure}

\noindent
{\it How important are spicules, and other jet-like phenomena, to the chromosphere-corona mass cycle?  What is the role of spicule heating in the coronal energy balance?  Can we characterize and model the coronal rain phenomenon well enough to understand its role as a return flow?}
\vskip 0.1in

\noindent
All coronal plasma has its origins in the lower solar atmosphere, with the coronal mass budget determined by a balance between upward flows (e.g., evaporative or eruptive), downward mass transport (e.g., coronal rain) and solar wind losses.  Many processes appear to be involved.  One  of the key outstanding challenges is determining which of them dominates the transfer of mass between the cool 
chromosphere and hot corona in regions of differing magnetic topology.  Previous observations have provided a partial view of the mass cycle, but the small spatial and temporal scales involved have significantly hindered advancement.  DKIST will provide a more complete view, especially when combined with coordinated observations using space-based observatories, such as IRIS (Interface Region Imaging Spectrograph), Hinode and Solar Orbiter, or the ground-based radio telescope array ALMA (Atacama Large Millimeter/submillimeter Array), which allows complementary temperature and soon polarization diagnostics of the solar chromosphere~\citep[see][]{2018ApJ...863...96Y}.  

Jets are an important dynamical process in the chromosphere-corona mass cycle.  They are observed to have a wide range of spatial scales, ranging from spicules (with widths of a few hundred kilometers), to larger chromospheric jets (~1,000 km widths) to coronal jets (widths of a few thousand kilometers).  Recent observations show that some jets are driven by the eruption of a small-scale flux rope~\citep{2015Natur.523..437S} and triggered by flux cancellation at the magnetic neutral line~\citep{2016ApJ...832L...7P, 2017ApJ...844..131P, 2018ApJ...868L..27P}.  However, the change in magnetic field that leads to the origin of jets over the range of their scales is not fully understood.  DKIST's spectropolarimetric capabilities will significantly contribute to our understanding of how jets are formed and how efficiently each type injects plasma into the solar atmosphere. 

Spicules appear to be the most ubiquitous jet-like feature in the solar chromosphere.  They are highly dynamic, vary on time scales of 10-30s, and are finely-structured, with widths $<300$ km and substructure at scales below that.  They transport plasma upwards into the solar atmosphere at speeds of 10-200 km/s and may, either alone~\citep{2011Sci...331...55D} or in combination with other slower jets~\citep[e.g.,][]{2012A&A...543A...6M}, play a significant role in the mass and energy balance of the corona and solar wind.   
Yet, despite having been observed over a wide range of wavelengths from EUV to visible, we do not understand
the mechanisms responsible for spicule formation, how they are heated, often to
transition region temperatures or higher~\citep{2014ApJ...792L..15P,  2016ApJ...820..124H, 2019Sci...366..890S}, the role hydromagnetic waves play in their dynamic evolution, or their full impact on the mass and energy balance of
the outer solar atmosphere~\citep{2019AdSpR..63.1434P}.  That spicule coronal signatures, have been observed at the smallest scales above both active regions~\citep{2011Sci...331...55D} and the quiet Sun~\citep{2016ApJ...820..124H} suggests that they may be important heat or mass conduits between the corona and the chromosphere across the Sun, with our current understanding of their importance limited by current instrumentation capabilities.  Spicule induced mass transport to coronal heights is estimated by some authors to be
two orders of magnitude larger than the solar wind mass flux~\citep{1968SoPh....3..367B, 2000SoPh..196...79S}, but others suggest that their role  in the mass cycle is rather limited compared to that of more uniform chromospheric evaporation due to small scale flaring processes~\citep[e.g.,][]{2012JGRA..11712102K}.  In fact, there is still some debate about the fundamental nature of spicules, for example whether only one or multiple spicule types exist~\citep[][but see~\citeauthor{2012ApJ...759...18P}, \citeyear{2012ApJ...759...18P}]{2007PASJ...59S.655D,  2012ApJ...750...16Z}  or whether they are indeed jets of accelerated plasma or instead correspond to warped two-dimensional sheet-like structures~\citep{2011ApJ...730L...4J, 2012ApJ...755L..11J, 2014ApJ...785..109L}.  

DKIST observations, employing a wide range of chromospheric spectral lines, will provide revolutionary new views of the fine-scale structure, magnetic and electric fields~\citep{2014ApJ...786...94A} and thermodynamic evolution of spicules and other chromospheric jets.  
DKIST will be able to capture the dynamic evolution of jets at high cadence ($\lesssim3$s) and make simultaneous measurements (with $\lesssim15$s cadence)  of the chromospheric and photospheric magnetic fields and flows which underly jet initiation.  These will help clarify the roles and relative importances of reconnection (\S\ref{ss:reconnection} below), magnetic tension amplification by ion-neutral coupling~\citep{2017Sci...356.1269M}, micro-filament eruption~\citep{2016ApJ...828L...9S} 
and other processes in jet initiation, dynamics and evolution.  

While jet-like features are perhaps the most prominent, they may not be the most important component of the chromosphere-corona mass cycle.  Perhaps more gentle processes play an important role?  One sensitive signature of mass transport, momentum flux and wave heating is the degree of elemental fractionation, i.e. the degree to which elements of low (lower than about 10eV) first ionization potential (FIP) are enriched or depleted compared to other elements~\citep[the FIP or inverse FIP effects; ][]{1985ApJS...57..173M, 1985ApJS...57..151M, 1992PhyS...46..202F, 2000PhyS...61..222F, 2015LRSP...12....2L}.  Because FIP fractionation is sensitive to the thermodynamic and electromagnetic plasma environment, it depends on the atmospheric height of the region being considered, the magnetic field geometry (open or closed) and connectivity, the heating mechanisms at play, the bulk plasma flow speed and dominant magnetohydrodynamic (MHD) wave modes present~\citep{2015LRSP...12....2L}.  In fact, spicules themselves may be responsible for the absence of the FIP effect at temperatures below $10^6$ K on the Sun~\citep[][and references therein]{2015LRSP...12....2L}.  These sensitivities, when understood well, allow FIP fractionation measurements to be used to constrain the solar wind source region and thus the origin of plasma sampled during in-situ heliospheric measurements~\citep[e.g.,][]{1995SSRv...72...49G, 2000A&A...363..800P, 2011ApJ...727L..13B, 2015NatCo...6.5947B, 2015ApJ...802..104B}.  Coordinated observations (\S\ref{ss:insitu}), combining chromospheric measurements by DKIST with in-situ measurements of the heliospheric plasma properties by Solar Orbiter and Parker Solar Probe, will allow unprecedented identification and characterization of the sources of the fast and slow solar wind.  

Another key aspect of the chromosphere-corona mass cycle is the return flow from the corona to the lower atmosphere.  One component of this is prominently visible at the solar limb as coronal rain, a finely structured and multi-thermal flow that appears to be driven by cooling instabilities~\citep[e.g.,][and references therein]{2012ApJ...745..152A, 2015ApJ...806...81A, 2019ApJ...874L..33M}.  The descending material is visible in the off-limb active corona at transition region and chromospheric temperatures, and can be observed in what are typically chromospheric optical spectral lines~\citep{1970PASJ...22..405K, 2001SoPh..198..325S, 2010ApJ...716..154A}.  Current coronal rain observations have limited spatial resolution and the peak in the coronal rain elemental width distribution remains unresolved~\citep{2014ApJ...797...36S}.  Thus we do not yet have a complete accounting of the total rain drainage rate.   Measurements of loop oscillations can be used to determine the thermally unstable mass fraction in a coronal loop system, and thus provide further constraint on that rate.  Since coronal rain condensations are coupled to the magnetic field lines they track the oscillatory motion of the loops~\citep{2016ApJ...827...39K} and the evolution of the oscillations can be used to deduce the fraction of the loop plasma mass that becomes thermally unstable and drains with time.  This seismic estimation provides more than a consistency check because it allows measurement of the unresolved coronal rain mass fraction~\citep{2018ApJ...855...52F}, which in turn constrains the fundamental spatial scales of the rain and the fine-scale structure of coronal loops.  DKIST observations will push these inferences even farther.

One-dimensional hydrodynamic loop models~\citep{2003A&A...411..605M} produce catastrophic cooling events that generate intermittent and repeating rain-like downflows even when steady, exponentially decaying with height, foot-point concentrated heating is employed.  Similarly, 2.5-dimensional and recent three-dimensional simulations form fine-scaled localize rain elements even when a spatially smooth but localized foot-point concentrated heating function is applied \citep{2013ApJ...771L..29F, 2015AdSpR..56.2738M}.  These numerical studies, along with observational evidence for foot-point concentrated heating in active regions~\citep{2001ApJ...560.1035A}, suggest that studies of coronal rain are important, not just in the context of the mass cycle, but also in constraining heating mechanisms.  The statistical properties of the rain depend on the spatial distribution of the heating because the heating location influences the thermal stability of the plasma in a coronal loop system~\citep{2010ApJ...716..154A}.  

Observational contributions to our understanding of coronal rain depend on leveraging the high resolution and multi-wavelength capabilities of DKIST to unravel its evolving, complex, multi-thermal behavior.  With a mix of hot (ionized) and cold (neutral) gas, coronal rain is an ideal context within which to study ion-neutral interaction effects, such as ambipolar diffusion, that are expected to play key roles more broadly in the dynamics of the partially ionized chromosphere.  But the rain's thermal state is complex and difficult to characterize.  Hydrogen is likely out of ionization equilibrium and other elemental ionization ratios are poorly constrained~\citep{2010ApJ...716..154A, 2015ApJ...806...81A}.  DKIST will also be able to provide spectropolarimetric measurements from which the vector magnetic field within rain producing loop systems can be inferred.  Such inferences are fundamental to characterizing the underlying flow instabilities~\citep[e.g.,][]{2020A&A...634A..36M}, but they will require careful theoretical underpinning.  Recent studies have been able to exploit the Zeeman effect in the CaII 854.2 nm chromospheric line to assess the presence of strong fields (100 - 300 G) in bright post-flare loops systems~\citep{2019ApJ...874..126K}, but this approach may be difficult in more quiescent cases which are characterized by much weaker fields and lower photon counts.  Significant progress may be possible if Hanle diagnostics in lines of neutral Helium such as the He I 1083 and 587.6 nm multiplets are employed~\citep{2018ApJ...865...31S}, and joint polarimetric observations of these two He I multiplets with DKIST may provide the first view of sub-arcsecond scale weak magnetic fields (with sensitivity of a few Gauss) in quiescent loop systems.

\subsection{Coronal Heating, Solar Wind Origin and Acceleration}
\label{ss:wind}

\vskip -0.2in
\begin{figure}[h] 
\includegraphics[width=0.9\textwidth,clip=]{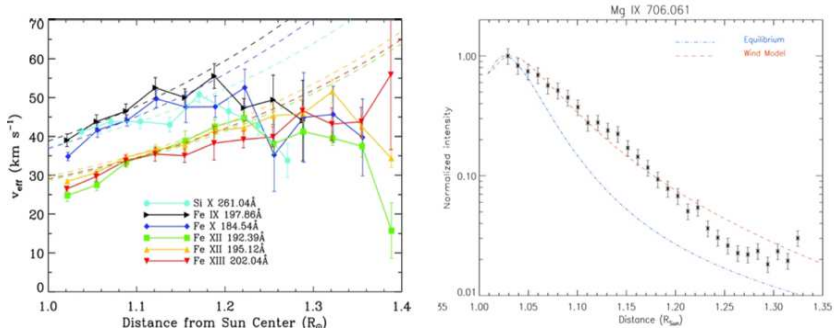}
\caption{Hinode/EIS line width measurements in an off-disk coronal hole as a function of height~\citep[{\it left}, from][]{2012ApJ...753...36H}.  The decrease in width with distance off-limb beyond 1.1 $R_\odot$ suggests wave damping.  Mg IX 706 off disk line intensity in a polar coronal hole~\citep[{\it right}, from][]{2012ApJ...750..159L}, measurements from SOHO/SUMER, estimates obtained by assuming ionization equilibrium (blue curve) and using non-equilibrium ion fractions (red curve).  The red curve was obtained choosing wind speed vs. height profile that allowed predicted ion fractions follow observed line intensities.}
\label{f:fig14}
\end{figure}

\noindent
{\it Waves or nanoflares?  What are the relative importances of different proposed coronal heating mechanisms?  What are the solar wind momentum sources?  What role does the chromosphere play in coronal heating?}
\vskip 0.1in

\noindent
The steep transition from the cool chromosphere to the million degree and higher corona is the direct result of the cooling catastrophe that results when hydrogen in the solar atmosphere becomes fully ionized so that radiative recombination can no longer cool the optically thin plasma ~\citep{1990ApJ...355..295W, 1990ApJS...73..489W}.  The plasma temperature climbs high enough so that electron conduction back down to the chromosphere, and resulting radiative losses from there, are sufficient to balance the heating above.  Solving the coronal heating problem requires identifying the heat source which in the statistically steady state balances thermal conduction to the chromosphere and any other smaller direct energy losses from the corona by radiation or advection.  Significant progress has been made in the last decades, and it is now apparent that no single heating mechanism is likely universally dominant~\citep[e.g.,][]{2012RSPTA.370.3217P, 2015RSPTA.37340269D}.  The importance of mechanisms, such as reconnection, MHD or plasma wave dissipation, or turbulent 
dissipation, likely vary depending on coronal conditions, particularly the magnetic field configuration.  Moreover, the mechanisms are highly intertwined and interdependent; dissipation of current sheets can produce waves and waves in magnetically structured media can induce current-sheets~\citep[e.g.,][]{2015RSPTA.37340262V}, and observations indicate that even on very large scales eruptive flares can trigger oscillations in coronal loops and filament oscillations can induce eruption~\citep[e.g.,][and references therein]{2015SSRv..190..103J, 2015A&A...581A...8R} and associated flaring.  Turbulence is similarly likely ubiquitous, and theoretical work suggests an important role for Alfv\'en wave induced turbulence~\citep[e.g.,][]{2011ApJ...736....3V, 2013ApJ...773..111A, 2014ApJ...782...81V}. 

Thus, while a number of general properties of coronal heating have been observationally established~\citep{2015RSPTA.37340269D}, coronal heating is unsteady (impulsive), coronal magnetic fields store energy which can be dissipated via reconnection, the corona supports a rich wave field, the corona can only be understood in conjunction with its coupling to the chromosphere, details are less certain~\citep[e.g.,][]{2015RSPTA.37340256K, 2015RSPTA.37340257S}.  Questions remaining include: What is the relative importance of different heating mechanisms?  Do current sheets in the corona play an important role in heating, and if so, what plasma processes are involved in their dissipation?  What is the fundamental scale of the substructure in multi-thermal coronal loops, and what processes determine this?  During small scale reconnection events, how much energy is dissipated directly, how much is radiated as waves, and how much goes into acceleration of non-thermal electrons~\citep{2014Sci...346B.315T}?  How does magnetic reconnection heat the plasma~\citep{2015RSPTA.37340263L}?  How and where are the ubiquitous MHD waves in the corona dissipated~\citep[e.g.,][]{2002ESASP.505..273P, 2017ApJ...836....4G}?  How far out in the solar wind does heating extend~\citep[][and references therein]{2019ApJ...879...43M}?  What are the characteristic scales and magnitudes of the heating events?  What triggers them?  

DKIST observations will be fundamental in answering these questions.  DKIST will enable careful, repeated and frequent measurements of the plasma properties of the inner corona at multiple heights, measurements essential to quantitative evaluation of suggested heating processes in varying magnetic environments.  Measurement of the line-continuum and line-line intensity ratios of ionic species, such as Fe IX to XV, that are also found in in-situ measurements of the fast and slow solar wind, will allow determination of the electron density, temperature and charge state evolution of the solar wind plasma, informing our understanding of the acceleration and heating processes~\citep{2012ApJ...750..159L, 2016JGRA..121.8237L, 2018ApJ...859..155B}.  Comparison between theoretical studies of fast and slow magneto-acoustic wave mode conversion, shock formation and dissipation~\citep[e.g.,][]{1995A&A...300..302Z, 1997ApJ...481..500C} and observations will help constrain the role of these processes as a function of height in the chromosphere.  For example, in models, high frequency ($>10$mHz) propagating acoustic waves develop into radiatively damped weak shocks within the first few hundred kilometers above the photosphere~\citep{2002ESASP.505..293C} while lower frequency waves ($\sim4 - 10$mHz) develop into strong shocks in the chromosphere (above 1 Mm) where radiative damping is less effective~\citep{2000smh..book.....P}.  Multi-height DKIST observations can assess this frequency dependence in regions of differing magnetic field configuration.  
Moreover, previous high-resolution observational work has reported evidence for shock induced turbulence~\citep{2008ApJ...683L.207R}, and such turbulence may provide a mechanism for the dispersal of the wave energy beyond the local shock region itself.  Turbulence may also play a role in wave mode conversion, coupling the compressive motions to Alfv\'enic fluctuations, which can continue to propagate outward, transmitting energy to higher layers of the solar atmosphere.  Alternatively, counter propagating Alfv\'en waves may nonlinearly interact to produce MHD turbulence, dissipating the waves and 
heating the plasma~\citep{2011ApJ...736....3V}.  DKIST observations can differentiate these processes in the solar atmosphere and determine their occurrence frequency.  Further, while direct observation of individual nanoflare heat events lies beyond the capabilities of DKIST, DKIST observations may help distinguish between specific reconnection heating mechanisms.  Heating models based on flux cancelation have been previously motivated by high resolution IMAX data from Sunrise~\citep{2018ApJ...862L..24P} and heating events in simulations of three dimensional kink-unstable flux ropes may be diagnosable using DKIST coronal lines~\citep{2018ApJ...863..172S}.  With DKIST, high resolution chromospheric observations of intensity fluctuations and motions at the foot-points of hot coronal loops should reveal key telltale signatures of nanoflares in the overlying corona~\citep{2014Sci...346B.315T}.  When coupled with radiative hydrodynamic modeling~\citep[e.g.,][]{2016ApJ...827..101K, 2018ApJ...856..178P}, such observations may be able to constrain the properties of the non-thermal particles or waves generated at nanoflare sites.  Finally, though the details are uncertain, it has been suggested that the chromosphere may play an important role in coronal heating~\citep[e.g.,][]{1977ARA&A..15..363W, 1999ApJ...521..451S, 2017ApJ...845L..18D} since chromospheric plasma can be heated to transition region temperatures and higher and carried via jets to coronal heights.  If so, the jet studies outlined in Section~\ref{ss:masscycle} are highly relevant to this research topic as well.

The hot outer solar corona escapes the Sun as the solar wind.  The solar wind carries plasma out into the heliosphere, driving interactions with planetary space environments and influencing CME arrival time and geo-effectiveness, yet there is no consensus understanding of where the solar wind originates or how it is accelerated.  The fast and slow solar wind show differing physical properties~\citep[e.g.,][]{2005JGRA..110.7109F, 2009JGRA..114.1109E}, likely come from different source regions and are possibly subject to different acceleration mechanisms.  The fast wind originates in coronal holes, but how the detailed properties of the field and plasma within a coronal hole lead to the observed properties of the wind is not clear.  Additional sources of momentum, beyond Parker's original gas pressure gradient mechanism ~\citep{1958ApJ...128..664P, 1963idp..book.....P}, are required for the plasma to reach the observed fast wind speeds.  Possible momentum sources included large-amplitude MHD waves~\citep{1971A&A....13..380A, 1977ApJ...215..942J, 2007Sci...318.1574D, 2014ApJ...790L...2T}, Type-II spicules~\citep[e.g.,][]{2009ApJ...701L...1D, 2011ApJ...731L..18M}, resonant interactions with ion cyclotron waves~\citep{2002JGRA..107.1147H} and others~\citep[e.g.,][and references therein]{2019ARA&A..57..157C}, but all are to date poorly constrained by observations.  The slow wind, has variously been proposed to originate from coronal hole edges, closed-open field boundaries within and bordering active regions, streamers, particularly streamer tops, or small coronal holes.  These locations are all associated with magnetic reconnection between closed magnetic flux systems and open ones that connect to the wind, but, as with the fast wind, there is no consensus on the dominant underlying physical configuration or mechanism~\citep[e.g.,][]{2005JGRA..110.7109F, 2009LRSP....6....3C}.  In fact, some of the observed differences between the fast and slow solar wind may have more to do with the expansion properties of the background magnetic field along which they are streaming than differences between their source regions~\citep{2003ApJ...587..818W}.  

The solar wind can be studied using either remote sensing or in-situ techniques.  In-situ measurements typically provide direct information on plasma properties only after the plasma has undergone much of its evolution, though Parker Solar Probe (PSP) is revolutionizing these measurements, aiming to sample, during its closest perihelia, regions of solar-wind heating and acceleration directly~\citep[e.g.,][]{2018A&A...611A..36V}.  Remote sensing observations, on the other hand, allow frequent multiple measurements of the solar wind source regions, but can be difficult to interpret.  Combining these types of measurements~\citep{2012ApJ...750..159L} is already yielding exciting results in the current early PSP era~\citep{2020ApJS..246...37R}.  DKIST will make significant contributions to these efforts (\S\ref{ss:insitu}).  
Regular off-disk coronal measurements of the magnetic field and the plasma properties will capture the solar wind as it emerges from its source up to a height of 0.5 solar radii above the limb.  Since the visible and infrared spectral lines observed by DKIST largely result from photoexcitation, their intensities decrease more slowly with height (closer to being in proportion to the electron density rather than its square) than do those of EUV spectral lines which result largely from collisional excitation~\citep[e.g.,][]{2016JGRA..121.8237L, 2018ApJ...852...52D}.  This makes plasma and magnetic field diagnostics to the edge of the DKIST field of view possible.  Importantly, lower down, in the chromosphere and low transition region, where current space based EUV observational efforts are focused, DKIST will allow inference of the vector magnetic field via spectropolarimetric signatures in lines such as He I 10830 \AA.  

DKIST's capabilities in combination with those of IRIS and Hinode will enable studies of solar wind acceleration physics from the chromosphere through transition region and into the corona.  Through coordinated efforts, these observatories will be able to address critical issues such as how the physical properties of the solar wind change with height, how that profile depends on position from the center to the edge of coronal holes, what defines coronal hole streamer boundaries, and how does reconnection at the boundary between closed and adjacent open field in active regions yield the observed wind properties.  Plasma diagnostics will allow estimates of the mass and energy flow along magnetic field lines, and coronal line width studies will help in understanding wave propagation and damping~\citep{2012ApJ...753...36H}.  Combining magnetic field, electron density and temperature measurements will make possible charge state evolution modeling of the accelerating solar wind and aid in the development of empirical models of the solar wind speed between 1 and 1.5 solar radii~\citep{2012ApJ...750..159L}.  Extrapolation to the freeze in height, with field extrapolations no longer strictly dependent on the photospheric field, will provide further links to in-situ instrumentation on PSP, SO and ACE and constrain the solar wind source locations (\S~\ref{ss:insitu}).  

\subsection{Magnetic Reconnection in the Solar Atmosphere}
\label{ss:reconnection}

\vskip -0.15in
\begin{figure}[h] 
\includegraphics[width=0.95\textwidth,clip=]{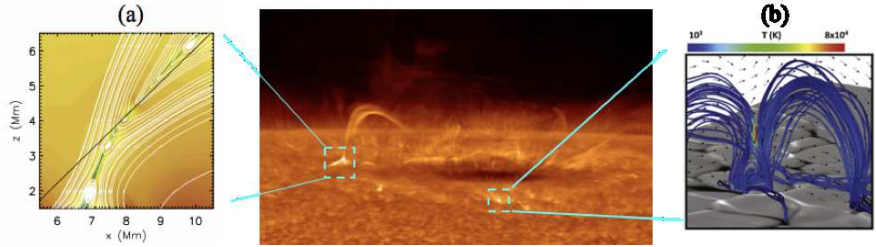}
\caption{The chromosphere near a sunspot as observed with the Hinode Solar Optical Telescope through the Ca II H broadband filter, showing transient brightenings, likely due to heating by reconnection events  ({\it middle}, from~\cite{2015PhPl...22j1207S}, fiducial lines added).  ($a$) Numerical simulation of a jet in which intermittent reconnection results from plasmoid instability.  From~\cite{2017ApJ...851L...6R}.  (b) Numerical simulation of an Ellerman-bomb, in which reconnection occurs between emerging bipolar fields.  From~\cite{2017ApJ...839...22H}.}
\label{f:fig15}
\end{figure}

\noindent
{\it What is the three-dimensional geometry of the magnetic field near reconnection sites?  What roles do ion-neutral collisions, current sheet instabilities and plasmoid ejection play?  How efficiently does the reconnection heat and accelerate the solar plasma?}
\vskip 0.1in

\noindent
Magnetic reconnection is a fundamental process that transforms magnetic energy into kinetic and thermal energies in astrophysical plasmas. The magnetic energy is stored, maintained and amplified over extended periods of time (minutes to hours in magnetic network and hours to weeks in sunspots and active regions) before being released suddenly during reconnection events, with flares occurring on small spatial scales (kilometers) over very short times (minutes).  Reconnection events can accelerate plasma in jet-like structures and induce local heating in the chromosphere and above. 

In addition to the ubiquitous spicules (\S\ref{ss:masscycle}), jets are observed in sunspot penumbra~\citep{2007Sci...318.1594K, 2018ApJ...869..147T}, at the edges of sunspots \citep{2012A&A...543A...6M}, in light bridges~\citep{2015ApJ...811..137T, 2018ApJ...854...92T}, and in the plage regions surrounding sunspots~\citep{2011ApJ...731...43N}.  These jets often display morphologies reminiscent of reconnection
sites \citep[e.g.,][]{2007Sci...318.1591S, 2012ApJ...759...33S}, but there are few direct measurements of the local magnetic field and its reconfiguration by reconnection.  Localized heating is observed, as Ellerman bombs and UV bursts, point-like brightenings in the wing of chromospheric lines such
as H I alpha and Ca II~\citep[e.g.,][]{2015ApJ...812...11V, 2016ApJ...823..110R, 2017ApJ...851L...6R, 2017ApJ...836...63T, 2018SSRv..214..120Y}.  These brightenings are often associated with bipolar moving magnetic features near sunspots or colliding bipolar structures in regions of emerging flux, similar to magnetic cancellation events associated brightenings with  in the quiet-Sun ~\citep{2016A&A...592A.100R, 2017ApJ...845...16N}.  Reconnection is again suggested, e.g., through the identification of plasmoids that enable fast reconnection~\citep{2017ApJ...851L...6R}, but without direct measurements of field reconfiguration.   

An important goal of DKIST is to measure the magnetic fields associated with reconnection phenomena at high resolution and simultaneously over multiple heights in the solar atmosphere from the photosphere into the chromosphere.  This will enable diagnosis of the field configuration at the reconnection sites before and after the reconnection events, facilitating reconstruction of the magnetic and thermodynamic history of the plasma in the reconnection volume.  It will allow detailed assessment of the total field annihilation, reconnection rate, total magnetic energy release and local heating induced.  Together with recent in-situ measurements of the underlying microphysical processes in the Earth's magnetosphere~\citep[e.g.,][]{2016Sci...352.2939B, 2018Sci...362.1391T, 2020ApJ...888L..16C, 2020JGRA..12525935H} and laboratory experiments~\citep[e.g.,][]{2012PhRvL.108u5001D, 2016RScI...87b5105G, 2016PhRvL.116y5001O, 2018PhPl...25e5501H, 2019PFR....1401054T, 2020PhPl...27b2109S}, such studies will advance our fundamental understanding of the energetics of solar and astrophysical reconnection, determining how the energy is partitioned between bulk flows and random motions, between plasma acceleration and heating~\citep{2011PhPl...18k1207J, 2016Sci...352.1176C, 2016PhPl...23e5402Y}.   

A key aspect of DKIST's contribution to reconnection studies, is the plasma environment which will be sampled. The lower solar atmosphere (photosphere and chromosphere) is relatively dense and weakly ionized, with an ionization fraction on the order of $10^{-4}$ at the height of the temperature minimum~\citep[e.g.,][]{2017PPCF...59a4038K}.  Weakly ionized plasmas are found over a wide range of astrophysical settings, including the atmospheres of other cool stars, the warm neutral interstellar medium ~\citep[ionization fraction of $10^{-2}$,][]{2013ApJ...764...25J}, dense cores of molecular clouds~\citep[ionization fraction of $10^{-7}$,][]{1998ApJ...499..234C}, and protostellar and protoplanetary disks~\citep[ionization fraction $10^{-10}$ or less, e.g.,][]{2019SAAS...45....1A}.  
Importantly, in both the solar and astrophysical contexts, low ionization fractions (in the range of those found in the solar chromosphere) increase reconnection rates.  Ion-neutral interactions can lead to increased resistivity and an increase in the effective ion mass, with consequent reduction in the Alfv\'en speed, steepening of current sheets, heating of the inflow and exhaust regions of reconnection sites, and enhancement of the plasmoid instability~\citep[e.g.,][]{1989ApJ...340..550Z, 1998ApJ...494...90C, 2011PhPl...18k1211Z, 2012ApJ...760..109L, 2012MNRAS.425.2824M, 2012ApJ...751...56M, 2015ASSL..407..285Z, 2020Ni}.  These in-turn have important implications for solar phenomena, likely being responsible for increased damping of MHD waves~\citep{2001ApJ...558..859D}, increased current dissipation and heating of the solar chromosphere~\citep{2012ApJ...747...87K, 2012ApJ...753..161M}, increased flux emergence rates into the corona, reduced Alfv\'en wave flux from photospheric foot-point motion, and changes in the structure of MHD shocks, prominences and quiet-Sun magnetic features~\citep[see][and references therein]{2017A&A...601A.103A}.   Observational verification the reconnection implications of partial ionization is more readily achieved in these solar contexts than in more distant astrophysical settings.

\subsection{Waves in the Solar Atmosphere}
\label{ss:waves}

\vskip -0.15in
\begin{figure}[h] 
\includegraphics[width=0.95\textwidth,clip=]{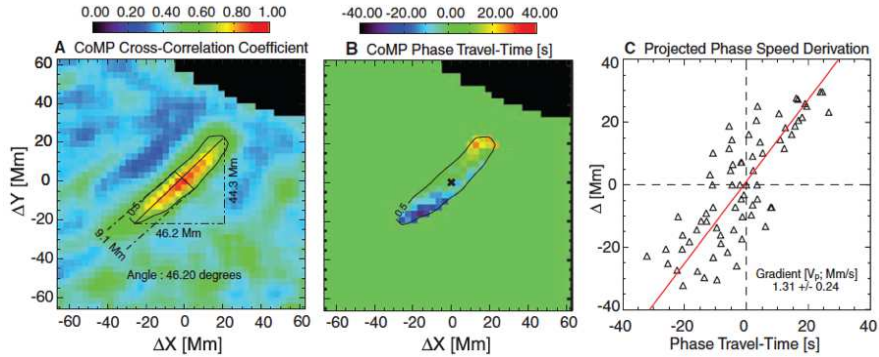}
\caption{Travel-time analysis of CoMP Doppler velocity measurements. In (A), a map of the cross correlation between the Fourier filtered time series (Gaussian filter centered on 3.5 mHz) at the reference pixel (marked with an x in B) and the filtered time series of neighboring pixels.  Contour of 0.5 defines region within which the phase speed of the waves is measured, with the phase travel times in that region shown in (B).  In (C), the relationship between the phase travel time and the distance to the reference pixel, yielding the phase speed of the analyzed region.  From~\cite{2007Sci...317.1192T}.}
\label{f:fig16}
\end{figure}

\noindent
{\it What wave modes are present at what heights in the solar atmosphere?  What are their sources?  What role do waves play in chromospheric and coronal heating?  How do the answers to these questions depend on the local magnetic field structure?}
\vskip 0.1in

\noindent
At least two implications promote the use of DKIST to study magnetohydrodynamic waves in the solar atmosphere~\citep[e.g.,][]{2013A&A...559A.107K}:  1) MHD waves carry energy into the solar atmosphere and, if dissipated at the correct heights, may provide at least a partial solution to the longstanding coronal and chromospheric heating problems, and 2) the presence of magnetic fields in the chromosphere and corona modify the waves observed, providing a possible opportunity to use them as diagniostics of the conditions there.  The chromosphere is a particularly important region of the solar atmosphere, as it modulates wave transmission into the corona; understanding the solar chromosphere is critical to constraining the mechanisms of wave energy transfer between the photosphere and the corona~\citep[see the recent review by][]{2015SSRv..190..103J}.  Moreover, while it is much cooler than the corona, the chromosphere's relatively high density and the efficiency of the cooling pathways available there, mean that high energy input is required to sustain radiative losses.  Typical radiative losses are estimated to be on the order of $10^6-10^7{\rm erg\ cm^{-2}\ s^{-1}}$ in the chromosphere compared to $10^4-10^6{\rm erg\ cm^{-2}\ s^{-1}}$ in the solar corona~\citep{1977ARA&A..15..363W, 1989ApJ...336.1089A}.  The solution to the coronal heating problem may well depend on solution of the chromospheric heating problem~\citep[e.g.,][]{2019ARA&A..57..189C}. 

The possibility that the Sun's chromosphere and the corona are heated by the dissipation of MHD waves, has led to a substantial body of research, starting over seventy years ago with suggestions that acoustic waves generated by convection are responsible for heating the solar atmosphere~\citep{1946NW.....33..118B, 1948ApJ...107....1S}.   With the observational discovery~\citep{1960IAUS...12..321L, 1962ApJ...135..474L} and correct theoretical interpretation~\citep{1970ApJ...162..993U} of the resonant solar acoustic oscillations, the diagnostic potential of the solar p-modes modes in the study of the solar interior was realized with helioseismology.  This was closely followed by an understanding of the observational and theoretical distinction between those modes and propagating atmospheric waves~\citep{1974ARA&A..12..407S} for an early summary), and in the intervening years, studies of chromospheric and coronal waves have leveraged increasingly sophisticated space and ground based instrumentation, and the consequent ever-increasing spatial, spectral and temporal resolution observations, to advance our understanding of chromospheric and coronal wave behavior.  Yet fundamental questions about wave energy transport and wave heating of the solar atmosphere persist~\citep[e.g.,][]{2007Sci...318.1572E}.  What processes dominate wave generation? What modes are excited?  How does the energy generated propagate into the solar corona?  What is the role of mode conversion?  What are the wave dissipation mechanisms that allow the solar corona to maintain its multi-million-kelvin temperature?  

Answering these questions requires the tracking of waves with height in the solar atmosphere, while simultaneously diagnosing changes in wave energy and the corresponding localized atmospheric heating~\citep{2015SSRv..190..103J}.   In combination with space-based UV observations, DKIST's  unprecedented multi-height spectropolarimetric measurements will be ideally suited to unraveling the nature of the waves, the energy propagation channels accessed and the atmospheric heating that results.

In a uniform plasma under the continuum approximation, there are three distinct types of MHD wave modes:  the slow and fast magnetoacoustic waves and Alfv\'en waves.  Solar observations are often interpreted in terms of these, but the manifestation of these modes in the stratified and highly magnetically structured solar atmosphere is much more complex~\citep[e.g.,][]{2007SoPh..246....3B, 2007Sci...317.1192T, 2007Sci...318.1574D, 2009Sci...323.1582J, 2012NatCo...3.1315M}.  In simple isolated magnetic geometries such as slabs or flux tubes, modes of the magnetic structures themselves (surface, body, kink, sausage, etc.) can be identified~\citep[e.g.,][]{2014masu.book.....P}, but in general, with space filling magnetic fields, the modes are mixed and coupled, and the waves are subject to resonant absorption, phase mixing and guided propagation~\citep[e.g.,][]{2000SoPh..192..373B, 2005LRSP....2....3N, 2014masu.book.....P, 2015ApJ...809...71O, 2015ApJ...809...72A}.  This makes observations difficult to interpret, but also consequently rich in diagnostic potential.  
Much of the complexity originates in the solar chromosphere.  While the coronal plasma can be treated as a single-fluid, low plasma-beta fully-ionized plasma, the chromosphere is a multi-fluid, partially ionized medium with a finite spatially varying plasma-beta coupled to a radiation field that is out of thermodynamic equilibrium~\citep[e.g.,][]{2007ASPC..368..107H}.  Additionally, waves in the solar chromosphere are subject to a highly structured (on sub-arcsecond to global scales) evolving field and flow~\citep[e.g.,][]{2009SSRv..144..317W}.  Since the manifestation of magnetohydrodynamic waves differs in different magnetic and plasma regimes (high-beta vs. low-beta, active regions vs. coronal holes, structured (on scale of wave) vs. unstructured field, as examples), wave signatures change as the waves propagate upward into the solar atmosphere, making tracking the wave energy from its source to the site of energy deposition challenging. 

Despite these difficulties, significant progress has been made and is anticipated with future observations~\citep[e.g.,][]{2007SoPh..246....3B}.   Convective motions in the photosphere excite both longitudinal magnetoacoustic and transverse Alfv\'enic perturbations.  Of the compressive wave field excited in the photosphere, only waves above the temperature minimum acoustic cut-off frequency are expected to propagate into the atmosphere above.  These magneto-acoustic compressive waves steepen, shock, and dissipate as they propagate into the chromosphere~\citep{1995A&A...300..302Z, 1997ApJ...481..500C}.  On the other hand, transverse Alfv\'enic perturbations can propagate through the chromosphere and into the corona with only weak damping.  The weak damping allows the waves to reach greater heights but also makes their role in plasma heating problematic.  One possibility is that the waves undergo mode conversion from Alfv\'enic to compressive, with the compressive motions providing an avenue for dissipation and heating~\citep[e.g.,][]{1982SoPh...75...35H, 1991A&A...241..625U, 1997ApJ...486L.145K}.  The height of mode conversion then becomes critical.  

Observations of compressive wave motions support this general picture with some modification.   High frequency power (three-minute period range) dominates the chromospheric wave field as expected, but it does so only in limited internetwork regions devoid of strong photospheric or chromospheric canopy fields~\citep{2007A&A...461L...1V, 2009A&A...494..269V}.  In plage regions, compressive wave power in the chromosphere reaches its maximum well below the acoustic cutoff frequency~\citep[e.g.,][]{2009ApJ...692.1211C}, leading to the formation of dynamic fibrils driven by magneto-acoustic shocks~\citep{2006ApJ...647L..73H}.  Low frequency magnetoacoustic waves also propagate upward from the photosphere in regions surrounding network elements~\citep{2006ApJ...648L.151J, 2007A&A...461L...1V}.  The presence of magnetic field thus appears to dramatically influence the frequency of the waves propagating into atmosphere.  Two mechanisms have been proposed to facilitate this:  inclined magnetic field creates wave-guides that effectively reduce the cut-off frequency~\citep{1973SoPh...30...47M, 2004Natur.430..536D} and radiative cooling of the wave temperature fluctuations can change the wave behavior, converting it from evanescent to propagating~\citep{1983SoPh...87...77R, 2008ApJ...676L..85K, 2009ApJ...692.1211C}.  These two mechanisms can be distinguished by careful measurement of the background magnetic field and wave temperature-velocity phase relations ~\citep[][but cf.~\citeauthor{2009ApJ...702....1H},~\citeyear{2009ApJ...702....1H}]{2013A&A...559A.107K}.  More generally, MHD wave properties depend in detail on the magnetic structure of the region within which they are propagating, and the propagation and amplitude of waves of different frequencies in the solar atmosphere depends on the partial ionization state of the plasma~\citep{2018A&A...618A..87K}.  

Direct detection of Alfv\'en wave perturbations is even more challenging than detection of compressive waves, but Alfv\'en waves have been successfully observed in the chromosphere~\citep{2007Sci...318.1574D, 2009Sci...323.1582J}, transition region~\citep{2014Sci...346D.315D} and solar corona, both as time varying nonthermal linewidths~\citep[see review of][]{2013SSRv..175....1M} and directly as linear polarization and Doppler velocity fluctuations~\citep[Figure 23, ][but cf.~\citeauthor{2008ApJ...676L..73V},~\citeyear{2008ApJ...676L..73V}]{2007Sci...317.1192T}.  The spectrum of the motions observed in the corona is similar to the solar p-modes, suggesting they have their origin in the solar photosphere~\citep{2009ApJ...697.1384T, 2019NatAs...3..223M}, though other evidence indicates that they may be more closely related to the ubiquitous Alfv\'en waves observed on spicules~\citep{2007Sci...318.1574D, 2011Natur.475..477M}.  Observational evidence for mode coupling in both directions has also been reported.  Upward propagating transverse motions coupled to longitudinal motions, subsequently dissipated, were identified in quiet-sun network bright points by careful cross correlation of wavelet identified wave packets at multiple heights~\citep{2003ApJ...587..806M, 2004ApJ...604..936B}, and evidence for photospherically generated longitudinal magnetoacoustic oscillations propagating upward before undergoing mode conversion to predominantly transverse motions were found in observations of spicules~\citep{2012ApJ...744L...5J}. 

While these observations are compelling, to fully disentangle the signal of the different wave modes as they travel upward through the complex solar atmosphere from the photosphere to corona requires simultaneous intensity, polarimetry (magnetic field) and Doppler measurements at many heights in order to deduce the background magnetic field, thermodynamic state of the plasma and wave perturbations.  Such measurements are required for many different magnetic field regimes. The high-resolution limb spectropolarimetry needed to achieve these measurements is difficult because low photon statistics off of the solar limb limit the accuracy of the Stokes profiles deduced and because low photon counts make the use of adaptive optics, needed to achieve the high resolutions required, challenging.  DKIST is poised to meet these challenges.  Its large aperture and coronographic capabilities, will allow spectropolarimetric measurements of chromospheric and coronal waves on unprecedentedly small spatial scales and at high cadence.

\vfill\eject
\subsection{Flux Emergence into the Non-Eruptive Solar Atmosphere}
\label{ss:emergence}

\vskip -0.15in
\begin{figure}[h] 
\includegraphics[width=0.95\textwidth,clip=]{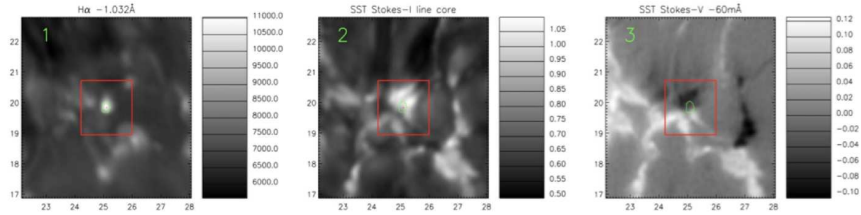}
\caption{Swedish Solar Telescope observations of an Ellerman bomb in H$\alpha - 1.032 {\rm \AA}$ (left), co-spatial Fe I 6302$\rm{\AA}$ line core Stokes I (middle) and co-spatial Fe I $6302{\rm\AA} - 60{\rm m\AA}$ Stokes V (right).  Ellerman bombs occur near solar active regions or areas of enhanced photospheric magnetic activity, when emerging flux interacts with pre-existing opposite polarity field.  From~\cite{2016ApJ...823..110R}.}
\label{f:fig17}
\end{figure}

\noindent
{\it How does magnetic flux emergence impact energy storage and release in the chromosphere and corona?  How are the underlying magnetic reconnection geometries and heights reflected in observations of different types of small-scale flux emergence/cancelation events?}
\vskip 0.1in

\noindent
Magnetic flux emergence allows for mass, energy and magnetic field to flow from the solar interior through the photosphere and into the chromosphere and corona above.  The impact of flux emergence depends on the amount of field emerging, its spatial distribution and the preexisting structure of the atmosphere into which it is emerging.  A small bi-pole will have vastly different impact when emerging into a coronal hole, the quiet Sun or adjacent to a delta-spot active region.  The emergence of magnetic field through the solar photosphere and its reconnection with pre-existing field at different heights thus leads to a variety of dynamic phenomena on different temporal and spatial scales:  global magnetic field restructuring~\citep{2014ApJ...782L..10T}, flaring active regions~\citep{2019LRSP...16....3T}, emerging flux regions or arch filament systems~\citep[e.g.,][]{2017ApJS..229....3C, 2018ApJ...855...77S}, Ellerman bombs and quiet-sun Ellerman bomb-like brightenings.  Larger scale field reconfiguration and destabilization was considered in Section~\ref{ss:bchange}.  The discussion in this section is focused on flux emergence on scales below those of an active region and the local response to that emergence.  Tracking the consequences of emergence allows us to understand the fundamental energy storage and release mechanisms which may be responsible for heating the chromosphere and corona.  
 
Ellerman bombs~\citep{1917ApJ....46..298E} are point-like brightenings seen in the wings of chromospheric lines (such as those of H I and Ca II).  They are often associated with bipolar moving magnetic features around well-developed sunspots or as colliding bipolar structures in emerging flux regions~\citep[e.g.,][]{2013JPhCS.440a2007R, 2016ApJ...823..110R}.  The atmospheric heating and associated bi-directional flows observed are thought to be caused by magnetic reconnection near the temperature minimum~\citep[e.g.,][]{2008PASJ...60...95M}, with rapidly evolving flame-like features in the wings of the Balmer $\alpha$ line suggesting reconnection of small scale fields when observed at high resolution~\citep[such as with the one meter Swedish Solar Telescope;][]{2011ApJ...736...71W}.  While Ellerman bombs primarily occur in active regions, and are sometimes associated with arch-filament systems~\citep[e.g.,][but see~\citeauthor{2013JPhCS.440a2007R},~\citeyear{2013JPhCS.440a2007R}, who suggest these are distinct phenomena]{1987SoPh..108..227Z, 2002ApJ...575..506G, 2015A&A...583A.110M}, suggesting that larger scale flux eruption may underly their occurrence~\citep{2004ApJ...614.1099P}, recent observations indicate that smaller, shorter-lived and lower ${\rm H}\alpha$-wing-intensity-contrast events also occur in the quiet-sun.  These quiet-sun Ellerman-like brightenings~\citep{2016A&A...592A.100R, 2018MNRAS.479.3274S} are similar to Ellerman bombs, but with typical size scales of $\sim0.25 - 0.5$ arcsecs and lifetimes of less than a minute.  Bright flame-like emission in the wings of ${\rm H}\alpha$, similar to that observed for Ellerman bombs, suggests a common reconnection origin, but the heating profile and the characteristics of the magnetic field evolution observed may imply a somewhat different reconnection scenario~\citep{2016A&A...592A.100R}.  Similarly, UV bursts appear to be reconnection events that differ from both Ellerman bombs and quiet-Sun Ellerman-like brightenings in their magnetic topology, atmospheric penetration height and energy, with the plasma in these events heated to transition region temperatures~\citep{2017ApJ...845...16N, 2018SSRv..214..120Y, 2020A&A...633A..58O}.

This full range of small-scale flux emergence events is well suited for study with the planned DKIST instrument suite.  High-resolution multi-thermal moderate field-of-view observations can resolve the finely structured thermal properties of the plasma with height, and high sensitivity spectropolarimetric observations can be used to determine the height-dependent vector magnetic field and Doppler velocities.  These quantities in turn can be used to estimate the local electric fields and energy fluxes.  Electric field measurements, derived from time-series of vector magnetic field and Doppler velocity maps~\citep[e.g.,][]{2012SoPh..277..153F, 2014ApJ...795...17K}, are important for determining the rate of electromagnetic energy transport into the solar atmosphere, the Poynting flux through the photosphere.  To date, such measurements have been made only in strong field regions due to the limited reliability of vector magnetic field deductions in weak field regions~\citep{2015ApJ...811...16K}.  Using DKIST's unprecedented vector magnetic field measurement capabilities, electric field determination can be dramatically improved, and when combined with transition region and coronal observations, can be used to address a range of open questions about the net transfer of magnetic energy into the solar atmosphere:  How much magnetic energy reaches the chromosphere?  Why are active region cores the sites of the hottest and most dense coronal loops?  Is there a measurable correlation between the input of energy at the photosphere and consequent emission in the chromosphere, transition region and corona?  Is the injected energy dissipated immediately, or stored with some typical latency time?  Importantly, with DKIST these questions can be addressed as a function of solar activity to uncover the underlying energetics of the solar cycle. 

A recent study of a UV burst with the Sunrise balloon-borne telescope~\citep{2018A&A...617A.128S} revealed dynamic substructure on scales of 75 km, and likely smaller, within a chromospheric heating site.  DKIST will not have UV capabilities, but the He I D3 and 10830 lines may serve as useful proxies when studying plasma at transition region temperatures~\citep{2017A&A...598A..33L}.  Coordinated observations with space-based UV assets are also anticipated.  Numerical models will play a critical role in data interpretation.  Radiative magnetohydrodynamic simulations of magnetic reconnection during flux emergence show reconnection events similar to Ellerman bombs and other burst events~\citep{2017A&A...601A.122D, 2017ApJ...839...22H}.  The occurrence of these in simulations of emerging active regions suggests a possible role for an underlying emerging  large-scale twisted loop structure~\citep[e.g.,][]{2007ApJ...657L..53I, 2009A&A...508.1469A}.  Detailed modeling of expected DKIST spectropolarimetric measurements in the photosphere and chromosphere will enable careful comparisons between the simulated and observed plasma flows and magnetic field evolution at the Ellerman bomb sites~\citep{2006SoPh..235...75S, 2016KPCB...32...13K} to determine if the existence of such structures is implicated by observations of the Sun.  Additionally, direct detection of the brightness temperature excess at bomb sites and in the surrounding atmosphere will be possible with coordinated observations using DKIST and radio instruments such as the ALMA.  These will allow careful assessment of magnetic reconnection heating efficiency in the partially ionized chromospheric plasma, and help clarify the overall importance of Ellerman bombs and other burst and localized brightening events to the energy budget.

\subsection{Multilayer Magnetometry and Magnetic Field Extrapolation}
\label{ss:multilayer}

\vskip -0.15in
\begin{figure}[h] 
\includegraphics[width=0.75\textwidth,clip=]{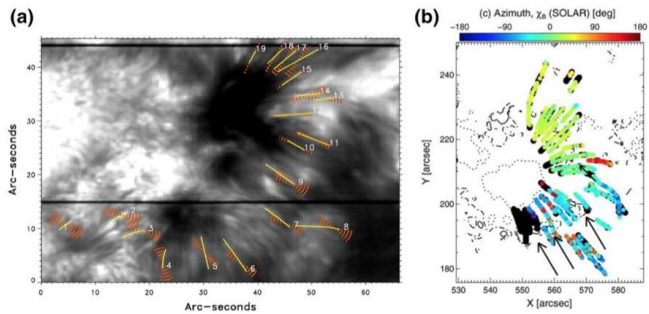}
\caption{Orientation and azimuthal field direction of super-penumbral fibrils.  In (a), yellow lines trace the core of Ca II $8542 {\rm\AA}$ intensity orange cones show the range of transverse magnetic-field azimuth compatible with the linear polarization measurements.  In some regions the fibril and horizontal field orientations appear aligned (fibrils 9 -19), while in others they do not (fibrils 1 - 5).   From ~\cite{2011A&A...527L...8D}.  In (b), magnetic field azimuth along selected super-penumbral fibrils (inferred from He I $10830 {\rm\AA}$ observations).  Field azimuth in chromospheric fibrils are generally consistent with those of the penumbral filaments below (not shown).  Black dots mark locations of significant deviation.  They are largely restricted to where fibrils are rooted.  From~\cite{2013ApJ...768..111S}.}
\label{f:fig18}
\end{figure}

\noindent
{\it How does the magnetic field change with height and evolve in time through different layers of the solar atmosphere?  How do we best use multi-layer magnetic field observations to constrain chromospheric/coronal field extrapolations?}
\vskip 0.1in

\noindent
The stratified solar atmosphere is threaded by magnetic field.  Most existing solar instruments employ one or few spectral lines at a time and thus simultaneously probe the magnetic field over a limited range of heights in the atmosphere.  Moreover, the rapid decrease of the field intensity with height implies very weak chromospheric polarization signals, making their measurement challenging with existing facilities ~\citep{2013ApJ...768..111S}.  A common need underlying much of the first critical science proposed for DKIST is thus multi-line high-sensitivity spectropolarimetry.  In meeting this need, DKIST will enable simultaneous multi-height measurements of the solar atmosphere that will revolutionize our understanding of the coupling between the different atmospheric layers.

The solar chromospheric plasma is highly dynamic, inhomogeneous and out of local thermodynamic equilibrium.  Magnetic fields play a central role in its behavior.  Observations of the upper chromosphere, particularly near active regions, are dominated by intricate filamentary structures called fibrils.  These fibrils are seen in images taken in the cores of strong chromospheric lines, such as H$\alpha$, Ca II K, the Ca II IR triplet~\citep[e.g.,][]{2006ApJ...647L..73H, 2008A&A...480..515C, 2009A&A...502..647P} and the He I 587.6 and 1083 nm lines~\citep{2013ApJ...768..111S}.  They are are assumed to be aligned with the magnetic field.   Fine scale chromospheric filamentary structure is detected even above sunspots, where the photospheric umbral field is strong, thought to be largely vertical and quite uniform; umbral flashes show filamentary fine-structure and apparent associated horizontal magnetic field, that appears to be at or below the scale of current resolution limits~\citep[e.g.,][]{2009ApJ...696.1683S}.  

The assumed alignment of chromospheric fibrils with the magnetic field is an important tool in active region field extrapolation~\citep{2008SoPh..247..249W, 2011ApJ...739...67J, 2012ApJ...752..126Y}, and accurate field extrapolation is critical to assessments of the free energy available for solar flares and eruptions, active region flaring potential and stability and models of chromospheric heating.  However, since the early conjecture by George Ellery Hale that fibrils around sunspots resemble lines of magnetic force~\citep{1908ApJ....28..100H}, a conjecture made even before his momentous measurement of the field using the then recently described Zeeman effect~\citep[][see~\citeauthor{1999ApJ...525C..60H},~\citeyear{1999ApJ...525C..60H}, for a brief history]{1908ApJ....28..315H}, the observational evidence for a direct association between fibrils and the local magnetic field direction has remained sparse.  This is largely due to the small amplitude of the polarized signals within the primarily horizontally oriented (relative to solar surface) chromospheric fibrils ($<0.1$\% in linear polarization), and conclusions are wide-ranging.  
Attempts to directly measure the alignment between the thermal and magnetic structure of super-penumbral fibrils have yielded results ranging from often but not always aligned~\citep{2011A&A...527L...8D} to aligned within $\pm10$ degrees with no evidence for misalignment~\citep{2013ApJ...768..111S}.  A recent Bayesian statistical analysis~\citep{2017A&A...599A.133A} finds penumbral and plage fibrils to be well aligned but with non-negligible dispersion.  That study concludes that higher signal-to-noise observations are needed to discern whether the misalignment seen in some simulations, particularly  those which include ion-neutral coupling~\citep{2016ApJ...831L...1M}, is compatible with that seen on the Sun.  More broadly, understanding the three-dimensional connectivity of the chromospheric field to the photosphere~\citep[at sites such as the outer foot-points of super-penumbral fibrils, Figure~\ref{f:fig18}; ][]{2013ApJ...768..111S} requires high-resolution high-sensitivity spectropolarimetric measurements.  DKIST will enable these.

The connectivity of the magnetic field through the atmosphere is an important issue outside of active regions as well.  In the quiet-Sun, the photospheric magnetic field is organized by supergranular motions into strong flux concentrations on the network scales and mixed-polarity internetwork magnetic field on the scale of granulation.  The field expands above the photosphere into the chromosphere and corona, and the presence of the weak small-scale internetwork magnetic field has a considerable effect on the overall field geometry with height, which deviates significantly from a simple funnel expansion model~\citep{2003ApJ...597L.165S, 2006ApJ...647L.183A, 2019ApJ...878...40M}.  This is critical because the magnetic field forms the underlying channel for energy transport into the solar chromosphere and corona, playing an important role in the acceleration of the solar wind~\citep[e.g.,][see also previous sections in this Research Area]{1976RSPTA.281..339G, 2005A&A...435..713A, 2007ApJ...654..650M, 2008A&A...478..915T}.  Beyond idealized potential or force free field extrapolations, the variation in field strength and topology with height is typically poorly known.

Force-free extrapolations can be improved.  Because of the availability of photospheric magnetograms,  field extrapolations generally depend on photospheric boundary conditions, but these boundary conditions are inconsistent with the force-free assumption because both gas pressure and gravity play important roles at photospheric heights.  Employing  chromospheric field measurements instead, measurements made at sufficiently great heights where the magnetic field is much more dominant and is consequently configured much closer to a force-free state~\citep{2016ApJ...826...51Z}, can significantly improve the reliability of field extrapolations~\citep{2019ApJ...870..101F}; when combined with photospheric measurements, even very limited chromospheric field measurements allow significant improvement~\citep{2019ApJ...870..101F}.   Additionally, careful comparison between independent extrapolations using photospheric and chromospheric field measurements can aid in determining relative line formation heights and in resolving the 180-degree field ambiguity~\citep{2012ApJ...748...23Y}, and reliable multi-height magnetic field measurements using DKIST will not only contribute to more reliable extrapolation of that field, but will also strengthen deductions of the local field at the site of measurement.

\vfill\eject
\subsection{Magnetic Topology, Helicity and Structures}
\label{ss:helicity}

\vskip -0.15in
\begin{figure}[h] 
\includegraphics[width=0.85\textwidth,clip=]{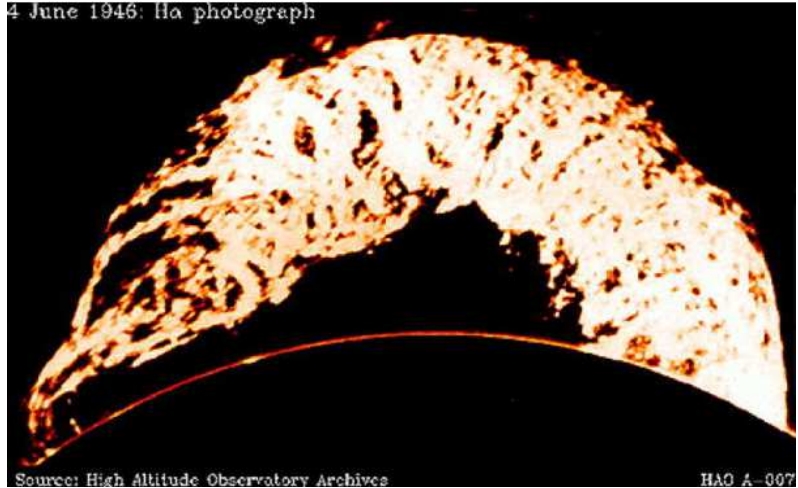}
\caption{The Grand Daddy Prominence.  Photographed by W.O. Roberts at Harvard College Observatory, Climax, Colorado on 4 June 1946 through a filter centered on H$\alpha$.  This prominence extends ~200000 km above the solar surface. Courtesy of the High Altitude Observatory/NCAR
(\url{https://www2.hao.ucar.edu/Education/Sun/grand-daddy-prominence}).}
\label{f:fig19}
\end{figure}

\noindent
{\it What role does the near conservation of helicity play in the structuring of the solar corona and coronal mass ejections?  Are measurements of helicity useful indicators of imminent eruption?  Do vortex tubes exist and do they act as portals for MHD wave propagation and energy transfer in the quiet solar atmosphere?  What are the magnetic field properties of vortex structures in the lower solar atmosphere?}
\vskip 0.1in

\noindent
Magnetic helicity is a property of the field which helps describe its topology, whether it is twisted or linked, writhes or is sheared~\citep[e.g.,][]{1969JFM....35..117M, 1999PPCF...41B.167B, 2014PNAS..111.3663M}.  It is strictly conserved in ideal MHD~\citep{1958PNAS...44..833W} and during two-dimensional reconnection, and approximately conserved after three-dimensional reconnection \citep{1974PhRvL..33.1139T, 1984GApFD..30...79B, 1997AdSpR..19.1789H}.  Magnetic helicity cascades to larger scales \citep{2006ApJ...640..335A} and is converted from one form to another as it moves to larger scales. This means that as magnetic fields reconfigure in the solar atmosphere, magnetic helicity is lost only slowly.

Many solar magnetic structures contain self-helicity (internal twisting) and/or mutual helicity (tangling about each other), with helicity observed on the Sun on scales ranging from the largest global to the smallest quiet Sun magnetic fields~\citep[e.g.,][and references therein]{2003AdSpR..32.1867P, 2003ApJ...588..620W}.  The intense magnetic field structures that form within and rise through the Sun's convection zone, and are thought to be responsible for sunspots and most solar activity, are likely highly twisted.  Untwisted, such tubes would lose their integrity as they ascend.  Observations of sunspots show that they rotate as they emerge~\citep[e.g.,][]{1909MNRAS..69..454E, 2003SoPh..216...79B}.  That rotation is likely associated with an underlying large-scale twisted flux tube rising through the photosphere~\citep[e.g.,][]{2015A&A...582A..76S}, and the helicity that enters solar atmosphere on all scales from below is important for the structure and behavior of the field there.  

Some coronal loops appear to be tangled about each other forming a braided pattern~\citep{1983ApJ...264..642P, 2013Natur.493..501C, 2017ApJ...837..108P}, and the degree of coronal loop braiding overall has been used to estimate the role of small scale reconnection in coronal heating~\citep{2007ApJ...662L.119S, 2017ApJ...835...85K}.  The appearance of braided structures, however, depends critically on the details of the field line windings within them.   Loop substructure can be difficult to distinguish in observations~\citep{2009ApJ...705..347B, 2017ApJ...837..108P, 2019A&A...626A..98L}, so careful high-resolution spectropolarimetric observations are vital.  Further, contrary to expectation, coronal loops have quite uniform width along their length~\citep{2000SoPh..193...53K, 2000SoPh..193...77W}.  Explanations for the observed lack of expected field expansion rely on such loop substructure, either to provide magnetic tension~\citep[e.g.,][]{2006ApJ...639..459L} or to allow fine scale interchange reconnection that enables cross field loss of the hot loop plasma~\citep{2007ApJ...662L.119S, 2009ApJ...706..108P}.  Distinguishing these observationally is important in understanding the thermodynamic structure of the corona and its maintenance.

Beyond loop substructure and heating, the accumulation of magnetic helicity in the corona appears to be of key importance.  Magnetic helicity accumulation accompanies the magnetic energy build-up that precedes the loss of stability when a coronal mass ejection is initiated~\citep{2005ARA&A..43..103Z, 2006ApJ...644..575Z, 2016A&A...594A..98Y}.  The precise stability implications of the helicity accumulation are still somewhat uncertain~\citep{2003ApJ...585.1073A, 2005ApJ...624L.129P}, and some measures of helicity may be more reliable instability indicators than others~\citep{2017A&A...601A.125P}.  Independent of the exact triggering mechanisms, coronal mass ejections associated with filament eruptions often reveal large-scale helical magnetic structures that partially unwind during an eruption~\citep[e.g.,][]{1987SoPh..108..251K, 2016NatCo...711837X}.  Coronal mass ejections may play an essential role in relieving the solar atmosphere of accumulated helicity~\citep{2006ApJ...644..575Z}, and detailed observational assessment of the coronal helicity budget and its role in coronal mass ejection initiation are crucial.  DKIST will significantly enhance our ability to deduce the magnetic helicity in pre-and post-eruptive structures. Moreover, highly twisted structures typically have a high magnetic energy, and that twist can lead to local instability (e.g., the kink instability), reconnection, flaring and small scale eruptions beyond coronal mass ejections proper.

Prominences (or filaments on the disk) are cool plasma structures (at chromospheric temperatures) embedded into the hot corona~\citep[e.g.,][]{2014LRSP...11....1P, 2018LRSP...15....7G}.  They are observed in emission off disk and as filaments in absorption when observed on-disk.  These helical structures are central to those coronal mass ejections associated with filament eruptions, but direct measurements of their magnetic field topology are limited~\citep[e.g.,][]{2003ApJ...598L..67C, 2012ApJ...750L...7X, 2012A&A...539A.131K, 2014A&A...561A..98S}.  Based on magnetostatic models and plasma support and stability considerations, the magnetic field in prominences is thought to be fundamentally tangential to the solar surface.  This may be true even in the feet (barbs) of the prominence, which can be the sites of swirling motions sometimes called solar tornados~\citep[e.g.,][]{2013A&A...549A.105P, 2016ApJ...826..164L, 2016ApJ...818...31L, 2015ApJ...810...89M, 2018ApJ...861..112M}. 
Sometimes plasma flows in opposite directions (known as counter-streaming/bi-directional flows) along the spine of filaments, as well as in barbs.  The source/origin of these bi-directional flows is still not well understood~\citep{1998Natur.396..440Z, 2020ApJ...897L...2P}.
They may reflect the rise and expansion of a twisted flux rope into the corona or the presence of a large vortex flow in the photosphere ~\citep[see][and references therein]{2015ApJ...810...89M}.  It is unclear which interpretation is correct, but the motions are implicated in prominence stability~\citep{2018ApJ...861..112M}.  Understanding this complex evolving dynamics is critical to assessing the role of prominences in the solar mass cycle (\S\ref{ss:masscycle}) and coronal mass ejection initiation (\S\ref{ss:precursors}).  

Magnetic helicity in the solar atmosphere has two sources, the emergence of helical field through the photosphere from below~\citep[e.g.,][]{1996ApJ...462..547L, 2008ApJ...673..532T} and field foot-point motions due to photospheric flows~\citep[e.g.,][]{1999GMS...111..213V, 2001ApJ...560L..95C} including differential rotation~\citep[e.g.,][]{1999GMS...111..213V, 2000ApJ...539..944D}.  Observations aimed at understanding the atmospheric helicity budget can either focus on these sources of helicity or attempt a direct measurement of the helicity in the solar atmosphere itself~\citep[see reviews][or more recent references in~\citeauthor{2018ApJ...865...52L},~\citeyear{2018ApJ...865...52L}]{2003AdSpR..32.1855V, 2007AdSpR..39.1674D, 2009AdSpR..43.1013D}.  To date, the latter has relied on field extrapolation while the former has been built on measurement of the photospheric field and flows and models of their implication for helicity injection into the atmosphere. DKIST observations will contribute to the improvement of both of these techniques.  DKIST's on-disk, multi-layer magnetometry (\S\ref{ss:multilayer}) will allow for more direct inference of the sheared and twisted field emerging through the photosphere and present in the chromosphere~\citep[e.g.,][]{2012A&A...539A.131K}, while DKIST coronal field measurements will help verify coronal field extrapolation models \S 3.3.6) and constrain magnetic field models~\citep[e.g.,][]{2011ApJ...731L...1D, 2013ApJ...770L..28B}.  DKIST observations will be regular and sustained allowing study of active region evolution and filament formation~\citep{2012ApJ...748...77S}.

For decades, braiding and twisting of magnetic field rooted within solar surface convection has been thought to be an efficient mechanism for solar atmospheric heating~\citep[e.g.,][]{1972ApJ...174..499P, 2006SoPh..234...41K, 2015Natur.522..188A}, with granular downflows providing conditions for vorticity and consequent magnetic helicity production and setting the stage for reconnection and MHD wave generation~\citep{1975SoPh...42...79S, 1985SoPh..100..209N,2000SoPh..192...91S, 2011A&A...526A...5S, 2012Natur.486..505W, 2012PhyS...86a8403K}.  Vortical flows in the photosphere can produce coherent magnetic field structures, twisted magnetic flux tubes, and these are braided by the random walk of downflows as the granulation evolves.  In turn, photospheric vortex motion is inferred from the motions of the magnetic bright points coincident with the twisted field structures~\citep[e.g.,][]{2008ApJ...687L.131B}.  Those motions reveal rotating flows over a range of spatial scales.  In the chromosphere, the magnetic field form swirls, 0.4-2 Mm in diameter that last for 5-10 minutes~\citep[e.g.,][]{2012Natur.486..505W}.  These appear as rotating spirals in H$\alpha$ and in the near-IR Ca II 854.2 nm line.  The near IR signal is prominent for small-scale swirls near the solar disk center~\citep{2009A&A...507L...9W} in high-resolution ground-based observations (40 km per pixel @630 nm) using CRisp Imaging Spectro-Polarimeter ~\citep{2008ApJ...689L..69S} at the Swedish Solar Telescope~\citep{2003SPIE.4853..341S}.  Observations of heating~\citep{2016A&A...586A..25P}, swaying ~\citep{2018A&A...618A..51T} and wave-like motions~\citep{2019ApJ...881...83S} within the spiraling chromospheric swirl structures suggests a rich dynamics and underlying magnetic field structure that awaits further exploration.  

In addition, photospheric vortices have been recently detected in large numbers using local correlation tracking~\citep[e.g.,][and references therein]{2008ASPC..383..373F} techniques in images of photospheric continuum intensity~\citep{2018ApJ...869..169G, 2019ApJ...872...22L}.  
At any one time about $10^{6}$ such photospheric vortices may cover about 2.8\% of the solar surface.  Extension of these detections to chromospheric heights and measurement of the correlation between swirls at different heights~\citep{2019NatCo..10.3504L} suggests that they are associated with upward propagating Alfv\'en waves.  Associated energy fluxes are sufficient to support local chromospheric energy losses if the waves dissipated.  This conclusion depends on the correlation between the velocity and magnetic fields within the swirls, which has been investigated via numerical simulations~\citep{2019A&A...632A..97L}, and is a compelling target for DKIST observations.   While most of the observed photospheric  vortices have quite short lifetimes~\citep{2019NatCo..10.3504L}, the more persistent vortices may more generally serve as important energy portals because they support a wide range of MHD waves.  They may also be 
precursors of the so-called magnetic tornadoes~\citep{2013JPhCS.440a2005W} observed in the chromosphere, which in turn may be the source of twist at the foot points of coronal loops that power the continuous basal coronal heating~\citep{2012Natur.486..505W}.   Detecting small scale vortices in the chromosphere is a challenging, and the substructure within them has not yet been resolved even in the photosphere.  With DKIST, we have to opportunity to discover the frequency and amplitude of vortex motions in the photosphere and chromosphere  down to very small scales, clarify the dynamical and magnetic connectivity of vortical structures across atmospheric layers, and assess the collective contribution of these to solar atmospheric heating and MHD wave generation.
   
\section{Long-Term Studies of the Sun, Special Topics and Broader Implications }
\label{s:special} 

The Sun exhibits remarkable changes over decadal time scales, with the spatial distribution of active regions, sunspots, coronal holes, and prominences continuously changing along with the roughly 11-year polarity reversal.   The frequency and severity of solar events, such as flares and CMEs, are strongly dependent on the phase of the solar cycle, and the amplitude of the basal high-energy radiative output of the Sun (in the X-ray and EUV) is modulated by orders of magnitude over the course of a solar cycle.  These cycle-dependent changes and others hold clues about the underlying operation of the global solar dynamo.

Cycle-related changes have been observed on the Sun at scales as small as supergranulation; supergranules get larger when the Sun is more active.  It is possible that careful synoptic observations at higher resolution will reveal cycle-dependent dynamics at even smaller scales.  Assessing these variations may provide fundamental insights into the multi-scale turbulent dynamics of highly stratified convection.  Additionally, global modulation of solar activity occurs on time scales longer than the solar cycle as well.   Variation in the strength of solar maxima and the duration and depth of solar minima lead to a cascade of consequences from total and spectral irradiance variations to changes in the solar-wind ionization state and mass flux, to variations in the complexity of the interplanetary magnetic field and consequent modulation of the cosmic ray flux into the inner heliosphere.  These, in turn, impact the near-Earth and interplanetary space environments, the Earth's upper atmosphere and to some degree the Earth's climate.  Understanding, modeling and potentially forecasting these impacts requires long-term consistent monitoring of the Sun's magnetic and plasma properties.  This is most readily achieved by a ground-based facility, which can maintain the required spatial and temporal resolution and spectropolarimetric sensitivity over decadal time scales because it can be fine-tuned, repaired and upgraded as needed.  

DKIST's unique capabilities will allow for a broad range of investigations beyond the core solar physics areas which form the bulk of this paper.  For example, as with previous space based solar observatories (such as SOHO, SDO and STEREO), DKIST will make contributions to cometary science and the use of comets as probes of otherwise inaccessible regions of the solar corona, regions inaccessible to both remote sensing observations (too faint to be detected) and direct in-situ measurements (closer than Parker Solar Probe's closest approach to the Sun).  As another example, just as atomic physics has benefited from the high resolution spectrometers on board SOHO (CDS and SUMER) and Hinode (EIS), so too will it benefit from DKIST's advanced spectropolarimetric instrumentation and measurements.  Previous missions have provided the critical line intensity measurements necessary to benchmark theoretical predictions of atomic spectra.  These have led to major advances in atomic transition calculations and spectral synthesis codes, such as CHIANTI, and in turn have allowed for far more accurate predictions of solar X-ray, EUV and UV spectral irradiance and improved spectral plasma diagnostics.  DKIST will similarly make fundamental contributions to the quantum mechanical underpinnings of polarization spectroscopy over a broad range of wavelengths extending into the near IR.

Several critical science topics in this research area are discussed in detail below, including 1) long-term studies of the Sun; 2) Sun-grazing comets; 3) Mercury transit science; 4) turbulence and reconnection processes; and 5) synergies with in-situ measurements.

\vfill\eject
\subsection{Long-Term Studies of the Sun}
\label{ss:synoptic}

\vskip -0.3in
\begin{figure}[h] 
\includegraphics[width=0.8\textwidth,clip=]{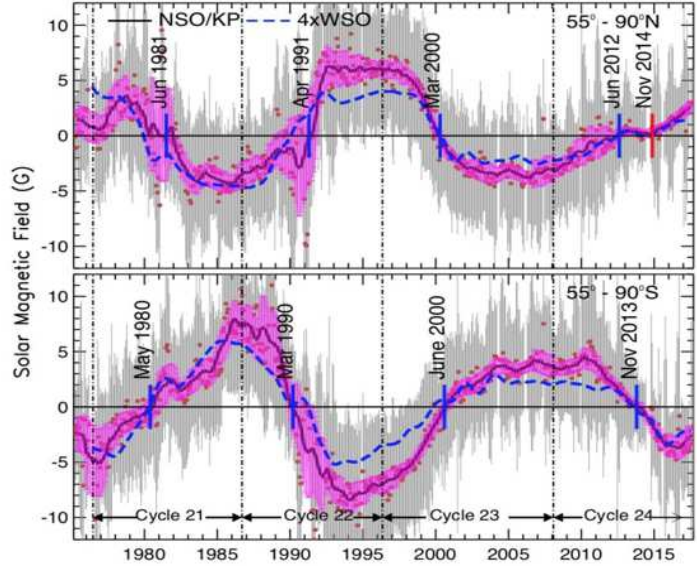}
\caption{Polar field strengths from NSO/KP averaged over the latitudes 55-90 degrees (red dots with $1\sigma$ variance in gray) in the solar northern (top) and southern (bottom) hemispheres.  Dashed blue curve over plots polar field measurements from the Wilcox Solar Observatory.  Field reversals, determined from NSO/KP smoothed data (purple curve), are indicated with blue vertical fiducial lines.  The red vertical fiducial line marks the completion of Cycle 24 northern polar reversal.  Cycle 24 field reversal was unambiguous in the southern but extended in time in the northern hemisphere.  From~\cite{2018A&A...618A.148J}.}
\label{f:fig21}
\end{figure}

\noindent
{\it How do the properties of the Sun's small-scale magnetic field depend on the phase of the solar cycle?  Does internetwork magnetism show cycle variations?  How do the polar magnetic fields and flows evolve with the solar cycle?  Are there systematic changes in prominence/filament fields which reflect a helicity cycle?}
\vskip 0.1in

\noindent
DKIST will provide regular, sustained and repeated photospheric, chromospheric and coronal measurements of specific targets using very similar instrumental configurations over many years.  The advantages of such repeated long-term observations of the Sun are most evident in the context of the large-scale solar dynamo.  Though the global magnetic field of the Sun evolves over the 11-year activity cycle, its detailed behavior likely depends on physical processes that occur on smaller spatial and shorter temporal scales.  Many questions regarding the connections between small-scale processes and long-term behavior remain unaddressed due to a lack of high-resolution high-sensitivity spectropolarimetric observations of the kind DKIST will make over an extended period of time.  DKIST will additionally make regular detailed maps of the coronal magnetic field, a unique observational capability currently missing, and together with coordinated in-situ observations by space missions such as the Parker Solar Probe and Solar Orbiter, these will be used to understand the Sun's cycle-dependent influence on the heliosphere.  Both types of synoptic observations can be used as benchmarks in the study of activity cycles of other stars~\citep[e.g.,][]{2017LRSP...14....4B} allowing further contextual understanding of the Sun's behavior.

Facular-scale magnetic elements are the building blocks of the magnetic field at the solar surface.  In some dynamo models they play an essential role in transport, flux cancelation and field reversal~\citep[e.g.,][and references therein]{2010LRSP....7....3C}, but systematic study of their motions and mutual interactions over time scales during which they are subject to differential rotation and meridional flow~\citep[such as those of][]{2017ApJ...836...10L} has not been undertaken in a latitude and cycle-dependent manner.  In particular, facular fields that survive cancellation converge by meridional circulation at polar latitudes~\citep[e.g.,][]{2008ApJ...688.1374T}.  There they form large-scale unipolar polar caps, with dynamo implications and global heliospheric influence~\citep{2015LRSP...12....5P}, but our knowledge of the details of their distribution, dynamics and behavior in the polar regions is limited by spatial resolution constraints associated with foreshortening~\citep{2017SoPh..292...13P}.  

The spatial resolution capabilities and polarimetric sensitivity of DKIST are essential in addressing this problem.  The multi-instrument capabilities of DKIST will enable polar field maps at complementary wavelengths (different heights in the atmosphere), over different fields of view and at cadences that maximize the coverage and resolution of long- and short-term polar field evolution.  These will enable the application of local helioseismic and local correlation tracking and structure tracking techniques at high latitudes.  With SDO/HMI (resolution 0.5 arcseconds per pixel), solar meridional and zonal flows can be recovered up to latitudes of about 75 degrees.  DKIST's better than 0.1 arcsec resolution and superior sensitivity will allow application of these techniques at latitudes reaching 90 degrees during March and September when the solar poles are most visible.  Making such high resolution near limb observations over a sufficiently large field of view will be challenging, but achieving them repeatedly over the course of a solar cycle will likely yield crucial insights into the time dependent nature of high latitude meridional and zonal flows, a critical missing piece in our understanding of global flux transport that is highly relevant to its role in the global dynamo process.

Separately, the contribution of small-scale magnetic structures (below the resolution element of current observations) to the total solar irradiance is still not fully understood (discussion in \S\ref{ss:irradiance}).  In particular, the radiative output of magnetic elements, and the spectral distribution of that output, is strongly dependent on the spatial substructure of the elements~\citep{2005A&A...439..323O, 2009A&A...495..621C, 2011ApJ...736...69U}.  Not knowing that substructure introduces significant uncertainty into present day irradiance models~\citep{2019ApJ...870...89P}.  Given that quiet sun covers approximately 90\% of the solar surface and contributes substantially to the disk-integrated magnetic surface flux, changes over the solar cycle in the size distribution or structure of small-scale network and internetwork magnetic elements could play a significant role in inferred irradiance trends~\citep{2009GeoRL..36.7801H, 2013ACP....13.3945E}.  Systematic, long term observations with DKIST, combining the highest spatial resolution imagery with the highest concomitant polarimetric sensitivity, will be essential in addressing this problem. 

Beyond individual magnetic element contributions, field induced variations of the temperature gradient in the deep solar photosphere (below 60 km above the 500 nm continuum) may contribute to cycle-dependent irradiance variability~\citep[e.g.,][and references therein]{2016A&A...595A..71F}.  Structural changes are suggested by observations of the frequencies of the acoustic p-modes, which show solar-cycle variation~\citep[e.g.,][]{1987A&A...177L..47F, 1990Natur.345..779L, 2015A&A...578A.137S}, and are evident in MHD simulations, which display decreasing temperature in the low photosphere with increasing internetwork magnetic field strength~\citep{2013ApJ...778...27C}.  Observational confirmation by direct measurement of the temperature gradient over the solar cycle is needed, particularly at high resolution so that the role of currently unresolved magnetic elements can be elucidated.  The method introduced by~\citep{2016A&A...595A..71F} derives the temperature gradient on an absolute geometrical scale based on spectroscopic observations at different heliocentric angles.  It is well tailored to exploit high resolution observations, suited to ground-based observations~\citep{2018A&A...616A.133F} and it can be readily extended to synoptic DKIST observations in multiple spectral lines.

Synoptic studies of prominences/filaments are also important.  Active region and quiet-sun prominences/filaments participate in and change with the solar cycle~\citep[e.g.,][]{2001ApJ...561..406Z, 2014SSRv..186..285P, 2018ApJ...869...62M}, and make a critical contribution to space weather~\citep[e.g.,][]{2019SpWea..17..498K}.  Filament models usually rely on measurements of the underlying photospheric field and/or emission patterns in the chromosphere, because, though filament magnetic fields have been intermittently measured in the past~\citep[e.g.,][]{1983SoPh...83..135L, 2012ApJ...750L...7X, 2012A&A...539A.131K, 2014A&A...561A..98S, 2019A&A...625A.128D, 2019A&A...625A.129D}, they have not been regularly measured.   We thus have only limited understanding of the range of field properties displayed or the variation of those with the solar cycle; no cycle-length synoptic program to directly measure chromospheric filament magnetic fields has yet been undertaken.  Such a program will be possible with DKIST.  DKIST's multi height capabilities can be used to clarify the three-dimensional helical magnetic structure of prominences/filaments (\S\ref{ss:helicity}) and its evolution, and will be able to supply space weather models with direct field measurements.  Additionally, it may be possible to extend and enhance space weather models and prediction by providing measurements of the underlying near-surface flows~\citep{2006ApJ...653..725H}.  Synoptic, continuous, high-cadence and high-resolution observations for several hours at a time, in conjunction with high-resolution local helioseismological analysis, will allow improved measurements of the local subsurface shear layer that is likely central to filament formation, dynamics and evolution.

More broadly, synoptic DKIST observations of the low solar corona will benefit collaborative science with the Parker Solar Probe and Solar Orbiter missions (\S\ref{ss:insitu} below).  The in-situ measurements these missions provide often require heliospheric field models for context and interpretation.  DKIST's Cryo-NIRSP instrument will be able to make direct coronal spectropolarimetric measurements spanning the spacecraft encounter windows.  These measurements can help provide constraints on the heliospheric field models.  While the best input for those models is still uncertain, and line of sight integration through the optically thin corona will likely pose difficulties, measurements of the forbidden Fe XIII line at 1075 nm will initially be used diagnose the magnetic field.  Measurements will include the Stokes V component, which is accessible to DKIST because of its anticipated exquisite polarization sensitivity and calibration accuracy.

\subsection{Turbulence and Reconnection Processes }
\label{ss:trp}

\vskip -0.15in
\begin{figure}[h] 
\includegraphics[width=0.65\textwidth,clip=]{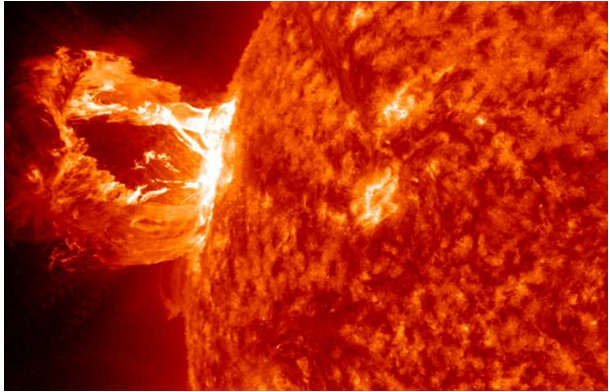}
\caption{Prominence eruption/coronal mass ejection observed on 16 April 2012 with SDO/AIA at 304\AA.  Accompanied by an M1.7  flare at 17:45 GMT.  Courtesy NASA/SDO/AIA.  \url{https://www.nasa.gov/mission_pages/sunearth/news/News041612-M1.7flare.html}.}
\label{f:fig22}
\end{figure}

\noindent
{\it Where is the turbulence and what causes it?  Is there evidence for magnetic island/plasmoid formation during reconnection in the solar atmosphere?  What role do ion-neutral collisions play in these processes?}
\vskip 0.1in

\noindent
The observational capabilities of the DKIST offer the opportunity to study the Sun as a plasma laboratory in order to learn more about the processes underlying reconnection, turbulence and dynamo action in the regime of strong nonlinearity, low molecular diffusivity and partial ionization.

Turbulence is a state of fluid motion that is characterized by unpredictable flow trajectories, a wide range of spatial and temporal scales, and a high degree of vorticity.  The fundamental aim of turbulence research is to understand its properties well enough to be able to predict the transport of scalar and vector quantities.  While, based on an assessment of molecular transport coefficients~\citep[][and references within these]{1979cmft.book.....P, 2005LRSP....2....1M, 2017PhPl...24d2120L}, the solar convection zone is very likely highly turbulent below the photosphere, granulation in the photosphere may not be~\citep{1959Natur.183..240L, 1997A&A...328..229N}.  Solar granulation is dominated by the local dynamics of a strongly radiatively cooled highly stratified boundary layer~\citep[e.g.,][]{1985SoPh..100..209N, 1995ApJ...443..863R, 1998ApJ...499..914S, 2009LRSP....6....2N}.  Upwelling fluid entering the photosphere from below is laminarized by rapid expansion due to the steep mean stratification, and downflowing plumes, initiated in the photosphere, advect flow instabilities out of the readily observable region~\citep{1998JFM...369..125R}.  While the two-dimensional horizontal transport properties of the observed photospheric flows~\citep[e.g.,][]{2011ApJ...743..133A, 2018ApJ...854..118A} are interesting, critical to some dynamo models~\citep[e.g.,][and references therein]{2010LRSP....7....3C}, and may contain clues about the convective driving scales at depth~\citep{2014ApJ...793...24L, 2016ApJ...829L..17C}, at the resolution of current observations these flows show little direct evidence of turbulence.  

It is very possible that at DKIST resolution the granular flows will appear significantly more structured than they do at lower resolution.  Very high resolution Doppler imaging with Imaging Magnetograph eXperiment (IMaX) instrument~\citep{2011SoPh..268...57M} on the first flight of the Sunrise stratospheric balloon~\citep{2010ApJ...723L.127S} revealed that, in small compact regions, granular upflows reach peak speeds approaching those found  in downflows~\citep{2019SoPh..294...18M}.  This substructuring may continue to even smaller scales.  Moreover, with DKIST spatial resolution and sensitivity it may be possible to resolve the flow gradient structure in the deep photosphere~\citep{2010ApJ...723L.159K} and/or the onset of turbulent instabilities at the shear interface between the granular upflows and the intergranular downflow lanes.  In numerical simulations these instabilities lead to recirculation of small scale mixed polarity magnetic field several minutes after new downflow plume formation~\citep{2018ApJ...859..161R}.    Observations of the properties of such recirculating flows and fields may provide key constraints on the relative importance of deep and shallow recirculation to the operation of the Sun's small scale dynamo (\S\ref{ss:smallscale}), and perhaps more broadly address the detailed structure of the photospheric boundary layer with implications for deep convection below~\citep{inprepR}.  Such work may be able to both leverage and augment the rich history of research focused on the Earth's convective planetary boundary layer~\citep[e.g.,][]{1976QJRMS.102..427W, 1976JAtS...33.2152K, 1986JAtS...43.1198L, 2009JAMES...1...16M, 2011JAtS...68.2395S, 2014JAtS...71.3975V}, to understand how the mean boundary layer structure is established at the interface between deep convection and discrete downflow (on Earth upflow) plume structures.  Additionally, though the concept of turbulent spectral line broadening~\citep[e.g.,][]{2005oasp.book.....G} may not be useful in high resolution photospheric observations~\citep{2000A&A...359..729A, 2010ApJ...723L.159K}, it is still used in the interpretation of stellar photospheric lines~\citep[e.g.,][]{2019KPCB...35..129S} as a way to capture unresolved motions.  It may also provide insight in the case of less than "perfectly" resolved solar observations~\citep{2020ApJ...890..138I}, and there is some evidence that a more careful assessment of the solar photospheric velocity field could improve stellar photospheric spectral line modeling~\citep{2017PASJ...69...46T}.  Confident progress toward elimination or accurate representations of non-thermal broadening parameterizations in spectral modeling of solar chromospheric, transition region and coronal lines awaits DKIST resolution of small-scale flow structure~\citep[e.g.,][]{2009A&A...503..577C, 2015ApJ...799L..12D, 2018A&A...612A..28L}.

In the so-called "local dynamo" scenario, convective turbulence in the quiet Sun is responsible for the creation and structuring of weak small-scale magnetic field (\S\ref{ss:smallscale}), with convective motions at the granular and subgranular scales determining the topology of the field and its degree of "entanglement."  The Hanle effect is a key tool for investigating this aspect of solar magnetism because depolarization of scattered photons is sensitive to the presence of small-scale tangled magnetic fields~\citep[e.g.,][]{2004Natur.430..326T}.  The Zeeman effect, by contrast, is blind to mixed polarity field on scales much smaller than those that can be resolved.  Since the amplitude of the Hanle scattering polarization signal is very low, the limited resolution and polarimetric sensitivity of current observations only allows determination of an upper limit to its value and spatial variation~\citep[e.g.,][]{2018A&A...619A.179Z}.  With the increased resolution and polarimetric sensitivity of DKIST, the degree to which the Zeeman and Hanle measurements differ will become a critical measure of the scale at which the field is generated.  However, fundamental advances in our understanding of the Hanle signal are required.  For example, interpretation of the scattering polarization signal at each wavelength depends on knowledge of the anisotropy of the illuminating radiation field.  This difficulty can be overcome either by using realistic radiative hydrodynamic models of the solar photosphere to constrain the radiation field~\citep{2004Natur.430..326T} or by employing differential measures of two or more spectral lines with similar formation properties and varying Hanle sensitivity~\citep[e.g.,][]{2010A&A...524A..37K, 2011A&A...536A..47K}.  The broad spectral coverage offered by the ViSP spectropolarimeter will enable the simultaneous monitoring of multiple lines with differing Hanle sensitivities, contributing to a deeper understanding of the processes that can affect scattering polarization measurements, and allowing the development of multiline scattering polarization inversion methodologies.  The consequent ability to deduce both the strength and direction of the weak fields pervading solar photosphere, and subsequent systematic monitoring of this "turbulent magnetic field" over a solar cycle, will yield evidence as to whether that field is of a local or larger-scale origin.

Outside of the photosphere, pre-DKIST evidence for turbulence has been found in the vicinity of chromospheric shocks which result from the steepening of acoustic waves as they propagate upward from the photosphere~\citep{2008ApJ...683L.207R, 2015ApJ...799L..12D}.  The generation of post-shock turbulence may provide a mechanism for the dispersal of the wave energy beyond the local shock region itself, and may thus be important for chromospheric heating.  Because the turbulent region is permeated by magnetic field, it may also play a role in wave-mode conversion, coupling the acoustic waves to Alfv\'enic motions that continue to propagate outward, transmitting energy to higher layers of the solar atmosphere~\citep{2008ApJ...683L.207R}.  While direct investigation of shock heating by plasma processes at the dissipative scale is out of reach, DKIST may be able to probe the larger-scale properties of the shock region to infer the MHD shock type~\citep[e.g.,][]{2011JPlPh..77..207D, 2014masu.book.....P} and the shock instability processes.  A number of shock instability mechanisms have been identified in other settings, MHD wave breaking~\citep{1987GeoRL..14.1007M}, radiative instability of the shock front~\citep[e.g.,][]{1989MNRAS.238..235S, 2005ApJ...626..373M} and shock front distortions due to  plasma inhomogeneities~\citep[][and references therein; ~\citeauthor{2018ITPS...46.2821M},~\citeyear{2018ITPS...46.2821M}]{2002AnRFM..34..445B, 2017PhR...723....1Z}.  In order for any of these to occur in the chromosphere, not only must the instability mechanism be feasible, but it must occur on a time scale short compared to shock-damping rates~\citep{1987ApJ...312..880H, 1988GeoRL..15..471L}.  Only preliminary numerical explorations of shock interactions within the complex solar atmosphere have been undertaken~\citep{2016A&A...590L...3S, 2019A&A...627A..25P, 2019A&A...630A..79P}.  Idealized studies of shock-instability mechanisms in a magnetized and partially ionized plasma constrained by observations at DKIST's spatial and temporal resolution, should lead to a more complete understanding of the physical process responsible for shock-induced turbulence in the solar chromosphere.  This in turn will help to constrain the underlying plasma processes that result in chromospheric heating and particle acceleration.  

Magnetic reconnection is similarly a process that occurs at fundamental scales well below DKIST resolution, but to the understanding of which DKIST can contribute.  Observations of macroscale flows associated with a canonical reconnection configuration in a flaring active region are quite convincing~\citep{2017ApJ...847L...1W}, but smaller scale evidence that can be used to constrain the physical processes involved are more elusive.  Many studies, starting with the pioneering work of~\cite{1983SoPh...84..169F, 1983JGR....88..863F}, indicate that magnetic reconnection is enhanced as a result of magnetic island formation in the plasma current sheet.  Magnetic island formation appears to be ubiquitous, occurring with or without ion-neutral collisions and with or without a guide field (which tends to reduce the importance of ambipolar diffusion)~\citep[e.g.,][]{2007PhPl...14j0703L, 2009PhPl...16k2102B, 2012ApJ...760..109L, 2012MNRAS.425.2824M, 2012ApJ...751...56M, 2015ApJ...799...79N, 2015ApJ...813...86I, 2019A&A...628A...8P}.  Further, while the magnetic islands likely lie well below current and future observational capabilities, bright localized plasmoid-like ejecta (sometimes distinguished as blobs) have been reported in post coronal mass ejection current sheets in the solar corona~\citep[e.g.,][]{2007ApJ...655..591R, 2007ApJ...661L.207W, 2012ApJ...745L...6T, 2013ApJ...771L..14G} and at smaller scales in chromospheric jets and Ellerman bomb events~\citep[e.g.,][]{1995Ap&SS.229..325H, 2007Sci...318.1591S, 2012ApJ...759...33S, 2017ApJ...851L...6R}.  It may be possible to use the statistics of these macroscale events to constrain the reconnection magnetic field and flow configuration and the underlying reconnection processes~\citep[e.g.,][]{2007ApJ...658L.123L,  2008JGRA..11311107L, 2012SoPh..276..261S, 2013ApJ...771L..14G, 2017ApJ...851L...6R}.  DKIST spatial and temporal resolution will be essential for this.  Further, recent work suggests that, while magnetic island substructure in the reconnection current sheet is not resolvable, it may be possible to assess the overall complexity of the underlying unresolved field by measuring a reduction in linear polarization~\citep{2019ApJ...887L..34F}.

\vfill\eject
\subsection{Sun-Grazing Comets}
\label{ss:comets}

\vskip -0.25in
\begin{figure}[h] 
\includegraphics[width=0.75\textwidth,clip=]{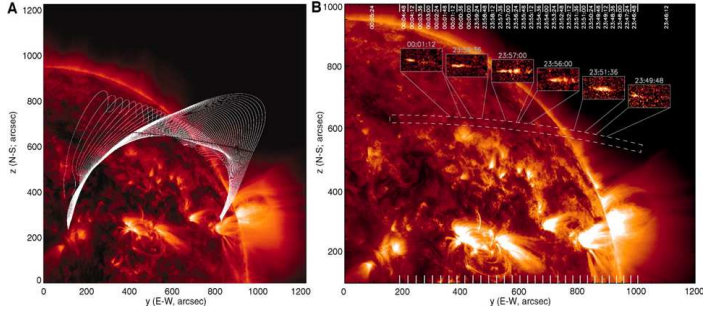}
\caption{(A) EUV (AIA 171\AA) image of the solar corona, overlaid in black with the projected orbit of the comet C/2011 N3.  Orbital positions marked by plus signs were used as starting points in a three-dimensional potential-field source-surface extrapolation of the Sun's magnetic field, shown in white.  (B)  Composite of AIA 171\AA images of the comet moving within the dashed region outlined. The six insets show an enlarged view of the comet at selected times in running difference images.  From~\cite{2012Sci...335..324S}.}
\label{f:fig23}
\end{figure}

\noindent
{\it What can we learn from Sun-grazing comets about the solar corona and cometary composition and structure?  Are Sun-grazing comets a significant source of solar wind pickup ions?}
\vskip 0.1in

\noindent
Comets are among the most pristine bodies within the solar system.  They provide critical clues about our solar system's formation and the origin of life on Earth.  Typically, $0.3-5$km in radius, comets are composed of a mixture of icy, organic and silicate materials.  Sun-grazing comets, those with perihelion distances of less than a few solar radii ~\citep[$ < 3.45 R_\odot$ from the Sun's center, within the fluid Roche limit, ][]{2018SSRv..214...20J}, are valuable tools in both cometary and coronal studies.  The intense solar radiation during their close perihelion passages evaporates thick layers of near-surface material, exposing their otherwise invisible pristine interiors, and their high-speed intrusion into the million-degree magnetized solar corona, in extreme cases skimming or plunging into the solar surface~\citep{2015ApJ...807..165B}, makes them natural probes of regions of the solar atmosphere inaccessible to human-made in-situ instruments.  DKIST is well poised to play a unique role in such studies over the coming decades~\citep{commR}.

Over the past two decades, solar space missions have contributed significantly to the study of Sun-grazing comets.  The LASCO white-light coronagraph onboard SOHO has observed more than 3200 Sun-grazing and near-Sun comets, with an average occurrence rate of one every two-to-three days~\citep{2017RSPTA.37560257B}.  A key advance over this past decade has been the detection of Sun-grazing comets, notably comets C/2011 N3 (SOHO) and C/2011 W3 (Lovejoy), in the lower solar corona, a region usually blocked by a coronagraph's occulting disk, using (E)UV instruments onboard SDO, STEREO, and SOHO.  Observations of these deeply penetrating comets have provided unique diagnostics of the coronal plasma and magnetic fields~\citep[e.g.,][]{2012Sci...335..324S, 2014ApJ...788..152R, 2013Sci...340.1196D}.  Ground-based solar telescopes have also observed Sun-grazing comets, including C/2012 S1 (ISON), which was observed with  NSO's Dunn and 
McMath-Pierce Solar Telescopes~\citep{2013AGUFM.P24A..07W} and with the Mees Observatory coronagraph on the summit of Haleakala, Maui, HI~\citep{2014ApJ...784L..22D}.  

Unlike virtually all other remote sensing diagnostics of the solar corona, which are subjected to either line-of-sight integration or height ambiguity, Sun-grazing comets take a very localized path through the corona, serving as probes.  The comets interact with the plasma producing observable signatures along specific paths through the three-dimensional corona.  To date, this has largely been exploited at UV and EUV wavelengths.  For example, O III and O VI emission from photo-dissociated cometary water has been used to diagnose the magnetic field direction and plasma density in the corona~\citep{2014ApJ...788..152R}.  Using DKIST's Cryo-NIRSP coronographic capabilities, comparable analysis may be possible by observing lines of photo-dissociated silicates such as Si IX.  Similarly, observations of Lyman-alpha emission during passage of Sun-grazing comets has been used to estimate the coronal density, temperature and solar wind velocity~\citep{2015AdSpR..56.2288B}.  The same techniques may be possible with DKIST using H-alpha or Paschen-alpha emission.  Moreover, ion tails have been seen in white light accompanying a few Sun grazing comets~\citep{2018SSRv..214...20J}, and their presence in exocomets is inferred from their signatures of Ca II absorption~\citep{2014Natur.514..462K}.  Observations of comet tail heliospheric current sheet interactions during cometary crossings may provide constraints on the current sheet morphology and the solar wind structure in the inner heliosphere~\citep[see][and references therein]{2018SSRv..214...20J}.  Such observations by DKIST would be groundbreaking. 

Beyond diagnostics of the solar corona, DKIST promises to contribute directly to the cometary science.  Most fundamentally, it will enable measurements of the size and composition of  cometary cores~\citep{2016ApJ...822...77B} as they are exposed very close to the Sun, too close for observation using night-time telescopes.  While some Sun-grazing comets, such as Lovejoy and ISON are discovered at great distances from the Sun and followed to their perihelia, many are not active enough to be observed at large distances and are first noticed close in to the Sun, within the field of view of the LASCO coronagraph.  Close to the Sun, cometary material  sublimates rapidly and is photo dissociated to form atomic species which are then ionized through successive ionization states~\citep{2012ApJ...760...18B, 2013ApJ...768..161M}, with optical and infrared line emission is largely confined to a small region surrounding the cometary nucleus.  The time-dependent emission allows characterization of both the cometary material and the local coronal plasma environment.  DKIST's unique capabilities will make these otherwise very difficult spectrographic measurements of pristine cometary material possible.  Moreover, DKIST infrared observations will allow measurements of cometary dust temperatures and dust sublimation rates.  These are particularly important in advancing our understanding of the origin of cometary neutral tails~\citep[e.g.,][]{2002AdSpR..29.1187C}, and may be critical in determining the inner source of pickup ions in the solar wind~\citep[e.g.,][]{2005A&A...435..723B}.  Finally, previous attempts to determine the tensile strength (if any) of Sun-grazing comets from their breakup~\citep{1966IrAJ....7..141O, 1989ESASP.302..197K} have been inconclusive.  This is an important measurement in the context of planetesimal formation, and DKIST will bring much higher spatial resolution to bear on the assessment of cometary fragmentation processes.

\subsection{Mercury Transit Science}
\label{ss:mercury}

\vskip -0.25in
\begin{figure}[h] 
\includegraphics[width=0.4\textwidth,clip=]{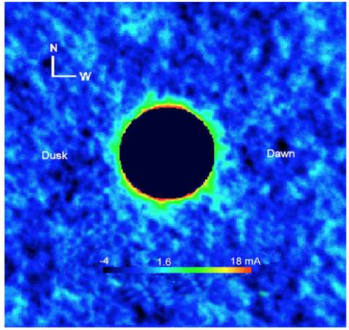}
\caption{Sodium absorption equivalent width around the disk of Mercury during the 8 November 2006 transit.  The noise outside the close region of sodium absorption is due to fluctuations in the solar intensity.  The strongest atmospheric absorption is found over Mercury's north and south poles.  Likely due to seasonal variations, the dawn/dusk terminator difference identified in the previous transit~\citep{2004A&A...425.1119S} is not seen during this one.  From~\cite{2013Icar..226..172P}.}
\label{f:fig24}
\end{figure}

\noindent
{\it What can we learn about Mercury's atmosphere and its seasonal variations from DKIST observations of Mercury's transit across the Sun?  Can the resonance absorption lines of K and Ca be observed at DKIST sensitivities?}
\vskip 0.1in

\noindent
There was some hope that early DKIST observations might overlap with the 11 November 2019 Mercury transit.  Unfortunately, it was not possible to meet that aggressive goal which lay outside of the nominal DKIST construction timeline.  We include a brief description of the science goals of Mercury transit observations here to illustrate the scientific flexibility of the DKIST observing system and to point to future possibilities within the DKIST lifetime.  The next partial Mercury transits visible from Haleakala are one ending in the early morning of 7 May  
2049 and one occurring midday to sunset on 8 November 2052. (\url{https://eclipse.gsfc.nasa.gov/transit/catalog/MercuryCatalog.html}, 
\hfill\url{https://www.timeanddate.com/eclipse/in/usa}).

Mercury has a non-spherical seasonally varying exosphere~\citep[e.g.,][]{2007SSRv..131..161D, 2008Sci...321...92M, 2015Icar..248..547C, 2016GeoRL..4311545V, 2017Icar..281...46M} and a dynamic magnetosphere with dayside reconnection and magnetotail activity resembling that found at Earth~\citep{2009Sci...324..606S, 2010Sci...329..665S, 2013pss3.book..251B}.  Full characterization of this unique atmosphere is challenging.  Traditional remote sensing techniques are difficult due to the Sun's proximity, and downlink limitations led to NASA's MESSENGER orbiter carrying a single point rather than slit scanning spectrometer.  To date, observations of absorption by Mercury's atmospheric constituents during rare solar transits have offered some of the highest spatial, temporal and spectral resolution measurements~\citep[Figure 30; ][]{2004A&A...425.1119S, 2013Icar..226..172P, 2018LPICo2047.6022S}.  So far only sodium in Mercury's atmosphere has been measured during transit, but with DKIST's sensitivity, observations of the K and Ca atomic resonance absorption lines may be possible.  The distribution of these in the Mercury atmosphere would constrain source and loss processes~\citep{2012JGRE..117.0L11B}.  Broad-band absorption measurements would additionally contribute to our understanding of the dust density distribution around the planet.  

The scientific use of Mercury transit measurements has progressed rapidly despite the rarity of transit events.  Measurements of Na I absorption by transit spectroscopy were first made the early years of this century~\citep{2004A&A...425.1119S}.  These provided convincing evidence for a dawn-side enhancement in Mercury's exosphere.  Subsequent measurements by~\cite{2013Icar..226..172P} were markedly different, showing no dawn/dusk terminator difference (Figure~\ref{f:fig24}).  That is now understood to be the result of seasonal variations.  These were also identified in MESSENGER orbiter data~\citep{2017Icar..281...46M}, and confirmed by later transit observations~\citep{2018LPICo2047.6022S}.  The 2018 transit measurements indicate nearly identical Na distribution as the earlier 2004 observations taken during the same Mercury season.  Further, the quality of the 2018 observations allowed measurements that spatially resolved the Mercury atmospheric scale height (~100 km) and enabled study of the exospheric time-dependence induced by solar wind interactions.  During future transits, DKIST capabilities will enable spectral analysis of Doppler velocities, ~90 km resolution of the atmospheric stratification, and temporal resolution of solar wind and interplanetary magnetic field angle influences.

\subsection{Synergistic Opportunities with In-Situ Measurements }
\label{ss:insitu}

\vskip -0.2in
\begin{figure}[h] 
\includegraphics[width=0.75\textwidth,clip=]{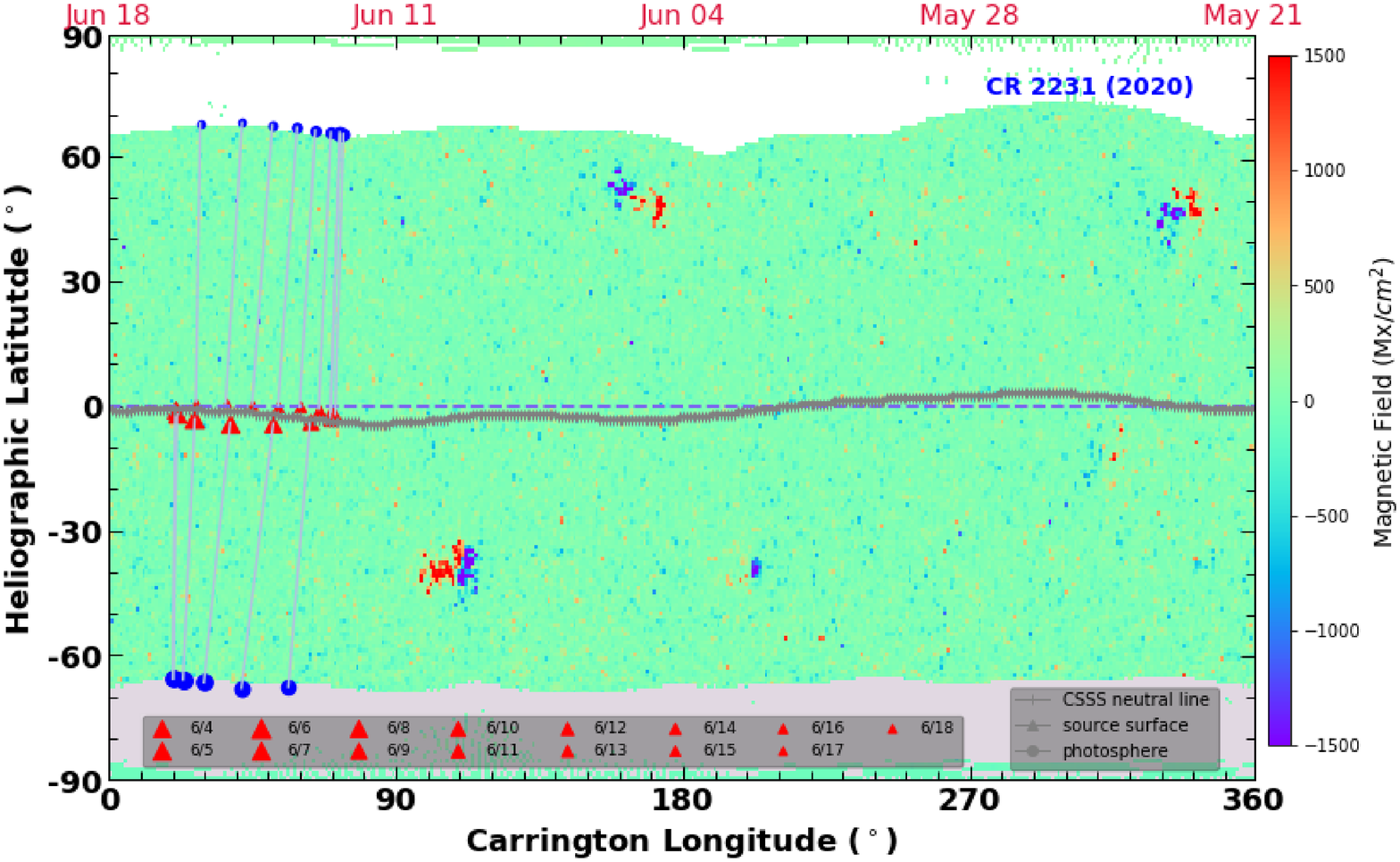}
\caption{The locations of Parker Solar Probe (4 - 18 June 2020), mapped back to the Sun using the Current Sheet Source Surface model~\citep[CSSS, developed by ][see also~\citeauthor{2014ApJ...782L..22P}, \citeyear{2014ApJ...782L..22P},~\citeauthor{2016ApJ...827L...6P}, \citeyear{2016ApJ...827L...6P},~\citeauthor{cha20}, \citeyear{cha20}]{1995JGR...100...19Z}, and superimposed on an SDO/HMI synoptic magnetogram (Carrington rotation 2231, downsampled to 1 degree resolution).
The {\it red triangles} are the map back locations at 15 $R_\odot$ (the CSSS source surface) assuming a constant radial solar wind speed of 395 km/s.  These are joined to their photospheric footpoints ({\it blue filled circles}) by {\it thin gray lines} to indicate connectivity.  
The CSSS deduced open magnetic field regions (coronal holes) are shown in {\it white} and {\it gray} in Northern and Southern hemispheres respectively, with the magnetic polarity inversion line (heliospheric current sheet) marked with {\it dark gray plus} symbols. 
Figure courtesy~\cite{inprepP}.}
\label{f:fig25}
\end{figure}

\noindent
{\it How well can the magnetic connectivity between the Sun and the in-situ measurements of Parker Solar Probe and Solar Orbiter be established?  Which aspects of the in-situ measurements originate at the Sun and which result from subsequent dynamics and instabilities in the solar wind?}
\vskip 0.1in

\noindent
DKIST, Parker Solar Probe and Solar Orbiter will together allow unprecedented synergistic study of the connectivity between the solar corona and the inner heliosphere.  Parker Solar Probe and Solar Orbiter will make in-situ measurements of the inner heliospheric electric and magnetic fields and plasma kinetic properties, while DKIST and Solar Orbiter will image and make high precision spectropolarimetric measurements of the solar atmosphere from the deep photosphere to the corona.  The combined capabilities of these assets will revolutionize our understanding of how stars create and control their magnetic environments~\citep{2020arXiv200408632M}.  

Parker Solar Probe's close approaches to the Sun and Solar Orbiter's inclined orbit with perihelia inside the orbit of Mercury
will allow direct sampling of the solar wind plasma before it has undergone extensive evolution and mixing.  The plasma sampled will preserve many of the signatures of its origins, allowing assessment of the acceleration mechanisms underlying the different solar wind components.  Moreover, the proximity of Parker Solar Probe's perihelia to the Sun allows periods of co-rotation with the solar surface and more importantly with the magnetic field that originates there.  This will enable studies of the relationship between temporal variations in the wind and short term changes at the source.  Separating these from the variations due to the spacecraft motion across solar wind structures is critical to understanding both the source behavior and the secondary development of the wind itself.  Similarly, periods of quasi-corotation by Solar Orbiter will enable extended observations of the same solar region~\citep{2013SoPh..285...25M}.   These periods will help connect solar activity evolution to changes in the wind over somewhat longer time periods.  Finally, Parker Solar Probe will spend about fifteen hours below ten solar radii, during which it will likely sample the sub-Alfv\'enic solar wind, allowing direct assessment of the conditions under which solar wind heating and acceleration likely occur~\citep[e.g.,][]{2005ApJS..156..265C}.   These scientific goals all require, or strongly benefit from, knowing how the regions of in-situ
heliospheric  measurements are connected to the magnetic field and particle source regions in the solar atmosphere.  
An example mapping is shown in Figure~\ref{f:fig25}.  It assumes a radial solar wind of constant speed between PSP and 15 $R_\odot$, and uses the Current Sheet Source Surface model~\citep[CSSS, developed by ][see also~\citeauthor{2014ApJ...782L..22P}, \citeyear{2014ApJ...782L..22P},~\citeauthor{2016ApJ...827L...6P}, \citeyear{2016ApJ...827L...6P},~\citeauthor{cha20}, \citeyear{cha20}]{1995JGR...100...19Z} to determine the corresponding photospheric footpoints.  Model improvements above the source surface, employing solar wind measurements from PSP/SWEAP, are anticipated as that data becomes publicly available~\citep[][]{inprepP}, and 
DKIST's ability to quantitatively map the magnetic field in the solar chromosphere and  low solar corona will significantly improve the0 ability to map the connectivity from there to source regions on the Sun (\S\ref{ss:wind}, \ref{ss:multilayer}).  Both field extrapolation~\citep[e.g.,][]{2020ApJS..246...23B} and magnetohydrodynamic modeling methods~\citep{2019ApJ...872L..18V, 2019ApJ...884...18R}, which currently employ moderate-resolution synoptic photospheric magnetograms, will significantly benefit from high-resolution multi-height DKIST data. 

Moreover, the chemical composition of the solar wind is a key indicator of its origin~\citep[e.g.,][]{1995SSRv...72...49G, 2000A&A...363..800P, 2011ApJ...727L..13B, 2015NatCo...6.5947B, 2015ApJ...802..104B}, and the First Ionization Potential (FIP) bias can be used to help trace that origin and establish magnetic connectivity with in-situ measurements (\S\ref{ss:masscycle}).  There is evidence that active regions emerge with photospheric abundances and develop a FIP bias in the chromosphere, which then propagates into the corona~\citep{2015LRSP...12....2L}.  This is supported by an observed lag between magnetic activity indices and abundance fluctuations in the solar wind at 1 AU~\citep[WIND spacecraft at L1; ][]{2019ApJ...879L...6A}).  With the synergistic capabilities of DKIST and Solar Orbiter the underlying causes for these correlations can be more directly examined.  In-situ instrumentation on Solar Orbiter is designed to measure abundance ratios, and the FIP bias measured in situ combined with chromospheric observations using DKIST will be used to more precisely assess the plasma's origin.  Similarly, there is a relationship between in-situ measurements of helium at 1 AU, solar activity and solar wind speed~\citep{2001GeoRL..28.2767A, 2007ApJ...660..901K, 2012ApJ...745..162K, 2019ApJ...879L...6A}.  The lowest helium abundances are observed during solar minimum and are correlated with regions of slower wind speed.  The helium abundance in the fast solar wind changes little with solar cycle.  The mechanisms underlying these correlations are unknown, but the combination of in-situ measurements and DKIST's unique capabilities will, again, allow new ways to investigate this fundamental problem.  Previous observations indicate that the depletion likely occurs below the solar corona~\citep{2001ApJ...546..552L, 2003ApJ...591.1257L}, and models suggest that the process is sensitive to the partially ionized state of the chromospheric plasma~\citep{2015LRSP...12....2L}.  Parker Solar Probe will make helium abundance measurements close to the Sun, reducing uncertainties in source region identification.  DKIST will not only assist with that mapping but enable key chromospheric observations to address the underlying helium depletion processes.

The dust content of the inner solar corona is also a question of significant importance.  It has practical importance both for the survival of the Parker Solar Probe and for coronal spectropolarimetric measurements using DKIST.  Observing the white-light scattered by dust (F-corona) and electrons (Thomson scattering), the WISPR camera on the Parker Solar Probe images the large-scale structure of the corona before the spacecraft passes through it.  As the orbit perihelion is reduced, WISPR will help determine whether a dust-free zone exists near the Sun.  Moreover, during the innermost perihelion passage of 2024, the boundary of the WISPR field-of-view will extend down to two solar radii above the solar photosphere, close to the outermost height that will be observable by DKIST (1.5 solar radii, 0.5 solar radius above the limb).  This proximity of the fields of view will provide a unique opportunity to test whether the diffuse coronal He I 1083 nm brightness reported by~\cite{1996ApJ...456L..67K, 2007ApJ...667L.203K} can be accounted for by helium neutralization on the surface of dust particles within the hot corona, as proposed by~\cite{2010ApJ...722.1411M}.  That hypothesis has been difficult to test, but it is important because the He I 1083 nm line is the only permitted infrared transition available to DKIST for spectropolarimetric observations of the corona.  

If the neutral helium signal indeed originates within the corona, new coronal Hanle magnetic field diagnostics are possible~\citep{2016FrASS...3...13D}.  Combined with linear polarization measurements in the Si X 1430 nm forbidden line, which is in the saturated Hanle regime under coronal conditions, polarization observations of the He I 1083 nm permitted line would enable inference of all three components of the coronal magnetic field.  This would critically constrain the coronal magnetic topology, and in turn would allow improvements in the accuracy of the coronal-heliospheric models used to predict the heliospheric magnetic configuration.  Such improvements are particularly important during periods in which the Sun is more active and the corona is consequently more complex than it is currently~\citep[e.g.,][]{2016FrASS...3...20R}, periods such as those likely to be encountered during the later perihelia of Parker Solar Probe.  DKIST's synoptic effort (\S\ref{ss:synoptic} above) to regularly measure the solar coronal magnetic fields in anticipation of Parker Solar Probe encounters will thus contribute to their scientific success.  The lines and techniques employed will likely evolve to generate the best input data for the coronal models, but in addition to the anticipated Si X 1430nm/He I 1083nm observations motivated above, early observations will include the forbidden Fe XIII line at 1075 nm, as the current HAO/CoMP instrument, but with the added capability of regularly measuring the Stokes V component.  That component, accessible to DKIST because of its anticipated polarization sensitivity, will allow robust inference of the line-of-sight magnetic field.  

Another important problem, the solution to which the synergistic capabilities of DKIST, Parker Solar Probe and Solar Orbiter may significantly contribute, is the so-called open flux problem~\citep{2017ApJ...848...70L, 2019ApJ...884...18R}.  A significant portion of the solar surface magnetic field opens out into the heliosphere, forming coronal holes.  This occurs where facular fields of a dominant polarity cover a large area of the solar surface, usually at the polar caps but also sometimes at low latitudes, and the solar wind ram pressure is sufficient to open much of the field at height.  The open flux problem describes the disparity between the total open magnetic flux at the Sun, as estimated from moderate-resolution magnetograms of coronal hole regions (defined as regions dark in EUV and x-ray emission), and the total open flux measured in situ at 1 AU~\citep{2017ApJ...848...70L}.  The former value falls significantly below the later.  This discrepancy is critical to solar wind and heliospheric modeling, which relies on the solar value, as determined from surface observations, for a boundary condition.  The mismatch in measured values implies either that a significant amount of open flux at the Sun lies below the sensitivity of current instrumentation, that the polar magnetic flux for example is significantly underestimated due to difficulties observing it~\citep{2019ApJ...884...18R}, or that the open flux measured at 1 AU does not map exclusively to coronal holes.  The later possibility connects the open flux problem to that of the origin of the slow solar wind, supporting suggestions that the slow solar wind arises from mixed open and closed field regions, possibly at the coronal hole boundaries~\citep[e.g.,][and references therein]{2017ApJ...848...70L, 2020SoPh..295...37O}.  However, it is currently uncertain what the true magnitude of the open flux problem is, whether it persists if the in-situ measurements are made in the inner heliosphere, or whether its origin lies with field reconfiguration in the solar wind inward of 1 AU.  Parker Solar Probe and Solar Orbiter field measurements will address the in-situ uncertainties directly, within the limitations of single-point measurements~\citep[][and references therein]{2008JGRA..11312103O}, and observations with the spatial resolution and polarization sensitivity available to DKIST will reveal the amount of small-scale open flux that currently remains undetected in less sensitive full-disk solar magnetograms.  Together these will either reconcile the in-situ and remote sensing open flux deductions or elucidate the magnitude and perhaps the source of the discrepancy.

Finally, one of the most prominent early results of the Parker Solar Probe mission has been observations of magnetic switchbacks~\citep{2019Natur.576..237B, 2019Natur.576..228K, 2020ApJS..246...39D, 2020ApJS..246...45H, 2020ApJS..246...68M}, rapid changes (over intervals ranging from seconds to tens of minutes, perhaps hours) in the radial magnetic field orientation away from and then back to its original orientation, with field deflections sometimes representing full reversal with respect to the Parker spiral. Such field switchbacks are accompanied by rapid enhancement of the radial wind speed, and they are often called velocity spikes for that reason.  The correlation between the magnetic field perturbations and jet like flows suggests that these events are large-amplitude Alfv\'enic structures being advected away from the Sun by the solar wind~\citep{2019Natur.576..237B}.  Their origin remains uncertain.  Clustering and correlation statistics of their occurrence during the first Parker Solar Probe perihelion encounter (10 day period, centered on perihelion at ~36 $R_\odot$) suggests that they are remnants of stronger events in the low corona~\citep{2020ApJS..246...39D}, but comparison between perihelion (one day period at ~36 $R_\odot$) and pre-perihelion (one day period five days earlier at ~48 $R_\odot$) intervals during the second encounter suggests the number of switchbacks and the magnetic field rotation angle of the field within the switchbacks increases with increasing distance from the Sun~\citep{2020ApJS..246...68M}.  Theoretical work is also inconclusive, indicating that switchback like structures which originate in  the lower solar corona can indeed survive in the solar wind out to Parker Solar Probe distances~\citep{2020ApJS..246...32T}, but also that such structures can form in the expanding solar wind itself as growing Alfv\'enic fluctuations ~\citep{2020ApJ...891L...2S}.  Sudden reversals of the magnetic field and associated jet like flows are not new to Parker Solar Probe.  They have been observed over several decades in earlier in-situ measurements~\citep[see references within][]{2020ApJS..246...45H, 2020ApJS..246...39D}.  What separates the Parker Solar Probe observations from those earlier measurements is the occurrence of switchbacks in relatively slow solar wind environments and the large number and magnitude of the events encountered.  Thousands of such events have been observed by the Parker Solar Probe, some in tight clusters separated by relatively event-free regions.  This clustering, the significant correlation of magnetic field deflection observed for clustered events, and the similarity of the flows observed to those deduced for other coronal jets all support a low coronal origin~\citep{2020ApJS..246...45H}.  This has led to the suggestion that they are folds in the magnetic field that originate as ubiquitous interchange reconnection events lower down, possibly driven by the global circulation of open flux~\citep{2020ApJ...894L...4F}.  DKIST's chromospheric and low corona observing capabilities will contribute to the assessment of the origin of switchbacks, both by improving the determinations of the connectivity to the source regions, as discussed earlier, and by directly looking for evidence of interchange reconnection heating or jets in the source regions.

%

%

%
\begin{acks}
This work rests on many years of sustained effort by NSO scientists and the DKIST Project and Instrument Teams.  It includes contributions from previous members of the DKIST Science Working Group and the DKIST Critical Science Plan Community, all of whom generously shared their experiences, plans, 
knowledge, and dreams.
It is impossible at this stage to fully acknowledge each individual contribution without error or error of omission, but all of them are greatly appreciated.  A living CSP document, including a list of people involved, can be found at \url{https://www.nso.edu/telescopes/dki-solar-telescope-2-2/csp/}.  If your name has been inadvertently left off the contributors list, please accept our apologies and contact the lead author of this paper to have it added.   
\end{acks}

%
%
\bibliographystyle{spr-mp-sola}
\bibliography{dkistcsp}  
%
%
%
%

\end{article} 
\end{document}